\documentclass[aps,super,
twocolumn,
reprint,
groupedaddress,superscriptaddress,
amsfonts,amssymb,amsmath,floatfix,
showkeys,showpacs,a4paper]{revtex4-2}

\usepackage{graphicx}
\usepackage[english]{babel}
\usepackage{hyperref}
\usepackage{upgreek}
\usepackage{amsmath}
\usepackage{natbib}
\usepackage{lineno}
\usepackage{color}
\usepackage{multirow}
\usepackage{gensymb}
\usepackage{csquotes}

\begin{document}

\title{Comprehensive investigation of crystallographic, spin-electronic and magnetic structure of (Co$_{0.2}$Cr$_{0.2}$Fe$_{0.2}$Mn$_{0.2}$Ni$_{0.2}$)$_3$O$_4$: Unraveling the suppression of configuration entropy in high entropy oxides}

\author{Abhishek Sarkar}\email{abhishek.sarkar@kit.edu}
\affiliation{KIT-TUD-Joint Research Laboratory Nanomaterials, Technical University Darmstadt,  64287 Darmstadt, Germany}
\affiliation{Institute of Nanotechnology, Karlsruhe Institute of Technology, 76344 Eggenstein-Leopoldshafen, Germany}

\author{Benedikt Eggert}
\affiliation{Faculty of Physics and Center for Nanointegration Duisburg-Essen (CENIDE), University of Duisburg-Essen, Lotharstr. 1, 47048 Duisburg, Germany}

\author{Ralf Witte}
\affiliation{Institute of Nanotechnology, Karlsruhe Institute of Technology, 
 76344 Eggenstein-Leopoldshafen, Germany}

\author{Johanna Lill}
\affiliation{Faculty of Physics and Center for Nanointegration Duisburg-Essen (CENIDE), University of Duisburg-Essen, Lotharstr. 1, 47048 Duisburg, Germany}

\author{Leonardo Velasco}
\affiliation{Institute of Nanotechnology, Karlsruhe Institute of Technology, 76344 Eggenstein-Leopoldshafen, Germany}

\author{Qingsong Wang}
\affiliation{Institute of Nanotechnology, Karlsruhe Institute of Technology, 
 76344 Eggenstein-Leopoldshafen, Germany}
 
\author{Janhavika Sonar}
\affiliation{Institute of Nanotechnology, Karlsruhe Institute of Technology, 76344 Eggenstein-Leopoldshafen, Germany}
\affiliation{Nano Functional Material Technology Centre (NFMTC), Department of Metallurgical and Materials Engineering, Indian Institute of Technology Madras, 600036 Chennai, India}

\author{Katharina Ollefs}
\affiliation{Faculty of Physics and Center for Nanointegration Duisburg-Essen (CENIDE), University of Duisburg-Essen, Lotharstr. 1, 47048 Duisburg, Germany}

\author{Subramshu S. Bhattacharya}
\affiliation{Nano Functional Material Technology Centre (NFMTC), Department of Metallurgical and Materials Engineering, Indian Institute of Technology Madras, 600036 Chennai, India}

\author{Richard A. Brand}
\affiliation{Institute of Nanotechnology, Karlsruhe Institute of Technology, 
 76344 Eggenstein-Leopoldshafen, Germany}
\affiliation{Faculty of Physics and Center for Nanointegration Duisburg-Essen (CENIDE), University of Duisburg-Essen, Lotharstr. 1, 47048 Duisburg, Germany}

\author{Heiko Wende}
\affiliation{Faculty of Physics and Center for Nanointegration Duisburg-Essen (CENIDE), University of Duisburg-Essen, Lotharstr. 1, 47048 Duisburg, Germany}

\author{Frank M.F. de Groot}
\affiliation{Inorganic Chemistry and Catalysis, Utrecht University, Universiteitsweg 99, 3584 CG Utrecht, The Netherlands}

\author{Oliver Clemens}
\affiliation{Institute for Materials Science, University of Stuttgart, Heisenbergstr. 3
70569 Stuttgart, Germany}
\affiliation{KIT-TUD-Joint Research Laboratory Nanomaterials, Technical University Darmstadt,  64287 Darmstadt, Germany}

\author{Horst Hahn}\email{horst.hahn@kit.edu}
\affiliation{KIT-TUD-Joint Research Laboratory Nanomaterials, Technical University Darmstadt,  64287 Darmstadt, Germany}
\affiliation{Institute of Nanotechnology, Karlsruhe Institute of Technology, 
 76344 Eggenstein-Leopoldshafen, Germany}
 
 \author{Robert Kruk}\email{robert.kruk@kit.edu}
\affiliation{Institute of Nanotechnology, Karlsruhe Institute of Technology, 
 76344 Eggenstein-Leopoldshafen, Germany}

%

\begin{abstract}

High entropy oxides (HEOs) are a rapidly emerging class of functional materials consisting of multiple principal cations. The original paradigm of HEOs assumes cationic occupations with the highest possible configurational entropy allowed by the composition and crystallographic structure. However, the fundamental question remains on the actual degree of configurational disorder in HEOs, especially, in systems with a low enthalpy barriers for cation anti-site mixing. Considering the experimental limitations due to the presence of multiple principal cations in HEOs, here we utilize a robust and cross-referenced characterization approach using soft X-ray magnetic circular dichroism, hard X-ray absorption spectroscopy, M\"ossbauer spectroscopy, neutron powder diffraction and SQUID magnetometry to study the competition between crystal field stabilization energy and configurational entropy governing the cation occupation in a spinel HEO (S-HEO), (Co$_{0.2}$Cr$_{0.2}$Fe$_{0.2}$Mn$_{0.2}$Ni$_{0.2}$)$_3$O$_4$. In contrast to the previous studies, the derived complete structural and spin-electronic model, (Co$_{0.6}$Fe$_{0.4}$)(Cr$_{0.3}$Fe$_{0.1}$Mn$_{0.3}$Ni$_{0.3}$)$_2$O$_4$, highlights a significant deviation from the hitherto assumed paradigm of entropy-driven non-preferential distribution of cations in HEOs. An immediate correlation of this result can be drawn with bulk as well as the local element specific magnetic properties, which are intrinsically dictated by cationic occupations in spinels. The real local lattice picture presented here provides an alternate viewpoint on ionic arrangement in HEOs, which is of fundamental interest for predicting and designing their structure-dependent functionalities.

\end{abstract}



\maketitle
\section{Introduction}

The approach to design novel functional materials exploiting configurational disorder originated with the discovery of high entropy alloys (HEAs).\cite{Cantor2004} The approach was recently extended to oxide systems.\cite{Rost2015,Berardan2016c, Sarkar2018c,Oses2020,Anand2018} High entropy oxides (HEOs) can be classified as single phase oxide solid solutions with the cationic sub-lattice/s populated by multiple elements in equiatomic or near-equiatomic proportions.\cite{Sarkar2019,Oses2020} HEOs exhibit several appealing functionalities, which in many cases are the outcomes of the \enquote{entropy-based} design approach.\cite{Oses2020,Berardan2016c,Sarkar2018c,Witte2019b,Zhang2019,Xu2020,QingsongWang2019,Lun2021,Sharma2020} On the one hand, non-preferential, random occupation of the cations, most often resulting in the highest possible degree of configurational disorder allowed by the crystallographic structure, is considered common amongst the different classes of HEOs reported on until the present.\cite{Oses2020,Sarkar2019} On the other hand, deviations from the complete disorder or non-preferential elemental occupation, possibly stemming from enthalpy effects, are inherently very difficult to study. A possibility or presence of short-range elemental correlations remain a subject of fundamental interest for all classes of high entropy materials. Recently, this paradigm has been experimentally verified in the case of HEAs and medium entropy alloys (MEAs), where certain degree of chemical short range order or elemental fluctuations are now reported.\cite{Ding2019HEA, Zhang2020Nat} Importantly, a direct impact of the non-random chemical distribution leading to enhanced mechanical properties has been shown, prompting further research on enthalpy/entropy aspects in all HEAs/MEAs.\cite{Ding2019HEA, Zhang2020Nat} However, these kind of phenomena are largely unexplored in the case of HEOs and other groups of non-metallic high entropy materials. From a unit cell perspective, spinel type HEO (S-HEO) is an ideal candidate for evaluating the influence of configurational disorder on a fundamental level as extreme scenarios of either complete disorder or order are possible due to the low enthaply barrier for cation antisite mixing in spinels \cite{NAVROTSKY1967}. Nevertheless, unravelling the exact cationic distribution even in conventional spinel oxides remains challenging, which in the case of S-HEOs is further aggravated by the presence of multiple principle cations with possibilities of their different oxidation and spin states. However, such an endeavor is key to future applications, as the cationic occupations govern a majority of functional properties in spinels, a primary one being the magnetic properties.

\begin{figure*}[!htb]
\[\includegraphics[width=0.85\textwidth]{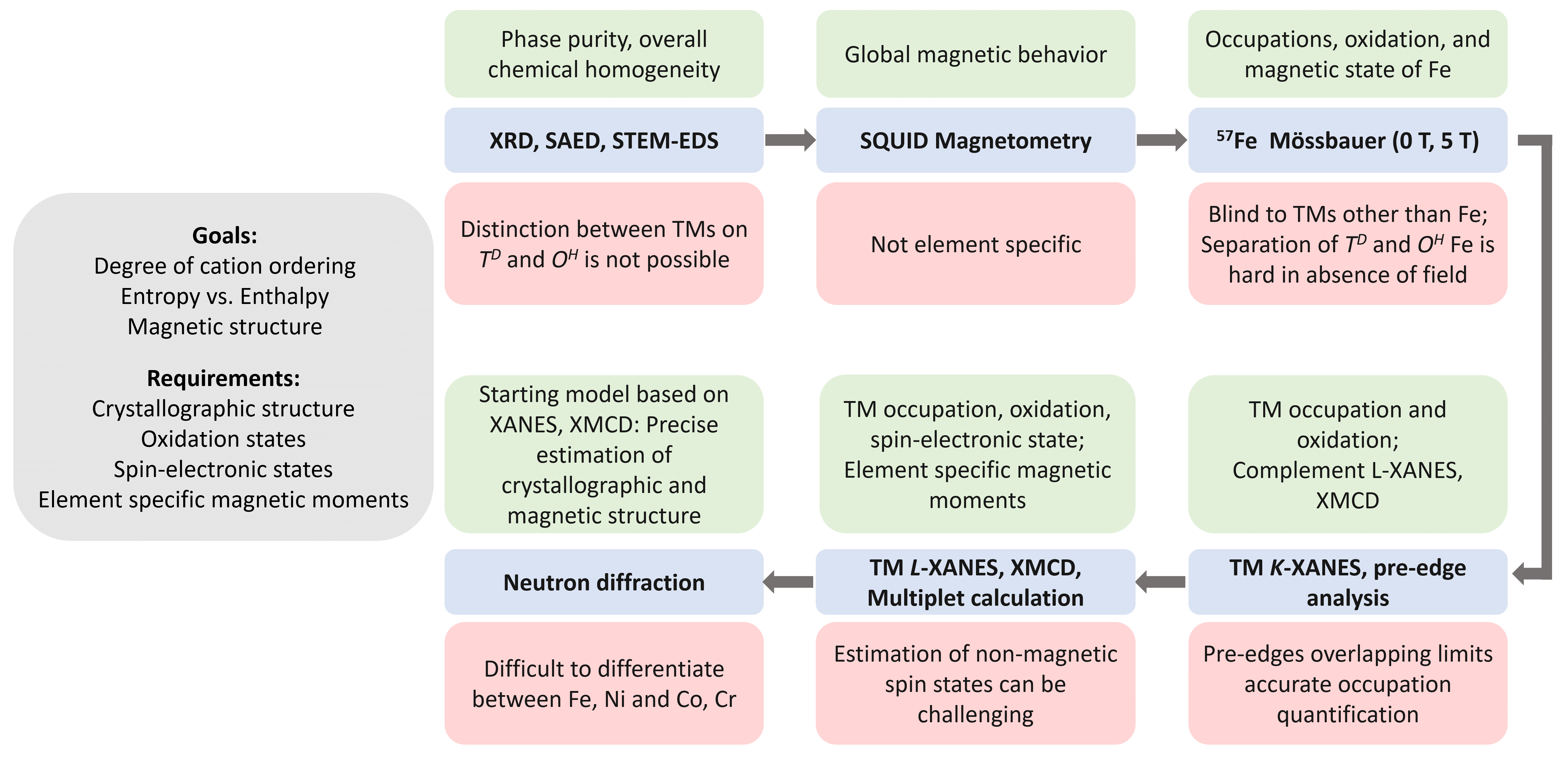}\]
\caption{The schematic represents the comprehensive experimental protocol used to achieve the picture of enthalpy/entropy balance and magnetic structure in the S-HEO. The limitations of the individual techniques to solely provide the full solution in the case of S-HEO are indicated in the light red boxes, which indicate the necessity of the combined and cross-correlated experimental used here. The conclusions that are drawn from each characterization set are highlighted in the light green boxes.}
\label{Exp}
\end{figure*}

Since the first report in 2018 \cite{Dabrowa2018}, S-HEOs have gained considerable research attention, which has resulted several compositions exhibiting a variety of functional properties (see review articles \cite{ALBEDWAWI2021109534,Salian21}). Much research efforts have been devoted to the applied perspective of S-HEOs, such as investigation of the functional properties, discovery of new compositions and exploration of alternate synthesis routes. On the other hand, rigorous studies \cite{FRACCHIA2020} focusing on the fundamental local structural and spin-electronic evaluation, which greatly influences their functional properties remains sparse. This study focuses on the initial S-HEO system, (Co$_{0.2}$Cr$_{0.2}$Fe$_{0.2}$Mn$_{0.2}$Ni$_{0.2}$)$_3$O$_4$ \cite{Dabrowa2018}, which has been extensively studied owing to its relevant magnetic, electrochemical and catalytic properties.\cite{Dabrowa2018,Mao2020,Musico2019,Wang2020s,Nguyen2020,Lin_cryst21,Nguyen20_Comb,TALLURI2021103004,HUANG2021129838,MAO201911,Cieslak2021,GRZESIK2020835,Nguyen2021, STYGAR20201644} Previous reports on this system have always assumed a completely disordered structural model for the (Co$_{0.2}$Cr$_{0.2}$Fe$_{0.2}$Mn$_{0.2}$Ni$_{0.2}$)$_3$O$_4$, where all the cations are expected to equally and randomly occupy the octahedral (O$^H$) and tetrahedral (T$^D$) sites.\cite{Dabrowa2018,Mao2020,Musico2019,Wang2020s,Nguyen2020,Lin_cryst21,Nguyen20_Comb,TALLURI2021103004,HUANG2021129838,MAO201911,GRZESIK2020835,Nguyen2021,STYGAR20201644}. In a recent study, Cieslak et al.\cite{Cieslak2021} used $^{57}$Fe M\"ossbauer spectroscopy for the first time along with Korringa-Kohn-Rostoker (KKR)-coherent potential approximation (CPA) approach, where they observed certain degree of site preferences. Nevertheless, the proposed occupation model, (Co$_{0.05}$Cr$_{0.35}$Fe$_{0.35}$Mn$_{0.05}$Ni$_{0.2}$)
(Co$_{0.275}$Cr$_{0.125}$Fe$_{0.125}$Mn$_{0.275}$Ni$_{0.2}$)$_2$O$_4$, is still very close to the random configuration. \cite{Cieslak2021} However, the experimental limitation for precisely determining the cation occupation in this study \cite{Cieslak2021} was the use of Cu X-Ray diffraction (XRD) and spectroscopy sensitive to one element only. The constituent transition metals (TM) cations have very close atomic scattering factors and cannot be easily distinguished in XRD, while $^{57}$Fe M\"ossbauer spectroscopy is exclusively specific to Fe and does not provide information about any of the other cations. Moreover, $^{57}$Fe M\"ossbauer spectrscopy was also performed in absence of an external magnetic field, which often makes it challenging to separate the effect of Fe occupying the T$^D$ and O$^H$ sites.

In this work, several element-specific techniques, such as hard and soft X-ray absorption near edge spectroscopy (XANES), soft X-ray magnetic circular dichroism (XMCD) and $^{57}$Fe M\"ossbauer spectroscopy (both in presence and absence of an external magnetic field) have been used. These are  complemented  with neutron powder diffraction (NPD), XRD, transmission electron microscopy (TEM) and magnetometry, to unravel the details ionic arrangements and magnetic structure of (Co$_{0.2}$Cr$_{0.2}$Fe$_{0.2}$Mn$_{0.2}$Ni$_{0.2}$)$_3$O$_4$. Each of these techniques possess certain limitations due to the presence of several cations with close atomic and neutron/atomic scattering factors, multivalency and multisite occupations (i.e., overlap of O$^H$ and T$^D$ pre-edge transitions in $K$-edge XANES), making it challenging to extract all the information individually from one technique. Hence, the combination of complementary characterization techniques becomes extremely critical in case of a complex scenario such as in S-HEO, as it substantially lowers overall experimental inaccuracies that can arise from the limitation of the individual ones. The schematic in \textbf{Figure \ref{Exp}}, provides an overview of the comprehensive experimental protocol used in this study. The conclusions drawn from each set of characterization technique and the limitations faced to provide the complete information of the S-HEO individually from these techniques are also highlighted in \textbf{Figure \ref{Exp}}. Consequently, in contrast to the existing reports on S-HEO, very strong preferences in the cation occupation resulting in the least possible configurational disorder allowed by the composition and crystal structure have been observed. Importantly, the broader objective of the study (Figure 1) is to provide a correlation between the long-range magnetic structure and element specific magnetic characteristics with the crystallographic structure and cation spin-oxidation states of (Co$_{0.2}$Cr$_{0.2}$Fe$_{0.2}$Mn$_{0.2}$Ni$_{0.2}$)$_3$O$_4$.

\section{Experimental Section}
\subsection{Synthesis}\label{syn}
An aerosol based nebulized spray pyrolysis (NSP)\cite{Sarkar2018c} technique was utilized for synthesis of nanocyrstalline S-HEO. Aqueous based precursor solution was prepared by dissolving stoichiometric proportion of the nitrate salts of the corresponding metals in deionized water. The aerosol was formed using a piezo-driven nebulizer that was carried to the hot-wall using N$_2$. The reactor was maintained at a temperature of 1050 $^{\circ}$C under a pressure of 900 mbar. The powders so formed were collected using a filter-based collector. The formed single-phase HEO after the NSP synthesis is nanocrystalline in nature. Hence, in order to obtain a sample representative of the bulk and as close as possible to the equilibrium ground state, the obtained powder was annealed at 1000 $^{\circ}$C for 10 hours in an air atmosphere with a ramp rate of 5 $^{\circ}$C per minute.

\subsection{Structural, electronic and magnetic characterization}

\begin{enumerate}
     
\item \textbf{X-Ray and neutron powder diffraction}

\textbf{High resolution X-Ray  diffraction (XRD)}. XRD patterns were recorded at room temperature, using a STOE Stadi P diffractometer, equipped with a Ga-metal jet X-ray source (Ga-K$\beta$ radiation, 1.2079 \AA). Patterns were collected  between 10$^{\circ}$ and 90$^{\circ}$ 
with a step size of 0.05$^{\circ}$ at a scan rate of 4 s per step. 

\textbf{Neutron powder diffraction (NPD)}: The time-of-flight (TOF) NPD pattern was measured on POWGEN at the Spallation Neutron Source (SNS) at the Oak Ridge National Laboratory. The sample was loaded into a vanadium can with a inner diameter of 6 mm, which  was sealed  with a copper gasket and aluminum lid. The measurement was performed at 300 K with neutrons of average 
wavelengths 0.8 \AA, covering a d-spacing range of 0.48 - 10 \AA.

Rietveld analysis of the XRD and NPD patterns was done using TOPAS V.5.0\cite{Topas2015}. The instrumental intensity distribution of the XRD and NPD instruments were determined empirically using reference scan of LaB$_6$ (NIST 660a) and Si, respectively. The microstructural parameters (crystallite size and strain broadening) were refined to adjust the peak shapes. Thermal displacement parameters were constrained to be the same for all atoms on a specific site. For oxygen, anisotropic displacement parameters were refined. For the determination of the magnetic structure, the reader is referred to  Sec. III C.

\item \textbf{Transmission electron microscopy (TEM) and energy dispersive X-ray spectroscopy (EDX)}: Specimens for TEM and EDX were prepared by  dispersing the finely ground powders onto a standard carbon coated copper grid. A FEI Titan 80-300 aberration (imaging $C_s$) corrected TEM equipped with an EDS detector and a Gatan Tridiem 863 image filter operated at 300 kV was used to examine the specimens.

\item\textbf{Superconducting Quantum Interference Device (SQUID)}. Magnetic characterization was performed using a Quantum Design MPMS3 SQUID vibrating sample magnetometer (VSM). All magnetization measurements were done in VSM mode. Temperature dependent measurements between 5  and 400\,K were performed following a zero-field cooled (ZFC) - field cooled (FC) protocol: the sample was cooled in zero magnetic field down to 5\,K. Then the external field $\upmu_0 H$ was applied and the magnetization then measured during warming up to 380\,K (ZFC branch). Subsequently, the magnetization was measured with the magnetic field applied from 380\,K to 2\,K (FC branch). Magnetic field dependent $M$($\upmu_0 H$) measurements at different temperatures were also performed after cooling in zero magnetic field. 

\item\textbf{M\"ossbauer spectroscopy}.
$^{57}$Fe M\"ossbauer spectroscopy both in absence ($H_{ext}$ = 0\,T) and presence ($H_{ext}$ = 5\,T) of external magnetic field was performed employing a $^{57}$Co:Rh source in transmission geometry (bulk sample) using a triangular sweep of the velocity scale. As conventionally done, all center shifts are given relative to $\alpha$-Fe at room temperature. A flow type cryostat operating with liquid He was used for low temperature measurements. The spectra are fitted using the WinNormos (from R.A. Brand, WISSEL) for Igor software.

\item \textbf{X-Ray absorption near edge spectroscopy (XAS) and X-ray magnetic circular dichroism (XMCD)}

\textbf{XANES in the hard X-ray regime} probes the K-edges of the transition metals (TM) corresponding to the 1$s \rightarrow$ 4$p$ transitions. Measurements were performed at beamline P65 (PETRA III, Hamburg) \cite{Welter2019}. The S-HEO powder was pressed into a pellet, while the absorption measurements were performed in transmission mode at 10 K. 

\textbf{XANES and XMCD in soft X-Ray regime} reveal the transition from 2$p \rightarrow$ 3$d$ corresponding to the L$_{2,3}$ edges of the TM cations and the related dichroic response of the 3$d$ states. Measurements were performed at the UE46$\_$PGM-1, BESSY II of the Helmholtz-Zentrum Berlin \cite{Weschke2018}. The measurements were conducted at 4 K in an external magnetic field of 6\,T. Unlike XAS in the hard X-ray regime, these measurements were not performed in transmission geometry. Instead, absorption signals were detected using a total electron yield technique. The samples were pressed onto an Indium foil to ensure electrical grounding. A broad feature corresponding to the In $M_2$ edge was observed at 700.5 eV. Muliplet calculations were performed with CRISPY and CTM4XAS programs.\cite{retegan_crispy, CTM4XAS}

\end{enumerate}

\section{Results}
\subsection{Initial structural characterization: XRD and TEM}

\begin{figure}[!h]
\[\includegraphics[width=0.99\columnwidth]{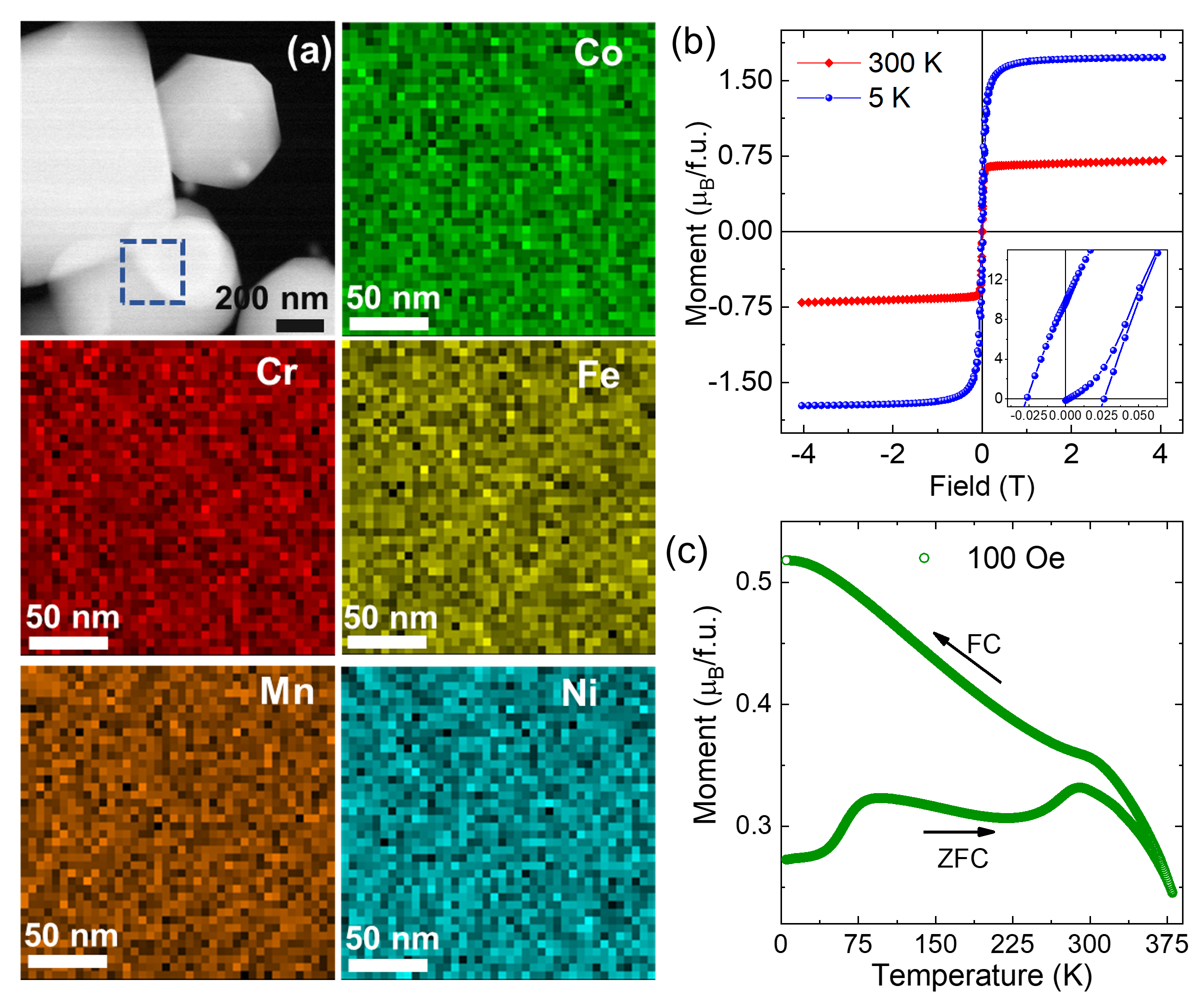}\]
\caption{(a) Scanning TEM micrograph and the corresponding elemental distribution maps for Co, Cr, Fe, Mn and Ni. (b) M-H plots of S-HEO at 300\,K and 5\,K with the inset showing the virgin curve measurement at 5\,K. (c) ZFC-FC data in the M-T plot at 100\,Oe indicate a series of magnetic transition in S-HEO with the ferrimagnetic transitions occurring above 380\,K.}
\label{MH_MT_MS}
\end{figure}

The S-HEO crystallizes in a phase pure spinel ($Fd\bar3m$) structure with lattice parameter $a$ = 8.3480(2)\,\AA\, as revealed by the high resolution Ga metal-jet XRD and TEM micrograph shown in Supplementary Figure SI1a and SI1b, respectively. The negligible peak broadening in the XRD pattern indicates a microcrystalline nature of the sample with minimal microstrain effects. The presence of micrometer sized crystallites is further corroborated by the transmission electron microscopy (TEM) studies (\textbf{Figure \ref{MH_MT_MS}}a). The energy dispersive X-ray spectra (EDS) acquired from the scanning TEM mode indicate a correct stoichiometry and the absence of phase segregation in S-HEO (\textbf{Figure \ref{MH_MT_MS}}a). It should be noted that the elemental distinction between the T$^D$ and O$^H$ cations in the studied S-HEO is not possible from the XRD or the EDS mapping. Hence, from XRD, TEM and EDS analysis, information on the phase purity of the S-HEO along with its lattice parameter, microstructural features and overall chemical homogeneity can be reliably ascertained.

Often HEOs are investigated to check for the signature of entropy driven single phase stabilization. In addition to calorimetric studies, a case of a clear entropy driven phase stabilization is typically revealed by phase segregation at lower temperatures followed by reversible transition to single-phase at higher temperature, as is observed in rocksalt-HEOs \cite{Rost2015}. Hence, low temperature heat treatments were also performed on S-HEO (Supplementary Figure SI2). Unlike entropy stabilized rocksalt-HEO, the XRD patterns indicate the stability of the single phase S-HEO even at lower temperatures making the effect of entropy in phase stability of S-HEO rather unclear.

\subsection{Bulk magnetic behavior: SQUID magnetometry}

The field dependent magnetization measurements (M-H plots) at 300\,K and 5\,K are shown in \textbf{Figure \ref{MH_MT_MS}}b. The M-H plot at 300\,K exhibits an extremely soft magnetic behaviour with negligible coercivity ($H_c$). Nevertheless, the considerable saturation ($M_s$ = 0.67 $\mu_B$/f.u.) can be achieved in fields  above 0.8\,T, while $M_s$ = 1.73 $\mu_B$/f.u. and $H_c$ of 230 Oe are observed at 5\,K. These values are in agreement with those for  bulk S-HEO prepared by solid state synthesis.\cite{Musico2019,Cieslak2021}

The temperature dependent magnetization measurements (M-T plots) from 5 - 380\,K in a field of 100\,Oe are shown in \textbf{Figure \ref{MH_MT_MS}}c (and at 50 and 500\,Oe  in Supplementary Figure SI3). 
Considerable splitting between the field cooling (FC) and zero field cooling (ZFC) curves can be observed even at 380\,K which decreases for larger magnetic fields. 
This indicates that the magnetic transition temperature ($T_c$) of S-HEO is above 380\,K, in agreement with $T_c$ $\sim$ 425\,K reported by Musico et al.\cite{Musico2019,Cieslak2021} It is known that magnetic ordering can also effect the stability of certain phases, e.g. in $\alpha$-Fe. However, such a scenario is unlikely in case of the S-HEO as the $T_c$ (425\,K or 152 $^{\circ}$C) is considerably lower than the synthesis temperature. Several inflection points can be seen in the M-T plots measured at 100\,Oe and 50\,Oe, similar to the observations in \cite{Musico2019} where M-T was measured at 100 Oe. The features at around 242\,K and 286\,K are relatively weak and cannot be distinguished at 500\,Oe or higher fields (Supplementary Figure SI3). Conversely, ZFC feature at $\sim$ 75 K is enhanced while measuring at higher magnetic field. The probable reasons for these magnetic features can be related to locally varying inter-site magnetic exchange interactions, whose strength can have different temperature dependencies. A consequence of this can be observed in spatial fluctuations of local magnetic order which might influence the movement of the domain walls in the external magnetic field. The virgin M-H magnetization measurements at different temperatures (see Supplementary Figure SI4) indicate a pinning type mechanism is prevalent in microcyrstalline S-HEO. Thus, the initial increase of the magnetization (ZFC) with temperature up to 75\,K, observed in all measurements, can be attributed to the movement of magnetic domains governed by a (de)pinning  of domain walls typical of many ferro/ferri-magnets. For a complete understanding of the exact nature of magnetization dynamics and switching, detailed microstructure studies and magnetic domain microscopy would be necessary,  outside of the scope of this current study. Additionally, for further information on the ion-ion magnetic exchange interactions, support from theoretical studies (such as Monte-Carlo or density functional theory \cite{Rak2020}) would be needed. Nevertheless, the aforementioned investigations confirm that the S-HEO sample studied here is phase pure and shows the prevalent magnetic features of bulk S-HEO.  

\subsection{$^{57}$Fe M\"ossbauer spectroscopy in absence and presence of external magnetic field}

$^{57}$Fe M\"ossbauer spectroscopy in absence of an external magnetic field ($H_{ext}$ = 0\,T) as well as in presence of an external magnetic filed ($H_{ext}$ = 5\,T)  was used to probe both the magnetic and charge state of Fe at a local level.

\begin{figure}[htb]
\[\includegraphics[width=0.99\columnwidth]{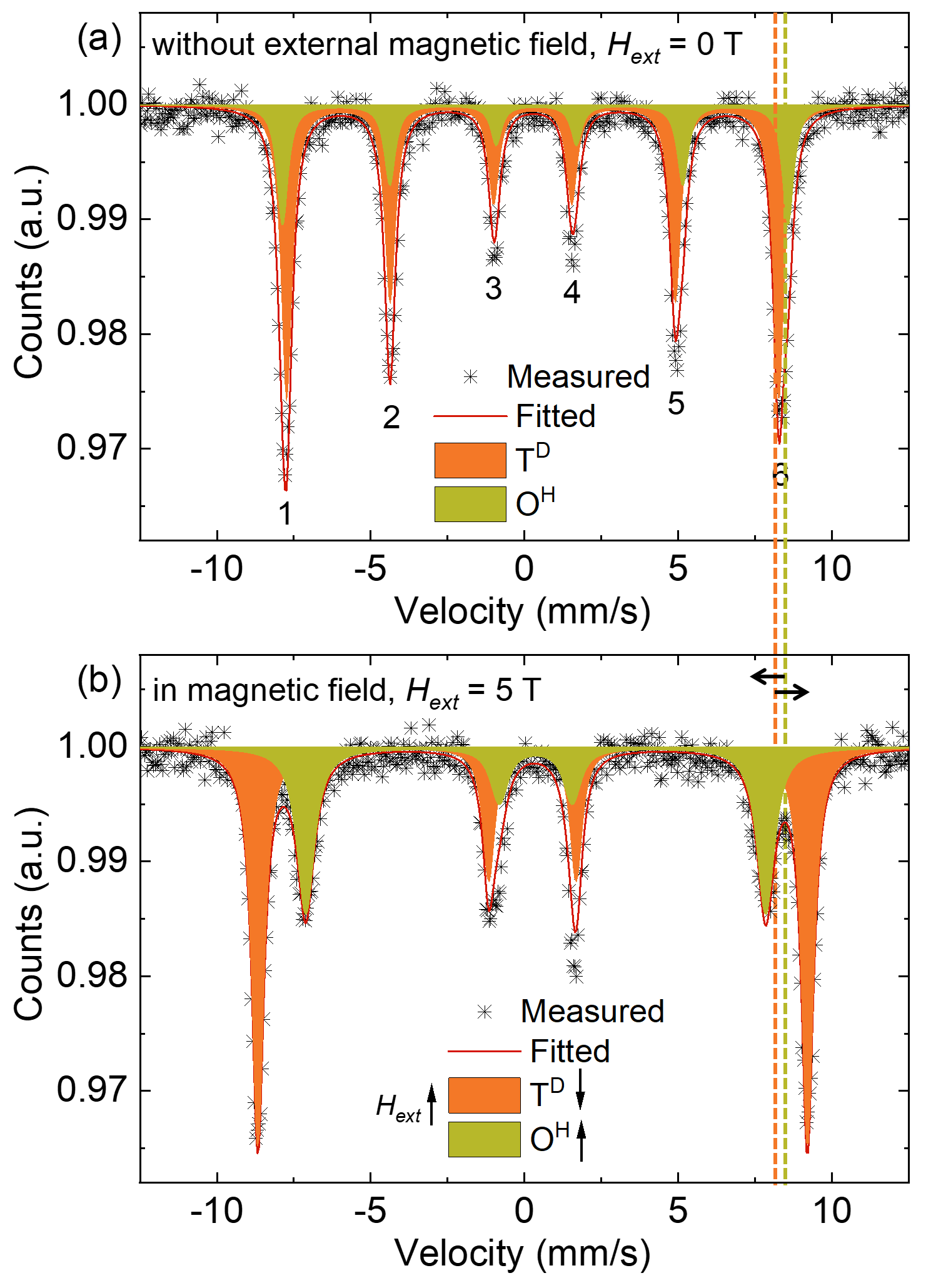}\]
\caption{Low temperature Mössbauer spectra of S-HEO (a) without magnetic field, i.e., 0 T (b) in-field external magnetic field of 5 T. The orange sub-spectrum indicates the Fe$^{3+}$ occupying the T$^D$ sites, while the green one corresponds to the O$^H$ sites. A clear distinction between the Fe$^{3+}$ occupying the T$^D$ and O$^H$ sites can be observed in (b), while strong overlap of the features is observed in (a). The disappearance of the inner lines 2 and 5 in the (b) indicates a collinear arrangement of the Fe spins in both the sites.}
\label{MS_LT}
\end{figure}

\textbf{Figure \ref{MS_LT}} presents the low temperature M\"ossbauer spectra (with $H_{ext}$ = 0\,T and 5\,T) of the S-HEO. The complete temperature series of $H_{ext}$ = 0\,T is presented in Supplementary Figure SI5a. A magnetic sextet is evident in all the spectra, supporting the magnetometry data that  $T_c$ is above room temperature. Supplementary Table SI1 includes detail of the hyperfine field ($B_{hf}$), isomer shift($\delta$) and the quadrupole line shift ($2\epsilon$) for S-HEO at different temperatures. A  deviation of the integral area ratio of the two outer sextet lines 1 and 2  to the innermost line 3 from the ideal values of 3:2:1  is observed in the room temperature spectrum Supplementary Figure SI5b, which can be ascribed to dynamic relaxation similar to that reported by Cieslak et al.\cite{Cieslak2021}. Completely static sextet with fitting of the 10\,K spectrum ($H_{ext}$ = 0\,T) is shown in \textbf{Figure \ref{MS_LT}}a, which reflects the two different Fe environments, both corresponding to high spin (HS) Fe$^{3+}$. The orange sub-spectrum, with a relative area of 27 \%, corresponds to the Fe$^{3+}$ occupying the O$^H$ sites with, $\delta$ = 0.50(3) mm/s (with respect to $\alpha$-Fe) and $B_{hf}$ = 51.2(8) T. The green sub-spectrum with a relative area of 73 \% isomer shift ($\delta$) of 0.39(4) mm/s (with respect to $\alpha$-Fe) and $B_{hf}$ = 49.6(5) T corresponds to T$^D$-Fe$^{3+}$. The quadrupole line shift (2$\epsilon$) is negligible in all cases below 225\,K, i.e., after the static sextet is observed. These observed values are in close agreement with literature for Fe$^{3+}$ occupying T$^D$ and O$^H$ sites.\cite{Demirci2017, Clemens2014}. The obtained amount of Fe occupying T$^D$ is slightly higher than the observation from \cite{Cieslak2021} corresponding to 60 \%. However, the aforementioned investigation, as well as the one from Cieslak et al.\cite{Cieslak2021}, were carried out in absence of magnetic field ($H_{ext}$ = 0\,T). Thus, complete separation of the contribution from the T$^D$ and O$^H$ is challenging, due their similar $B_{hf}$ values. Hence, in this study, in-field $^{57}$Fe M\"ossbauer spectroscopy with an external magnetic field ($H_{ext}$) of 5 T applied parallel to the $\gamma$-rays at 4.2\,K was additionally employed and the results are presented in \textbf{Figure \ref{MS_LT}}b.

\begin{table}[htb]
\caption{Parameters obtained from fitting of the in-field ($H_{ext}$ = 5\,T) $^{57}$Fe M\"ossbauer spectroscopy of S-HEO measured at 4.2\,K, where $B_{eff}$ is the effective hyperfine field, $B_{hf}$ is the actual hyperfine field, $\delta$ is the isomer shift, $2\epsilon$ is the quadrupole line shift, and $I$ is the \% of Fe on the respective sites. The canting angle $\Psi$ $\approx$ 0 (modulo $\pi$) for both T$^D$ and O$^H$.}
\begin{tabular}{cccccc}
\hline
Site & $B_{eff}$ (T) &  $B_{hf}$ (T) & $\delta$ (mm/s) & $2\epsilon$ (mm/s) & $I$ (\%)\\
\hline
T$^D$   &  55.4(2) & 50.4(2) & 0.38(0)   & 0.01(0)  &  68.1(6)\\
O$^H$   &  46.5(2) & 51.5(3) & 0.48(1)   & 0.01(0)  &  31.9(5)\\\hline
\end{tabular}
\label{MS_LT_tab}
\end{table}

In-field $^{57}$Fe Mössbauer measurements are extremely precise in separating the Fe-site occupancies in the case of spinels, due the antiferromagnetic coupling between the O$^H$ and T$^D$  sub-lattices \cite{ChappertPRL1967}. Consequently, \textbf{Figure \ref{MS_LT}}b show two well-resolved outer lines compared to the \textbf{Figure \ref{MS_LT}}a. Since the hyperfine field in $^{57}$Fe is antiparallel to the atomic moment, the $H_{ext}$ of 5 T gets subtracted from the $B_{hf}$ at the O$^H$-site, i.e., the direction of the primary magnetization of S-HEO. Conversely, the $H_{ext}$ = 5 T is added to $B_{hf}$ at the T$^D$-site i.e., originally anti-parallel to the net magnetization of the S-HEO. The details of the hyperfine parameters obtained from in-field $^{57}$Fe Mössbauer measurement are presented in \textbf{Table \ref{MS_LT_tab}}. It can be observed that actual $B_{hf}$ obtained from the in-field measurement are in close agreement with ones to the estimated from fitting of the Mössbauer spectra measured in absence of field ($H_{ext}$ = 0 T), \textbf{Figure \ref{MS_LT}}b. As expected, the isomer shifts ($\delta$) of Fe$^{3+}$ on either sites obtained from $H_{ext}$ = 5 T  (\textbf{Table \ref{MS_LT_tab}}) matches with the values obtained from the $H_{ext}$ = 0 T measurement. The relative occupation of Fe, estimated from the well-resolved in-field $^{57}$Fe Mössbauer spectra, indicates a 1:2 ratio of Fe$^{3+}$ on O$^H$:T$^D$, \textbf{Table \ref{MS_LT_tab}}.

Apart from the visible shift in the $B_{hf}$, another distinct difference between \textbf{Figure \ref{MS_LT}}a and \textbf{Figure \ref{MS_LT}}b is the almost complete disappearance 2nd and 5th inner line in the case of $H_{ext}$ = 5 T. This information is intimately related to the magnetic structure and the alignment of the (Fe) spins in the system. Intensity of the 2nd and 5th lines of the sextet in an in-field measurement depends on the square of the cosine of the canting angle $\Psi$ ($\cos^2(\theta)$), which can be defined as the angle between the direction of the effective hyperfine field ($B_{eff}$) and the direction of $H_{ext} \parallel \gamma$-ray. The absence of the 2nd and 5th lines in S-HEO dictates that the $\Psi$ for both O$^H$ and T$^D$ is close to 0$^{\circ}$  (or 180$^{\circ}$, modulo $\pi$). Note that this extremal value indicates that all moments are parallel or antiparallel to the $\gamma$-ray direction. This indicates a Néel-type collinear spin arrangement in S-HEO, similar to bulk NiFe$_2$O$_4$ \cite{ChappertPRL1967}. However, unlike NiFe$_2$O$_4$ the collinearity of the spins is not intuitive in S-HEO, as the presence of multiple cations often leads to non-collinear spin arrangements in other classes of HEOs \cite{Witte2019b}.

Thus, the M\"ossbauer spectroscopy provides a first evidence of the Néel-type ferrimangetic structure along substantial inversion and highly preferential cationic distribution in (Co$_{0.2}$Cr$_{0.2}$Fe$_{0.2}$Mn$_{0.2}$Ni$_{0.2}$)$_3$O$_4$.

\subsection{TM $K$-edge XANES and pre-edge analysis}

\begin{figure*}[htb]
\[\includegraphics[width=0.97\textwidth]{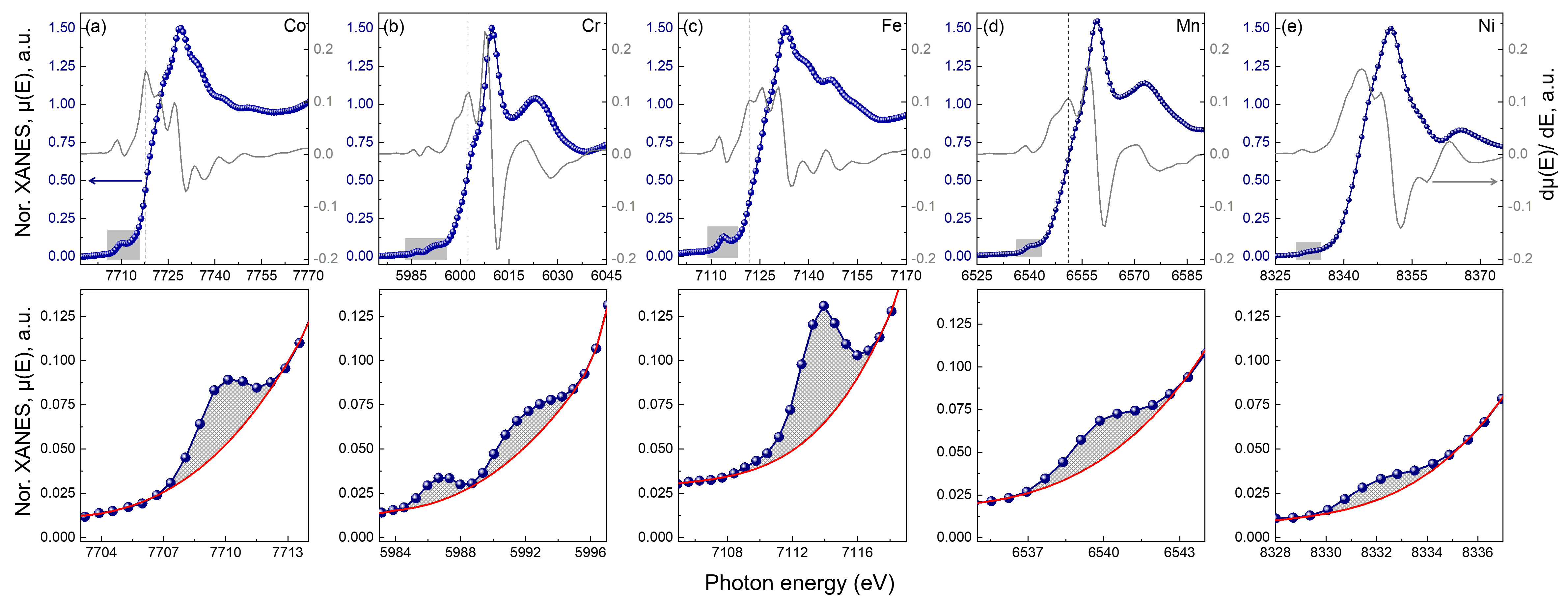}\]
\caption{The XANES spectra, $\mu(E)$ (blue), the first order derivative, $d\mu(E)/dE$ (grey) and the integrated intensity of the pre-edge region (bottom) at the TM $K$-edges of S-HEO. Cr, Mn and Ni exclusively occupy the O$^H$ sites, Co occupies the T$^D$ site and Fe occupies both with predominant T$^D$ occupancy.}
\label{XAS_K-edge}
\end{figure*}

\textbf{Figure \ref{XAS_K-edge}}a-e present the XANES ($\mu(E)$) and the first order derivative ($d\mu(E)/dE$) of transition metal (TM) $K$-edges measured in transmission mode on a pellet sample, hence, probing the bulk of the systems. The background corrected integrated intensity of the pre-edge feature in TM $K$-edges provides precious information about the site occupancy.\cite{Groot2001CR, de_Groot_2009} Typically, the integrated pre-edge intensity of a given 3$d$ TM cation is larger in the T$^D$ coordination than that of O$^H$ coordination, due to negligible $d–p$ hybridization in the latter. However, in comparing different TMs, the electronic configuration of 3$d$ state must also be considered.

The fine structure of the $K$-edge XANES spectrum of Co (\textbf{Figure \ref{XAS_K-edge}}a) along with the strong pre-edge feature centering at 7709.6 eV, closely corresponds to Co$^{2+}$ in a T$^D$ environment.\cite{Hunault2013} The edge energy of 7718 eV and the features of the $d\mu(E)/dE$ also lead to the same conclusion.\cite{Hunault2013,Moen1997} However, a fraction of Co$^{3+}$ is also likely, hinted by the the intensity ratio of the peaks centered at 7718\,eV and 7722\,eV observed in the $d\mu(E)/dE$ feature of Co $K$-edge XANES (\textbf{Figure \ref{XAS_K-edge}}a).\cite{Moen1997} $L_{2,3}$-edge XANES and XMCD provide further insight into the charge and occupancy of Co, which will be discussed later. The spectral features of the Cr $K$-edge XANES (\textbf{Figure \ref{XAS_K-edge}}b), $d\mu(E)/dE$ and edge energy (6007.1 eV) closely resemble those found for Cr$^{3+}$ in O$^H$, as is observed in normal ZnCr$_2$O$_4$ spinel and other Cr containing oxides.\cite{ChenZrCrO4, Miyano1997, Dubrail_2009} Although the pre-edge features are less intense compared to Co or Fe, the two peaks, a narrow one centered at 5986.6 eV and a broad one centered at 5991.7 eV, are distinct. These are indicative of the 1$s \rightarrow$ 3$d$ electric quadrupolar transitions, for Cr$^{3+}$ in an O$^H$ coordination.\cite{Miyano1997} It should be noted that a part of the broad pre-edge feature at 5991.7 eV can additionally stem from non-local transitions.\cite{ZIMMERMANN201874} The results obtained here for Cr (also for Ni and too some extent for Fe, see below) are similar to what has been observed from the $K$-XANES study on another S-HEO composition, (Co,Mg,Mn,Ni,Zn)(Al,Co,Cr,Fe,Mn)$_2$O$_4$ \cite{FRACCHIA2020}. For Fe, the edge energy (7123.5 eV) in the $K$-edge XANES (\textbf{Figure \ref{XAS_K-edge}}c) along with the line shape and the peak position of the $d\mu(E)/dE$ closely resembles that of the 3+ ferrites, where Fe occupies both T$^D$ and O$^H$ sites.\cite{Carta2007JPCC,Carta2008} The background subtracted pre-edge can be best fitted using a bimodal distribution with the dominant feature corresponding to the T$^D$ geometry centering around 7113.5 eV with a secondary peak stemming from the O$^H$ occupancy centering around 7115 eV. However, an accurate estimation of the Fe occupancy solely from $K$-edge XANES is not reliable as it is difficult to distinguish the weak O$^H$ contribution from the strong T$^D$ transition. Nevertheless, the support from M\"ossbauer spectroscopy and $L_{2,3}$- XANES and XMCD (discussed later) provide a precise estimate of the Fe occupancy. Mn $K$-edge XANES (\textbf{Figure \ref{XAS_K-edge}}d), is the most challenging spectrum where the support from the $L_{2,3}$-edge XANES and XMCD becomes more important. From the $K$-edge XANES analysis, we conclude that Mn is present  predominantly in a 3+ oxidation state in O$^H$ coordination.\cite{chalmin2009XANES} The primary reasoning behind assigning the O$^H$ occupancy relies on the fact that despite having fewer $d$ electrons, the integrated intensity of the pre-edge feature (centering at 6539.7 eV) is significantly lower in magnitude compared to Fe and Co. A closer inspection of the line shapes and the peak positions in the $d\mu(E)/dE$ (\textbf{Figure \ref{XAS_K-edge}}d) further support a conclusion that Mn$^{3+}$ predominates.\cite{Liu2013Cat, chalmin2009XANES} However, a small fraction of Mn$^{2+}$ or Mn$^{4+}$ cannot be excluded.\cite{Liu2013Cat,chalmin2009XANES} The $K$-edge XANES of Ni (\textbf{Figure \ref{XAS_K-edge}}e) is in good agreement with those found in NiO and inverse NiFe$_2$O$_4$ indicating the presence of Ni$^{2+}$ solely on the O$^H$ sites.\cite{ANSPOKS2011, Carta2008} The weak pre-edge feature and the edge energy of 8344\,eV further substantiate this finding \cite{FRACCHIA2020} Hence, from the analysis of the $K$-edge XANES, it can be concluded that the S-HEO exhibits substantial preferences in the cation occupations. However, as discussed earlier (and shown in \textbf{Figure \ref{Exp}}), the overlapping of these pre-edge features, makes it challenging to accurately estimate cation occupations, especially in the case of Co, Fe and Mn. Thus, $L_{2,3}$-edges XANES, XMCD and NPD have been utilized in order to synergistically better estimate the occupations.

\subsection{TM $L_{2,3}$-edges XANES and XMCD}\label{L-XMCD}

\begin{figure*}[htb]
\includegraphics[width=0.97\textwidth]{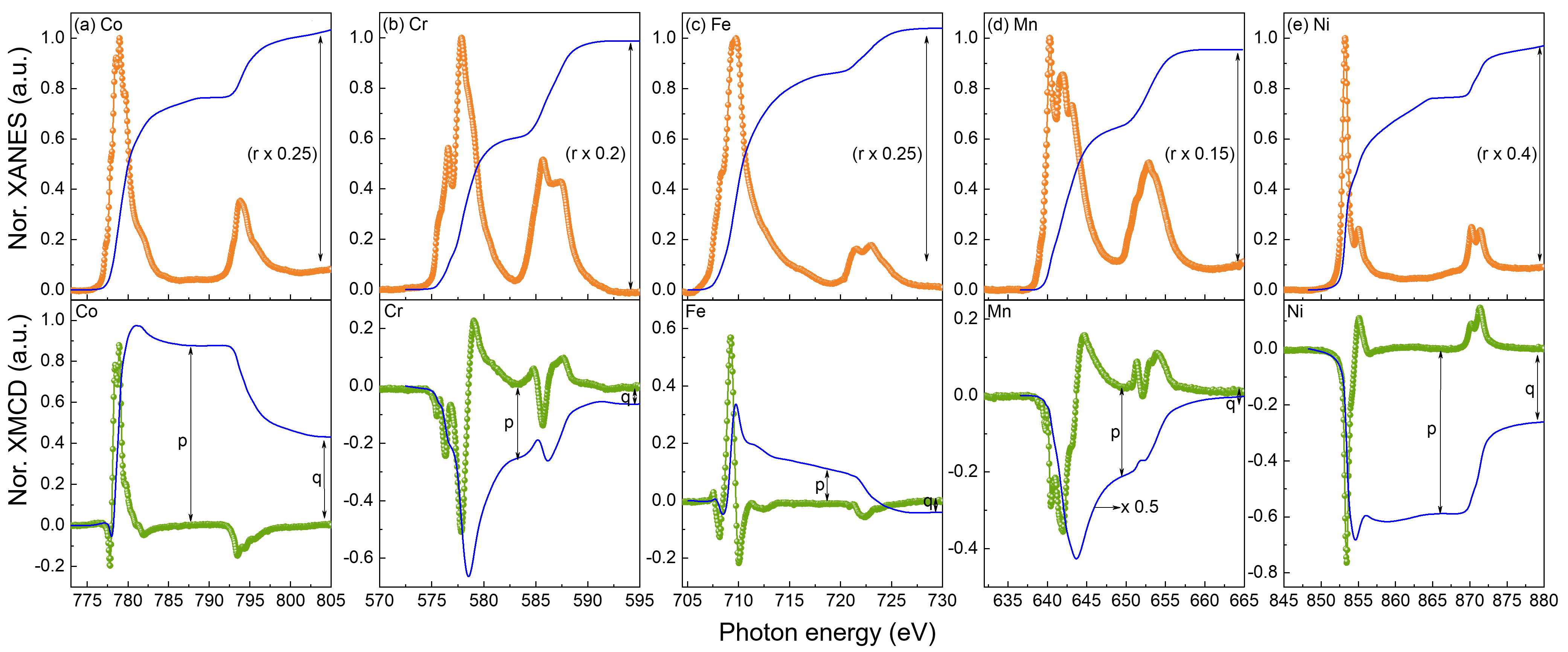}
\caption{XANES (upper, orange) and XMCD (lower, green) spectra at the TM $L_{2,3}$ edges of S-HEO measured with an external field of 6 T. The $r$ is the integration over the XANES spectra, while $p$ and $q$ are the integration of the $L_3$ edge and $L_3+L_2$ edges of the XMCD spectra. The magnitude and the sign of $p$ in the XMCD spectra provides an indication about the elemental occupation, the negative $p$ indicates predominant occupation parallel to the applied field, i.e., O$^H$ occupation in S-HEO, while vice-versa indicates the T$^D$ occupation.}
\label{XMCD_fig}
\end{figure*}

Next, we evaluate the $L_{2,3}$-edges XANES and XMCD data which are presented in \textbf{Figure \ref{XMCD_fig}}. Owing to the antiparallel coupling between the T$^D$ and O$^H$ sites in spinels, the shape and direction of XMCD spectra on the $L_{2,3}$-edges provide comprehensive information about the occupation, oxidation and spin-electronic state of the constituent elements. In addition, XMCD sum rules \cite{Chen1995, Carra1993} offer the possibility to estimate  the element specific spin ($m_S$), orbital ($m_L$) and total  $m_{tot}$ moments (Supplementary Eqn. 1). The moments, $m_S$, $m_L$, $m_{tot}$ are given in $\mu_B$/atom (Supplementary Table SI2). The effective magnetic moment ($m_{eff}$) in $\mu_B$/f.u. of the S-HEO is the $\sum m_{tot}\times0.6$. It should be noted that for the 3$d$ TM cations, sum rules are more reliable for iron and above, where the spin-orbit splitting, i.e., the distance between the $L_3$ and $L_2$ is considerable.\cite{Wende2010, Piamonteze2009PRB} Hence, in S-HEO, the inaccuracies in the moments estimated from the sum rules for Fe, Co and Ni are comparatively smaller, which typically arise only from the background corrections affecting the area defined under the XANES spectra. For Cr and Mn, the inaccuracies are larger \cite{Wende2010, Piamonteze2009PRB}. Nonetheless, even for Cr and Mn the sum rules are often used to provide a rough estimate and  relative orientation of  $m_S$ and $m_L$.\cite{Shengfu2017,Yang2020APL, Garcia2010} Here, we utilize multiplet calculations, in addition, for a more precise evaluation of the $L_{2,3}$ XANES and XMCD spectra, which are presented in the Supplementary Figure SI6a-e and Supplementary Table SI3. 

The $L_{2,3}$-edge XANES spectrum (\textbf{Figure \ref{XMCD_fig}}a) of Co in S-HEO appears to be a mixture of the 2+ and 3+ oxidation states, with clear predominance of the 2+ oxidation state as was also concluded from the the $K$-edge XANES. The XMCD spectrum (\textbf{Figure \ref{XMCD_fig}}a) is nearly identical to high spin (HS)-Co$^{2+}$ in T$^D$ coordination (3$d^7$: ${e}^4$ ${t_{2}}^3 \downarrow$).\cite{Daffe2018} It should be noted that all the spin states are presented with respect to the magnetization and the applied field. However, the additional presence of Co$^{3+}$ contributions on the T$^D$ site cannot be ignored. Hence, in combination with $K$-edge XANES and its first derivative, it can be estimated that roughly 60 \% of Co in HS-Co$^{2+}$ and 40 \% of Co in HS-Co$^{3+}$ is present in S-HEO. It is important to note that the intense and positive $L_3$ edge ($p$) in the XMCD (\textbf{Figure \ref{XMCD_fig}}a) almost completely rules out any possible magnetic Co O$^H$ occupation. Another possibility, i.e., the presence of low spin(LS)-Co$^{3+}$ (3$d^6$: ${t_{2g}}^6$) on the O$^H$, for which $m_S$ is zero (therefore no XMCD signature), can also be discarded because it will then lower the overall $m_{tot}$ for Co significantly. Further, the addition of Co on the O$^H$ site worsens the neutron powder diffraction (NPD) fit that will be discussed later. Hence, the only likely possibility is the presence of HS-Co$^{3+}$ on the T$^D$ site 3$d^6$: ${e}^2 \downarrow$ ${e}^1 \uparrow$ ${t_{2}}^3 \downarrow$, which also supports the high negative value of the Co-$m_{tot}$ or $m_{S}$ (Supplementary  Table SI3). The multiplet calculation also indicates a predominance of Co$^{2+}$ and the fit with 60 \% HS-Co$^{2+}$ and 40 \% HS-Co$^{3+}$ is shown in Supplementary Figure SI6a and Table SI3. A good match to the experimental and fitted XANES and XMCD fine structure could be obtained, supporting the presence of the two sub-spectra. Importantly, the presence of Co solely in T$^D$ is affirmed. In addition, the overall oxidation state of Co is estimated to be 2.4+. However, an inaccuracy of 2.4+ $\pm$ 0.2, but mostly  slightly more Co$^{2+}$ can be expected. This inaccuracy can plausibly be the reason for the width misfit in the calculated XMCD spectra (Supplementary Figure SI6a). 

In the case of Cr (\textbf{Figure \ref{XMCD_fig}}b), the XANES spectrum, especially the line shape and the relative peak intensities closely match to Cr$^{3+}$.\cite{Yang2020APL,Vinai2015} Likewise, the fine structures of the XMCD spectra along with the highly negative $p$ (\textbf{Figure \ref{XMCD_fig}}b) and therefore, high and positive $m_{tot}$ (Supplementary Table SI2) as a consequence, resemble the occupation of Cr$^{3+}$ on the O$^H$ sites with a HS electronic configuration: 3$d^3$, ${t_{2g}}^3 \uparrow$.\cite{Vinai2015, Yang2020APL} The result is further supported by the good agreement of the multiplet calculations with the experimental data, as shown in Supplementary Figure SI6b and Table SI3. 

The spectral features of the Fe $L_{2,3}$ XANES and XMCD measurements (\textbf{Figure \ref{XMCD_fig}}c), specifically the peak at 708 eV in $L_3$ XANES and significantly low value of $m_L$/$m_S$ (Supplementary Table SI2) are in good agreement with the 3+ oxidation state.\cite{Pattrick2002} This further reinforces the conclusions reached for the oxidation state from M\"ossbauer spectroscopy and $K$-edge XANES. Focusing on the $L_3$ XMCD spectra of Fe, three distinct peaks can be observed which indicate  antiferromagnetic coupling between the cationic sub-lattices. The intense positive peak at around 709 eV corresponds to the high spin (HS) T$^D$ Fe$^{3+}$ (3$d^5$: ${e}^2 \downarrow$ ${t_{2}}^3 \downarrow$), while the two minor peaks at 708 and 710 eV correspond to the O$^H$ HS Fe$^{3+}$ (3$d^5$: ${t_{2g}}^3 \uparrow$ ${e_g}^2 \uparrow$).\cite{Yang2020APL} The experimental XANES and XMCD spectra can be adequately fitted using multiplet calculations for $\sim$33 \% Fe$^{3+}$ in O$^H$ and $\sim$67 \% Fe$^{3+}$ in T$^D$ as shown in Supplementary Figure SI6c. This quantitative finding is in close agreement with in-field M\"ossbauer analysis (\textbf{Table \ref{MS_LT_tab}}). The $m_S$, $m_L$ and $m_{tot}$ for Fe obtained from the sum rules and multiplet calculations are listed in Supplementary Table SI2 and Supplementary Table SI3. The high negative value of $m_{tot}$ indicates that Fe is predominantly present in anti-parallel to the net magnetization, i.e., on the T$^D$. 

For Mn, the XMCD and the highly negative $p$ feature strongly indicate that Mn exclusively occupies the O$^H$ sites (\textbf{Figure \ref{XMCD_fig}}d).\cite{Garcia2010, Gilbert2003} In addition, almost complete quenching of the orbital moment of Mn can be concluded from the negligible $m_L$/$m_S$ (Supplementary Table SI2). The $L_{2,3}$ XANES appears similar to a predominant of the 3+ oxidation state with a small fraction of 2+ resulting in the initial $L_{3}$ feature at 640 eV.\cite{Gilbert2003}. The line shape of the XMCD spectrum largely resembles Mn$^{3+}$ (${t_{2g}}^3 \uparrow$ ${e_g}^1 \uparrow$) on the O$^H$ site. Any presence of Mn$^{2+}$ in T$^D$ can be discounted  given that its 3$d^5$ electronic configuration would have led to an intense XMCD positive peak aligned anti-parallel to the applied field.\cite{Kang2008}
On the contrary, the related $L_3$ XMCD feature (640.4 eV) arising from Mn$^{2+}$ exhibits a maximum in the negative direction (moment parallel to the external field). The multiplet calculation is challenging for Mn \cite{Piamonteze2009PRB}, due to the Jahn-Teller effect of Mn$^{3+}$ on the O$^H$ site. As can be observed in Supplementary Figure SI6d, the feature arising form the O$^H$ Mn$^{2+}$ (640.4 eV) can be readily fitted using the multiplet calculation. A strong intensity mismatch for the other features, especially the one at 642 eV, arising from the Mn$^{3+}$ can be observed. Nevertheless, the corresponding XMCD features for all the $L_3$ XANES results are always parallel to the applied magnetic field, i.e., on the O$^H$ site.\cite{Garcia2010} In regards to the charge distribution, $\sim$90 \% of HS-Mn$^{3+}$: ${t_{2g}}^3 \uparrow$ ${e_g}^1 \uparrow$ and 10 \% of HS-Mn$^{2+}$: ${t_{2g}}^3 \uparrow$ ${e_g}^2 \uparrow$ can be estimated from the $K$- and $L_{2,3}$-edge XANES, resulting in a overall charge of 2.9+ $\pm$ 0.3. The slightly increased inaccuracy in the oxidation state, compared to other cations, might originate from a minor fraction of charge disproportionation \cite{POLLERT2000661,Ma2019Mn}, where O$^H$ HS-Mn$^{4+}$ and additional O$^H$ HS-Mn$^{2+}$ at the expense of O$^H$ HS-Mn$^{3+}$.

The line shapes and the peak position in the Ni $L_{2,3}$ XANES spectrum (\textbf{Figure \ref{XMCD_fig}}d) are similar to those for the Ni$^{2+}$ state, which supplement the information obtained from the $K$-edge XANES.\cite{Kang2011APLA} The XMCD fine structure along with intense negative $p$ at 853.4 eV corresponding to the $L_3$ edge and the positive $m_{tot}$ as a consequence (Supplementary Table SI2) indicate the presence of Ni$^{2+}$ in the HS O$^H$ environment, i.e., 3$d^8$: ${t_{2g}}^6$ ${e_g}^2 \uparrow$.\cite{Kang2011APLA} The experimental XANES and XMCD spectra and values obtained form sum rules are in good agreement with the multiplet calculations fit, as presented in Supplementary Figure SI6e and Table SI3. 

Thus, the $L_{2,3}$-edge XANES and XMCD along with the multiplet calculations, $K$-edge XANES, and ${57}$Fe M\"ossbauer spectroscopy provide a near complete model of the occupations and spin-electronic structure of the S-HEO. The overall average oxidation state of the cations is close 2.66+, as summarized in \textbf{Table} \ref{summary_tab}. This further supports the crystallization of the studied composition into a 1:2:4 oxide spinel structure. In addition, the element specific magnetic moments and the spin-orientations are also precisely estimated from the $L_{2,3}$-edge XANES and XMCD. The overall magnetic moment obtained from the experimental sum rule analysis is less than 15\,\% off compared to the bulk magnetic moment obtained from spatially averaging SQUID magnetometry, while the agreement between the bulk moment and the overall moment obtained from multiplet calculations is even better (\textbf{Table} \ref{summary_tab}). This correlation further strengthens the obtained ionic distribution model. Given this knowledge, NPD is used as the final complementary step to derive the complete crystallographic and magnetic structure of the S-HEO.

\begin{table}[htb]
\caption{The oxidation states (Ox.), spin-electronic states, average oxidation states (Avg. Ox.) and occupation of the different cations per formula unit (f.u.) of the S-HEO, (Co$_{0.2}$Cr$_{0.2}$Fe$_{0.2}$Mn$_{0.2}$Ni$_{0.2}$)$_3$O$_4$, obtained from the combination of spectroscopic and diffraction approach is summarized. It should be noted that the spin states are presented with respect to the magnetization and the applied field. As Fe$^{3+}$ occupies both sites, thus the effective spin-state is provided. The effective magnetic moments at 5 K per f.u. obtained from SQUID magnetometry ($m_{eff}$, SQUID), XMCD sum rules ($m_{eff}$, XMCD), multiplet calculations ($m_{eff}$, MC) and neutron diffraction ($m_{eff}$, NPD) are presented.}
\begin{tabular}{cccccc}
\hline
Elements & Ox. & Spin-state & T$^D$ & O$^H$ & Avg. Ox. \\\hline
\multirow{2}{*}{Co} & 2 & ${e}^4$ ${t_{2}}^3 \downarrow$  & 0.36  & -            & \multirow{2}{*}{2.4(2)}\\
                    & 3  & ${e}^2 \downarrow$ ${e}^1 \uparrow$ ${t_{2}}^3 \downarrow$ & 0.24         & -            & \\\hline
Cr   & 3 & ${t_{2g}}^3 \uparrow$ & -  & 0.6 & 3.0(0)\\\hline
Fe & 3 & ${e}^2 \downarrow$ ${t_{2}}^3 \downarrow$ & 0.4 & 0.2  & 3.0(0)\\\hline
\multirow{2}{*}{Mn} & 3 & ${t_{2g}}^3 \uparrow$ ${e_g}^1 \uparrow$ &     -    & 0.54         & \multirow{2}{*}{2.9(3)}\\
    & 2 & ${t_{2g}}^3 \uparrow$ ${e_g}^2 \uparrow$ & -  & 0.06  &\\\hline
Ni   & 2 & ${t_{2g}}^6$ ${e_g}^2 \uparrow$ &   -   & 0.6   & 2.0(0)\\\hline
$m_{eff}$, SQUID & & & & & 1.7(1) $\mu_B$/f.u.\\
$m_{eff}$, XMCD & & & & & 1.5(4) $\mu_B$/f.u.\\
$m_{eff}$, MC & & & & & 1.7(1) $\mu_B$/f.u.\\
$m_{eff}$, NPD & & & & & 1.5(2) $\mu_B$/f.u.\\\hline
\end{tabular}
\label{summary_tab}
\end{table}

\subsection{Complete structural and magnetic model: Neutron powder diffraction}

\begin{figure*}[htb]
\[\includegraphics[width=0.75\textwidth]{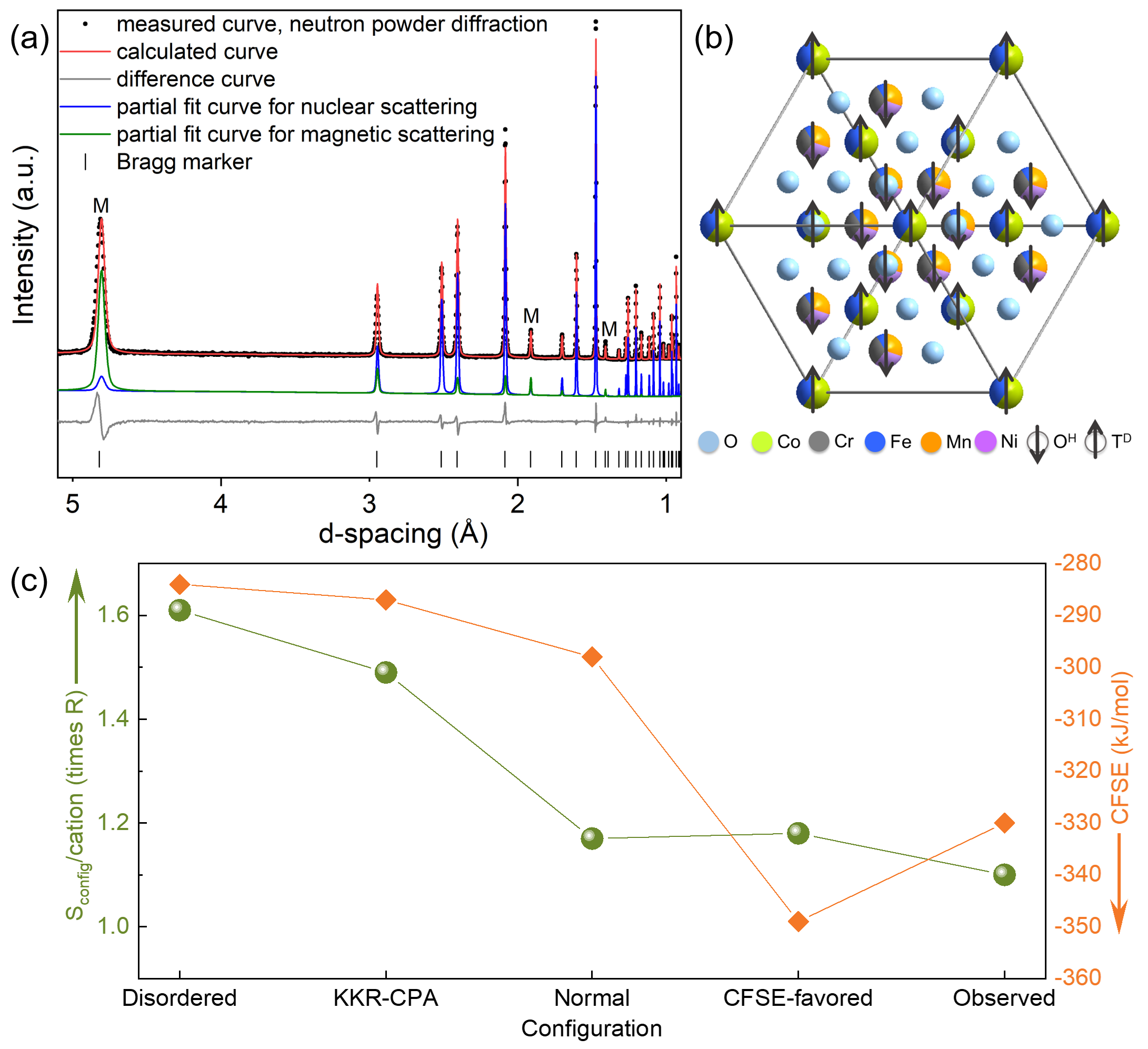}\]
\caption{(a) Neutron powder diffraction of S-HEO, where the reflections indicated by M are the ones strongly affected by the magnetic ordering. (b) Schematic of crystallographic and magnetic structure of S-HEO, along $[1 1 1]$ (parallel projection). (c) The configurational entropy ($S_{config}$) and enthalpy (CFSE) for different possible occupational models of S-HEO are schematically compared. It should be noted that the $S_{config}$ is presented independent of temperature, i.e. in units different from the CFSE.}
\label{NPD}
\end{figure*}

NPD is used as the final characterization tool for validation of the observed cationic distribution. The neutron scattering length of the constituent elements (in barn) are as follows: 2.49 for Co, 3.64 for Cr, 9.45 for Fe, -3.73 for Mn, 10.3 for Ni and 5.80 for O. Although neutrons provide the capability to distinguish between the cations and estimate the presence of oxygen vacancies, a fully unconstrained structural refinement is still not feasible as the scattering powers of Fe and Ni are too similar. Nevertheless, the combination of NPD with the results from $K$ and $L_{2,3}$ XANES and XMCD affords a great precision by overcoming the limitations of the individual techniques. Out of the several models tried that are close to the predictions made from XANES and XMCD, the best goodness of fit was obtained when Co was placed solely in T$^D$ and while the rest of T$^D$ was populated by Fe. Placing some of the Co on the O$^H$ site, mimicking possible LS-Co$^{3+}$, something difficult to judge from XMCD, while the remaining T$^D$ is occupied by Fe or Mn leads to a poorer fit of the results. Thus, the combination of these results obtained from several complementary techniques, allows for estimating structural model with great precision (i.e., occupation of cations) of the S-HEO, which is very close to (Co$_{0.6}$Fe$_{0.4}$)(Cr$_{0.3}$Fe$_{0.1}$Mn$_{0.3}$Ni$_{0.3}$)$_2$O$_4$. Using the structural model, the occupation of oxygen in $32e$ was also refined. Negligible variation in the oxygen occupation from NPD indicate stoichiometric oxygen content in S-HEO, which further supports the fact that the overall average cation charge in the system is close to 2.66+ (\textbf{Table \ref{summary_tab}}). The obtained structural model with the oxidation states is expected to be close to (Co$_{0.6}^{2.4+}$Fe$_{0.4}^{3+}$)(Cr$_{0.3}^{3+}$Fe$_{0.1}^{3+}$Mn$_{0.3}^{2.9+}$Ni$_{0.3}^{2+}$)$_2$O$_4$ (\textbf{Figure \ref{NPD}}). However, as earlier discussed minor inaccuracies $\sim$10 \% in the overall oxidation state of Co and Mn, resulting in a charge compensation between these two cations, remain a possibility where slightly more Co$^{2+}$ and correspondingly a fraction of Mn$^{4+}$ at the expense of Mn$^{3+}$ is possible. In addition to the nuclear scattering, we need to consider the existing magnetic structure, especially in order to refine the intensities of high $d$-spacing reflections correctly. No additional reflections apart from the ones allowed by a symmetry of $Fd\bar3m$ (227) could be observed in \textbf{Figure \ref{NPD}}a, indicating  the crystallographic commensurate ferrimagnetic structure of S-HEO, i.e., $k = [0 0 0]$. \textbf{Table} \ref{summary_tab} provides an overview of the results obtained from the combination of the different spectroscopic and diffraction techniques used in this study. The expected magnetic moment of the obtained from NPD (\textbf{Table} \ref{summary_tab}), further complements the obtained structural model: (Co$_{0.6}$Fe$_{0.4}$)(Cr$_{0.3}$Fe$_{0.1}$Mn$_{0.3}$Ni$_{0.3}$)$_2$O$_4$ (as shown in \textbf{Figure \ref{NPD}}b).

\section{Discussion}

Cation occupation in oxide spinels is largely dictated by two governing factors: one is the crystal field stabilization energy (CFSE) in the oxygen ligand field, while the other one is the configuration entropy ($S_{config}$) gain arising from the disordered arrangement due to antisite mixing. The ideal $S_{config}$ per mole of atom in a system, especially with multiple sub-lattices, can be calculated using the sub-lattice model (\cite{PARIDA2020513,NAVROTSKY1967,Seko2010,Miracle2017,DIPPO2021113974}) as presented in Equation \ref{config_atom}:

\begin{equation}
   \frac{S_{config}}{atom} = -R(\frac{\sum_{x=1}^{x} a^x\sum_{N=1}^{N}(f_i^x ln f_i^x)}{\sum_{x=1}^{x}a^x})    
\label{config_atom}
\end{equation}

where, R is the universal gas constant, $a^x$  is the number of sites on the $x$ sub-lattice, $f_i^x$ is the fraction of elemental species randomly distributed on the respective sub-lattice and $N$ is the number of elements in a given sub-lattice.

For 1:2:4 oxide spinel, the Equation 1 can be written as follows:

\begin{equation}
\begin{aligned}
   \frac{S_{config}}{atom} = {} & \-\frac{R}{7}[1*\sum_{P=1}^{P}(f_p ln f_p)_{T^D}+2*\sum_{Q=1}^{Q}(f_q ln f_q)_{O^H}\\
   &+4*\sum_{O=1}^{O}(f_o ln f_o)_O]
\label{configab}
\end{aligned}
\end{equation}

where total number of atoms per formula unit is $\sum_{x=1}^{x}a^x$ is 7, P, Q, O refer to the number of elements on the different sub-lattices, T$^D$,O$^H$ and oxygen (anion), respectively, while $f_p$, $f_q$ , $f_o$ refer to the atomic fractions of the elements on the respective sub-lattices. For instance, in case of (Co$_{0.6}$Fe$_{0.4}$)(Cr$_{0.3}$Fe$_{0.1}$Mn$_{0.3}$Ni$_{0.3}$)$_2$O$_4$, the $\frac{S_{config}}{atom}$ can be defined as follows:

\begin{equation}
\begin{aligned}
   \frac{S_{config}}{atom} = {} & \-\frac{R}{7}[1*(0.6 ln 0.6+0.4 ln 0.4)+\\
   &2*(0.3 ln 0.3+0.1 ln 0.1+0.3 ln 0.3+0.3 ln 0.3)\\
   &+4*(1 ln 1))]
\label{configspin}
\end{aligned}
\end{equation} 

As can be observed, the factor $ln f_o = 0$ due the stoichiometric oxygen content. In addition, the configuration entropy per mole of formula unit $\frac{S_{config}}{f.u.}$ and configuration entropy per mole of cation $\frac{S_{config}}{cation}$ in 1:2:4 spinel can calculated as follows:

\begin{equation}
   \frac{S_{config}}{f.u.} = 7* \frac{S_{config}}{atom}
\label{configfu}
\end{equation} 

\begin{equation}
   \frac{S_{config}}{cation} = \frac{7}{3}* \frac{S_{config}}{atom}
\label{configcat}
\end{equation} 

\textbf{Table \ref{table_entropy}} compares the $S_{config}$ values for the cation occupation model obtained from this study, with some of the extreme occupational scenarios and the model obtained from the KKR-CPA study\cite{Cieslak2021}. The $\frac{S_{config}}{cation}$ can be considered as a more reliable metric to compare the $S_{config}$ among HEOs,  consisting multiple cations on different cation sub-lattices. \textbf{Figure \ref{NPD}}c schematically compares the $ \frac{S_{config}}{cation}$ and CFSE for these different cationic occupation possible in the studied S-HEO. The details of the calculation for CFSE for the different occupation models are provided in Supplementary Table SI4-9.

\begin{table*}[htb]

\caption{The configuration entropy per mole of atom ($\frac{S_{config}}{atom}$), per mole of formula unit ($\frac{S_{config}}{f.u.}$) and per mole of cation ($\frac{S_{config}}{cation}$) for the different occupational models mentioned in Figure 6c are tabulated here. In addition, the $S_{config}$ for rocksalt-HEO (R-HEO) is included for comparison, where the relationships between $\frac{S_{config}}{atom}$, $\frac{S_{config}}{f.u.}$ and $\frac{S_{config}}{cation}$ are different compared to the spinels. It should be noted that in all cases stoichiometric oxygen content has been assumed, while for the observed scenario, i.e., (Co$_{0.6}$Fe$_{0.4}$)(Cr$_{0.3}$Fe$_{0.1}$Mn$_{0.3}$Ni$_{0.3}$)$_2$O$_4$, it has been  experimentally validated. R is the universal gas constant.}
\begin{tabular}{ccccc}
\hline
Composition & Short name & $\frac{S_{config}}{atom}$ & $\frac{S_{config}}{f.u.}$ & $\frac{S_{config}}{cation}$\\
\hline
(Co$_{0.2}$Cr$_{0.2}$Fe$_{0.2}$Mn$_{0.2}$Ni$_{0.2}$)(Co$_{0.2}$Cr$_{0.2}$Fe$_{0.2}$Mn$_{0.2}$Ni$_{0.2}$)$_2$O$_4$ & Disordered & 0.69 R & 4.83 R & 1.61 R \\
(Co$_{0.05}$Cr$_{0.35}$Fe$_{0.35}$Mn$_{0.05}$Ni$_{0.2}$)(Co$_{0.275}$Cr$_{0.125}$Fe$_{0.125}$Mn$_{0.275}$Ni$_{0.2}$)$_2$O$_4$ & KKR-CPA\cite{Cieslak2021} & 0.64 R & 4.48 R & 1.49 R \\
(Co$_{0.36}$Cr$_{0.06}$Ni$_{0.06}$)(Co$_{0.12}$Cr$_{0.3}$Fe$_{0.3}$Mn$_{0.3}$)$_2$O$_4$ & Normal & 0.500 R & 3.50 R & 1.17 R \\
(Co$_{0.34}$Mn$_{0.06}$Fe$_{0.6}$)(Co$_{0.13}$Cr$_{0.3}$Mn$_{0.27}$Ni$_{0.3}$)$_2$O$_4$ & CFSE-favored & 0.504 R & 3.52 R & 1.18 R \\
(Co$_{0.6}$Fe$_{0.4}$)(Cr$_{0.3}$Fe$_{0.1}$Mn$_{0.3}$Ni$_{0.3}$)$_2$O$_4$ & Observed & 0.47 R & 3.30 R & 1.10 R\\
(Co$_{0.2}$Cu$_{0.2}$Mg$_{0.2}$Ni$_{0.2}$Zn$_{0.2}$)O & R-HEO\cite{Rost2015,Berardan2016c}  & 0.80 R & 1.61 R & 1.61 R\\\hline
\end{tabular}
\label{table_entropy}
\end{table*}

The $\frac{S_{config}}{cation}$ with a complete "disorder" (i.e., equal distribution of cations in O$^H$ and T$^D$ site) would equal 1.61 R similar to that of a five cation rocksalt-HEO. This value is considerably higher than the "observed" scenario (\textbf{Figure \ref{NPD}}c) with $\frac{S_{config}}{cation}=$ 1.10 R. The deviation from the highest disordered state can be explained using the CFSE or the octahedral site preferential energy/enthalpy (OSPE) for a particular cation (Supplementary Table SI5). It is clear that CFSE/OSPE plays a dominant role over configurational disorder in determining the occupation of the cations. However, as \textbf{Figure \ref{NPD}}c and Supplementary Table SI5 indicate, CFSE/OSPE is not the sole player, otherwise 3$d^5$ Fe$^{3+}$ and Mn$^{2+}$, especially  Mn$^{2+}$ with comparatively larger cationic radii in a six-fold coordination, should be preferred on the T$^D$ over HS Co$^{2+}$/Co$^{3+}$. In fact, the enthalpy favorable structural model ("CFSE favored", \textbf{Figure \ref{NPD}}c), (Co$_{0.34}$Fe$_{0.3}$Mn$_{0.06}$)(Co$_{0.13}$Cr$_{0.3}$Mn$_{0.27}$Ni$_{0.3}$)$_2$O$_4$, should ideally result in higher $S_{config}$ of 1.18\,R. This certainly indicates a role of other competing thermodynamic features influencing the cation occupancy, exploration of which remains a subject for future endeavor. A possible influence of temperature, pressure and size/surface effect on the cationic distribution and antisite mixing will be worth exploring in the future. This is because even in the case of conventional spinel systems, it is known that the cationic distribution can vary as these aforementioned factors \cite{Rozenberg2007, ChenZrCrO4}. Endeavors in this direction will not only help to better understand the nature of cationic ordering in S-HEO as a function of temperature and/or pressure but can also be useful to predict their temperature/pressure dependent functional properties.

The ionic ordering due to the enthalpy factors (CFSE) observed in the S-HEO here is inherently correlated to the magnetic properties. As observed, the S-HEO exhibits a N\'eel type ferrimagnetic ground state, which means the magnetism is governed by the intersite AFM coupling. This means that any change in the cation occupation will have a direct impact on the overall ferrimagentic moment, which is verified utilizing multiple techniques used in this study (\textbf{Table} \ref{summary_tab}). Likewise, depending upon the cation distribution over the respective sites different magnetic ground states and transition temperatures can also be expected. In fact, the presence of cation order can also be anticipated in other HEO-classes, which can have a decisive impact on their functional properties. A few scenarios that can be of immediate interest for future explorations are considered here. Perovskite-HEOs exhibit a magnetic phase separation that manifests itself through a vertical exchange bias \cite{Witte2019b}. It is assumed that ferromagnetic clusters are present within the predominant antiferromagnetic lattice, which can perhaps be an indication of chemical short-range  ordering of cations that couple ferromagnetically via the bridging oxygen.  Likewise, even in the case of rocksalt-HEOs \cite{Zhang2019, Rak2020}, the anomalous heat capacity behavior around the N\'eel temperature hints towards short range magnetic correlations well above the N\'eel temperature that can also be an outcome of a certain of degree of local chemical ordering. In addition, the extent of cation order can also play a major role in the energy storage capabilities in the case of layered-HEOs \cite{Zhao2019HEO-Na_Lay, J_Wang2020}. For instance, in the case of conventional layered Li-delafossites an ordered Li sub-lattice is preferred for better performance \cite{KimEES2021}. Hence, the extent of cation ordering on a lattice level, as observed here, or chemical short-range ordering leading local lowering of symmetry can be extremely crucial for better understanding the unique properties exhibited by the HEOs, which warrant future studies along direction.  

\section{Conclusions}

This demonstration study focusses on a multi-functional spinel-HEO, (Co$_{0.2}$Cr$_{0.2}$Fe$_{0.2}$Mn$_{0.2}$Ni$_{0.2}$)$_3$O$_4$. To overcome the experimental limitations, stemming from the presence of multiple principle cations with different occupation and spin-electronic states, we devised and followed a cross-referenced experimental approach to determine the crystallographic, magnetic and spin-electronic structure of the spinel-HEO.  X-Ray magnetic circular dichroism (XMCD) along with X-Ray absorption spectroscopy (XANES) at the transition metal $L$-edges was used as the primary technique to unravel the cationic occupation along with the element specific magnetic behavior. TM $K$-edge XANES, $^{57}$Fe M\"ossbauer spectroscopy (in presence and absence of magnetic field), neutron diffraction and SQUID magnetometery have been further utilized to complement the obtained results. A $Fd\bar3m$ structure with N\'eel-type collinear ferrimagnetic spin arrangement and elemental distribution model close to (Co$_{0.6}$Fe$_{0.4}$)(Cr$_{0.3}$Fe$_{0.1}$Mn$_{0.3}$Ni$_{0.3}$)$_2$O$_4$ is observed. In contrast to the existing reports on the spinel-HEO, the observed structural model reveals significant preferences in cationic occupation resulting in the lowest configuration entropy allowed by the given composition in a spinel structure. The observed structure indicates the crucial role of the fundamental enthalpy factors, such as CFSE, in governing the cations occupancy in the spinel-HEO. On a practical level, the detailed structural and magnetic mapping can help to build a knowledge-sharing platform for elucidation and prediction of the properties in the multifunctional spinel-HEO that are governed by the cationic distribution. In the broader context, this initial study highlights significant deviation from a completely disordered configuration often assumed in HEOs, prompting the necessity of comprehensive and complementary studies to determine the actual contribution of configurational entropy. A precise determination of the actual degree of chemical ordering from a lattice perspective or even short-range chemical ordering, resulting in local lowering of crystallographic symmetry, can potentially open up further opportunities for property tailoring in HEOs.

\section*{Supplementary Information}
Supplementary Information includes images presenting the Rietveld refinements of the XRD patterns, HR-TEM micrographs, temperature magnetization data at different measuring fields, virgin magnetization plots at different temperatures, temperature dependent M\"ossbauer spectra and multiplet calculation plots. The numbers obtained from the temperature dependent M\"ossbauer spectra, XMCD sum rules, multiplet calculations, configurational entropy and CFSE calcualtions for the different occupational models are summarized in the tables.

\section*{Acknowledgements}
We acknowledge financial support from the Deutsche Forschungsgemeinschaft (DFG) project HA 1344/43-2 (A.S. and H.H.) and WE 2623/14-2 Project-ID 322462997 (B.E. and H.W.). L.V. acknowledges Karlsruhe Nano Micro Facility (KNMF) for the use of the TEM. J.S. acknowledges funding support from DAAD/IIT-Master-Sandwich-Programm 2018-19. We thank support from Eugen Weschke and Helmholtz-Zentrum Berlin for the beam time at the beamline UE46 PGM\_1 (proposal 192-08578-ST/R),  Qiang Zhang, Melanie Kirkham and Oak Ridge National Laboratory for beam time at BL-11A POWGEN (proposal IPTS-22136.1), Ruidy Nemausat and DESY (Hamburg, Germany) for the beam time at PETRA III P65 (proposal 20190485).


\begin{thebibliography}{78}%
\makeatletter
\providecommand \@ifxundefined [1]{%
 \@ifx{#1\undefined}
}%
\providecommand \@ifnum [1]{%
 \ifnum #1\expandafter \@firstoftwo
 \else \expandafter \@secondoftwo
 \fi
}%
\providecommand \@ifx [1]{%
 \ifx #1\expandafter \@firstoftwo
 \else \expandafter \@secondoftwo
 \fi
}%
\providecommand \natexlab [1]{#1}%
\providecommand \enquote  [1]{``#1''}%
\providecommand \bibnamefont  [1]{#1}%
\providecommand \bibfnamefont [1]{#1}%
\providecommand \citenamefont [1]{#1}%
\providecommand \href@noop [0]{\@secondoftwo}%
\providecommand \href [0]{\begingroup \@sanitize@url \@href}%
\providecommand \@href[1]{\@@startlink{#1}\@@href}%
\providecommand \@@href[1]{\endgroup#1\@@endlink}%
\providecommand \@sanitize@url [0]{\catcode `\\12\catcode `\$12\catcode
  `\&12\catcode `\#12\catcode `\^12\catcode `\_12\catcode `\%12\relax}%
\providecommand \@@startlink[1]{}%
\providecommand \@@endlink[0]{}%
\providecommand \url  [0]{\begingroup\@sanitize@url \@url }%
\providecommand \@url [1]{\endgroup\@href {#1}{\urlprefix }}%
\providecommand \urlprefix  [0]{URL }%
\providecommand \Eprint [0]{\href }%
\providecommand \doibase [0]{https://doi.org/}%
\providecommand \selectlanguage [0]{\@gobble}%
\providecommand \bibinfo  [0]{\@secondoftwo}%
\providecommand \bibfield  [0]{\@secondoftwo}%
\providecommand \translation [1]{[#1]}%
\providecommand \BibitemOpen [0]{}%
\providecommand \bibitemStop [0]{}%
\providecommand \bibitemNoStop [0]{.\EOS\space}%
\providecommand \EOS [0]{\spacefactor3000\relax}%
\providecommand \BibitemShut  [1]{\csname bibitem#1\endcsname}%
\let\auto@bib@innerbib\@empty
\bibitem [{\citenamefont {Cantor}\ \emph {et~al.}(2004)\citenamefont {Cantor},
  \citenamefont {Chang}, \citenamefont {Knight},\ and\ \citenamefont
  {Vincent}}]{Cantor2004}%
  \BibitemOpen
  \bibfield  {author} {\bibinfo {author} {\bibfnamefont {B.}~\bibnamefont
  {Cantor}}, \bibinfo {author} {\bibfnamefont {I.}~\bibnamefont {Chang}},
  \bibinfo {author} {\bibfnamefont {P.}~\bibnamefont {Knight}},\ and\ \bibinfo
  {author} {\bibfnamefont {A.}~\bibnamefont {Vincent}},\ }\bibfield  {title}
  {\bibinfo {title} {{Microstructural development in equiatomic multicomponent
  alloys}},\ }\href {https://doi.org/10.1016/j.msea.2003.10.257} {\bibfield
  {journal} {\bibinfo  {journal} {Materials Science and Engineering: A}\
  }\textbf {\bibinfo {volume} {375-377}},\ \bibinfo {pages} {213} (\bibinfo
  {year} {2004})}\BibitemShut {NoStop}%
\bibitem [{\citenamefont {Rost}\ \emph {et~al.}(2015)\citenamefont {Rost},
  \citenamefont {Sachet}, \citenamefont {Borman}, \citenamefont {Moballegh},
  \citenamefont {Dickey}, \citenamefont {Hou}, \citenamefont {Jones},
  \citenamefont {Curtarolo},\ and\ \citenamefont {Maria}}]{Rost2015}%
  \BibitemOpen
  \bibfield  {author} {\bibinfo {author} {\bibfnamefont {C.~M.}\ \bibnamefont
  {Rost}}, \bibinfo {author} {\bibfnamefont {E.}~\bibnamefont {Sachet}},
  \bibinfo {author} {\bibfnamefont {T.}~\bibnamefont {Borman}}, \bibinfo
  {author} {\bibfnamefont {A.}~\bibnamefont {Moballegh}}, \bibinfo {author}
  {\bibfnamefont {E.~C.}\ \bibnamefont {Dickey}}, \bibinfo {author}
  {\bibfnamefont {D.}~\bibnamefont {Hou}}, \bibinfo {author} {\bibfnamefont
  {J.~L.}\ \bibnamefont {Jones}}, \bibinfo {author} {\bibfnamefont
  {S.}~\bibnamefont {Curtarolo}},\ and\ \bibinfo {author} {\bibfnamefont
  {J.-P.}\ \bibnamefont {Maria}},\ }\bibfield  {title} {\bibinfo {title}
  {{Entropy-stabilized oxides}},\ }\href {https://doi.org/10.1038/ncomms9485}
  {\bibfield  {journal} {\bibinfo  {journal} {Nature Communications}\ }\textbf
  {\bibinfo {volume} {6}},\ \bibinfo {pages} {8485} (\bibinfo {year}
  {2015})}\BibitemShut {NoStop}%
\bibitem [{\citenamefont {B{\'{e}}rardan}\ \emph {et~al.}(2016)\citenamefont
  {B{\'{e}}rardan}, \citenamefont {Franger}, \citenamefont {Meena},\ and\
  \citenamefont {Dragoe}}]{Berardan2016c}%
  \BibitemOpen
  \bibfield  {author} {\bibinfo {author} {\bibfnamefont {D.}~\bibnamefont
  {B{\'{e}}rardan}}, \bibinfo {author} {\bibfnamefont {S.}~\bibnamefont
  {Franger}}, \bibinfo {author} {\bibfnamefont {A.~K.}\ \bibnamefont {Meena}},\
  and\ \bibinfo {author} {\bibfnamefont {N.}~\bibnamefont {Dragoe}},\
  }\bibfield  {title} {\bibinfo {title} {{Room temperature lithium superionic
  conductivity in high entropy oxides}},\ }\href
  {https://doi.org/10.1039/C6TA03249D} {\bibfield  {journal} {\bibinfo
  {journal} {Journal of Materials Chemistry A}\ }\textbf {\bibinfo {volume}
  {4}},\ \bibinfo {pages} {9536} (\bibinfo {year} {2016})}\BibitemShut
  {NoStop}%
\bibitem [{\citenamefont {Sarkar}\ \emph {et~al.}(2018)\citenamefont {Sarkar},
  \citenamefont {Velasco}, \citenamefont {Wang}, \citenamefont {Wang},
  \citenamefont {Talasila}, \citenamefont {de~Biasi}, \citenamefont
  {K{\"{u}}bel}, \citenamefont {Brezesinski}, \citenamefont {Bhattacharya},
  \citenamefont {Hahn},\ and\ \citenamefont {Breitung}}]{Sarkar2018c}%
  \BibitemOpen
  \bibfield  {author} {\bibinfo {author} {\bibfnamefont {A.}~\bibnamefont
  {Sarkar}}, \bibinfo {author} {\bibfnamefont {L.}~\bibnamefont {Velasco}},
  \bibinfo {author} {\bibfnamefont {D.}~\bibnamefont {Wang}}, \bibinfo {author}
  {\bibfnamefont {Q.}~\bibnamefont {Wang}}, \bibinfo {author} {\bibfnamefont
  {G.}~\bibnamefont {Talasila}}, \bibinfo {author} {\bibfnamefont
  {L.}~\bibnamefont {de~Biasi}}, \bibinfo {author} {\bibfnamefont
  {C.}~\bibnamefont {K{\"{u}}bel}}, \bibinfo {author} {\bibfnamefont
  {T.}~\bibnamefont {Brezesinski}}, \bibinfo {author} {\bibfnamefont {S.~S.}\
  \bibnamefont {Bhattacharya}}, \bibinfo {author} {\bibfnamefont
  {H.}~\bibnamefont {Hahn}},\ and\ \bibinfo {author} {\bibfnamefont
  {B.}~\bibnamefont {Breitung}},\ }\bibfield  {title} {\bibinfo {title} {{High
  entropy oxides for reversible energy storage}},\ }\href
  {https://doi.org/10.1038/s41467-018-05774-5} {\bibfield  {journal} {\bibinfo
  {journal} {Nature Communications}\ }\textbf {\bibinfo {volume} {9}},\
  \bibinfo {pages} {3400} (\bibinfo {year} {2018})}\BibitemShut {NoStop}%
\bibitem [{\citenamefont {Oses}\ \emph {et~al.}(2020)\citenamefont {Oses},
  \citenamefont {Toher},\ and\ \citenamefont {Curtarolo}}]{Oses2020}%
  \BibitemOpen
  \bibfield  {author} {\bibinfo {author} {\bibfnamefont {C.}~\bibnamefont
  {Oses}}, \bibinfo {author} {\bibfnamefont {C.}~\bibnamefont {Toher}},\ and\
  \bibinfo {author} {\bibfnamefont {S.}~\bibnamefont {Curtarolo}},\ }\bibfield
  {title} {\bibinfo {title} {{High-entropy ceramics}},\ }\bibfield  {journal}
  {\bibinfo  {journal} {Nature Reviews Materials}\ }\href
  {https://doi.org/10.1038/s41578-019-0170-8} {10.1038/s41578-019-0170-8}
  (\bibinfo {year} {2020})\BibitemShut {NoStop}%
\bibitem [{\citenamefont {Anand}\ \emph {et~al.}(2018)\citenamefont {Anand},
  \citenamefont {Wynn}, \citenamefont {Handley},\ and\ \citenamefont
  {Freeman}}]{Anand2018}%
  \BibitemOpen
  \bibfield  {author} {\bibinfo {author} {\bibfnamefont {G.}~\bibnamefont
  {Anand}}, \bibinfo {author} {\bibfnamefont {A.~P.}\ \bibnamefont {Wynn}},
  \bibinfo {author} {\bibfnamefont {C.~M.}\ \bibnamefont {Handley}},\ and\
  \bibinfo {author} {\bibfnamefont {C.~L.}\ \bibnamefont {Freeman}},\
  }\bibfield  {title} {\bibinfo {title} {{Phase stability and distortion in
  high-entropy oxides}},\ }\href
  {https://doi.org/10.1016/j.actamat.2017.12.037} {\bibfield  {journal}
  {\bibinfo  {journal} {Acta Materialia}\ }\textbf {\bibinfo {volume} {146}},\
  \bibinfo {pages} {119} (\bibinfo {year} {2018})}\BibitemShut {NoStop}%
\bibitem [{\citenamefont {Sarkar}\ \emph {et~al.}(2019)\citenamefont {Sarkar},
  \citenamefont {Wang}, \citenamefont {Schiele}, \citenamefont {Chellali},
  \citenamefont {Bhattacharya}, \citenamefont {Wang}, \citenamefont
  {Brezesinski}, \citenamefont {Hahn}, \citenamefont {Velasco},\ and\
  \citenamefont {Breitung}}]{Sarkar2019}%
  \BibitemOpen
  \bibfield  {author} {\bibinfo {author} {\bibfnamefont {A.}~\bibnamefont
  {Sarkar}}, \bibinfo {author} {\bibfnamefont {Q.}~\bibnamefont {Wang}},
  \bibinfo {author} {\bibfnamefont {A.}~\bibnamefont {Schiele}}, \bibinfo
  {author} {\bibfnamefont {M.~R.}\ \bibnamefont {Chellali}}, \bibinfo {author}
  {\bibfnamefont {S.~S.}\ \bibnamefont {Bhattacharya}}, \bibinfo {author}
  {\bibfnamefont {D.}~\bibnamefont {Wang}}, \bibinfo {author} {\bibfnamefont
  {T.}~\bibnamefont {Brezesinski}}, \bibinfo {author} {\bibfnamefont
  {H.}~\bibnamefont {Hahn}}, \bibinfo {author} {\bibfnamefont {L.}~\bibnamefont
  {Velasco}},\ and\ \bibinfo {author} {\bibfnamefont {B.}~\bibnamefont
  {Breitung}},\ }\bibfield  {title} {\bibinfo {title} {{High‐Entropy Oxides:
  Fundamental Aspects and Electrochemical Properties}},\ }\href
  {https://doi.org/10.1002/adma.201806236} {\bibfield  {journal} {\bibinfo
  {journal} {Advanced Materials}\ }\textbf {\bibinfo {volume} {31}},\ \bibinfo
  {pages} {1806236} (\bibinfo {year} {2019})}\BibitemShut {NoStop}%
\bibitem [{\citenamefont {Witte}\ \emph {et~al.}(2019)\citenamefont {Witte},
  \citenamefont {Sarkar}, \citenamefont {Kruk}, \citenamefont {Eggert},
  \citenamefont {Brand}, \citenamefont {Wende},\ and\ \citenamefont
  {Hahn}}]{Witte2019b}%
  \BibitemOpen
  \bibfield  {author} {\bibinfo {author} {\bibfnamefont {R.}~\bibnamefont
  {Witte}}, \bibinfo {author} {\bibfnamefont {A.}~\bibnamefont {Sarkar}},
  \bibinfo {author} {\bibfnamefont {R.}~\bibnamefont {Kruk}}, \bibinfo {author}
  {\bibfnamefont {B.}~\bibnamefont {Eggert}}, \bibinfo {author} {\bibfnamefont
  {R.~A.}\ \bibnamefont {Brand}}, \bibinfo {author} {\bibfnamefont
  {H.}~\bibnamefont {Wende}},\ and\ \bibinfo {author} {\bibfnamefont
  {H.}~\bibnamefont {Hahn}},\ }\bibfield  {title} {\bibinfo {title}
  {{High-entropy oxides: An emerging prospect for magnetic rare-earth
  transition metal perovskites}},\ }\href
  {https://doi.org/10.1103/PhysRevMaterials.3.034406} {\bibfield  {journal}
  {\bibinfo  {journal} {Physical Review Materials}\ }\textbf {\bibinfo {volume}
  {3}},\ \bibinfo {pages} {034406} (\bibinfo {year} {2019})},\ \Eprint
  {https://arxiv.org/abs/1901.02395} {arXiv:1901.02395} \BibitemShut {NoStop}%
\bibitem [{\citenamefont {Zhang}\ \emph {et~al.}(2019)\citenamefont {Zhang},
  \citenamefont {Yan}, \citenamefont {Calder}, \citenamefont {Zheng},
  \citenamefont {McGuire}, \citenamefont {Abernathy}, \citenamefont {Ren},
  \citenamefont {Lapidus}, \citenamefont {Page}, \citenamefont {Zheng},
  \citenamefont {Freeland}, \citenamefont {Budai},\ and\ \citenamefont
  {Hermann}}]{Zhang2019}%
  \BibitemOpen
  \bibfield  {author} {\bibinfo {author} {\bibfnamefont {J.}~\bibnamefont
  {Zhang}}, \bibinfo {author} {\bibfnamefont {J.}~\bibnamefont {Yan}}, \bibinfo
  {author} {\bibfnamefont {S.}~\bibnamefont {Calder}}, \bibinfo {author}
  {\bibfnamefont {Q.}~\bibnamefont {Zheng}}, \bibinfo {author} {\bibfnamefont
  {M.~A.}\ \bibnamefont {McGuire}}, \bibinfo {author} {\bibfnamefont {D.~L.}\
  \bibnamefont {Abernathy}}, \bibinfo {author} {\bibfnamefont {Y.}~\bibnamefont
  {Ren}}, \bibinfo {author} {\bibfnamefont {S.~H.}\ \bibnamefont {Lapidus}},
  \bibinfo {author} {\bibfnamefont {K.}~\bibnamefont {Page}}, \bibinfo {author}
  {\bibfnamefont {H.}~\bibnamefont {Zheng}}, \bibinfo {author} {\bibfnamefont
  {J.~W.}\ \bibnamefont {Freeland}}, \bibinfo {author} {\bibfnamefont {J.~D.}\
  \bibnamefont {Budai}},\ and\ \bibinfo {author} {\bibfnamefont {R.~P.}\
  \bibnamefont {Hermann}},\ }\bibfield  {title} {\bibinfo {title} {{Long-Range
  Antiferromagnetic Order in a Rocksalt High Entropy Oxide}},\ }\href
  {https://doi.org/10.1021/acs.chemmater.9b00624} {\bibfield  {journal}
  {\bibinfo  {journal} {Chem. Mater.}\ }\textbf {\bibinfo {volume} {31}},\
  \bibinfo {pages} {3705} (\bibinfo {year} {2019})},\ \Eprint
  {https://arxiv.org/abs/1902.00833} {arXiv:1902.00833} \BibitemShut {NoStop}%
\bibitem [{\citenamefont {Xu}\ \emph {et~al.}(2020)\citenamefont {Xu},
  \citenamefont {Zhang}, \citenamefont {Liu}, \citenamefont {Do-Thanh},
  \citenamefont {Chen}, \citenamefont {Xu}, \citenamefont {Lin}, \citenamefont
  {Jiao}, \citenamefont {Wang}, \citenamefont {Wang}, \citenamefont {Chen},\
  and\ \citenamefont {Dai}}]{Xu2020}%
  \BibitemOpen
  \bibfield  {author} {\bibinfo {author} {\bibfnamefont {H.}~\bibnamefont
  {Xu}}, \bibinfo {author} {\bibfnamefont {Z.}~\bibnamefont {Zhang}}, \bibinfo
  {author} {\bibfnamefont {J.}~\bibnamefont {Liu}}, \bibinfo {author}
  {\bibfnamefont {C.-L.}\ \bibnamefont {Do-Thanh}}, \bibinfo {author}
  {\bibfnamefont {H.}~\bibnamefont {Chen}}, \bibinfo {author} {\bibfnamefont
  {S.}~\bibnamefont {Xu}}, \bibinfo {author} {\bibfnamefont {Q.}~\bibnamefont
  {Lin}}, \bibinfo {author} {\bibfnamefont {Y.}~\bibnamefont {Jiao}}, \bibinfo
  {author} {\bibfnamefont {J.}~\bibnamefont {Wang}}, \bibinfo {author}
  {\bibfnamefont {Y.}~\bibnamefont {Wang}}, \bibinfo {author} {\bibfnamefont
  {Y.}~\bibnamefont {Chen}},\ and\ \bibinfo {author} {\bibfnamefont
  {S.}~\bibnamefont {Dai}},\ }\bibfield  {title} {\bibinfo {title}
  {{Entropy-stabilized single-atom Pd catalysts via high-entropy fluorite oxide
  supports}},\ }\href {https://doi.org/10.1038/s41467-020-17738-9} {\bibfield
  {journal} {\bibinfo  {journal} {Nature Communications}\ }\textbf {\bibinfo
  {volume} {11}},\ \bibinfo {pages} {1} (\bibinfo {year} {2020})}\BibitemShut
  {NoStop}%
\bibitem [{\citenamefont {Wang}\ \emph {et~al.}(2019)\citenamefont {Wang},
  \citenamefont {Sarkar}, \citenamefont {Wang}, \citenamefont {Velasco},
  \citenamefont {Azmi}, \citenamefont {Bhattacharya}, \citenamefont
  {Bergfeldt}, \citenamefont {D{\"{u}}vel}, \citenamefont {Heitjans},
  \citenamefont {Brezesinski}, \citenamefont {Hahn},\ and\ \citenamefont
  {Breitung}}]{QingsongWang2019}%
  \BibitemOpen
  \bibfield  {author} {\bibinfo {author} {\bibfnamefont {Q.}~\bibnamefont
  {Wang}}, \bibinfo {author} {\bibfnamefont {A.}~\bibnamefont {Sarkar}},
  \bibinfo {author} {\bibfnamefont {D.}~\bibnamefont {Wang}}, \bibinfo {author}
  {\bibfnamefont {L.}~\bibnamefont {Velasco}}, \bibinfo {author} {\bibfnamefont
  {R.}~\bibnamefont {Azmi}}, \bibinfo {author} {\bibfnamefont {S.~S.}\
  \bibnamefont {Bhattacharya}}, \bibinfo {author} {\bibfnamefont
  {T.}~\bibnamefont {Bergfeldt}}, \bibinfo {author} {\bibfnamefont
  {A.}~\bibnamefont {D{\"{u}}vel}}, \bibinfo {author} {\bibfnamefont
  {P.}~\bibnamefont {Heitjans}}, \bibinfo {author} {\bibfnamefont
  {T.}~\bibnamefont {Brezesinski}}, \bibinfo {author} {\bibfnamefont
  {H.}~\bibnamefont {Hahn}},\ and\ \bibinfo {author} {\bibfnamefont
  {B.}~\bibnamefont {Breitung}},\ }\bibfield  {title} {\bibinfo {title}
  {{Multi-anionic and -cationic compounds: new high entropy materials for
  advanced Li-ion batteries}},\ }\href {https://doi.org/10.1039/C9EE00368A}
  {\bibfield  {journal} {\bibinfo  {journal} {Energy Environ. Sci.}\ }\textbf
  {\bibinfo {volume} {12}},\ \bibinfo {pages} {2433} (\bibinfo {year}
  {2019})}\BibitemShut {NoStop}%
\bibitem [{\citenamefont {Lun}\ \emph {et~al.}(2021)\citenamefont {Lun},
  \citenamefont {Ouyang}, \citenamefont {Kwon}, \citenamefont {Ha},
  \citenamefont {Foley}, \citenamefont {Huang}, \citenamefont {Cai},
  \citenamefont {Kim}, \citenamefont {Balasubramanian}, \citenamefont {Sun},
  \citenamefont {Huang}, \citenamefont {Tian}, \citenamefont {Kim},
  \citenamefont {McCloskey}, \citenamefont {Yang}, \citenamefont
  {Cl{\'{e}}ment}, \citenamefont {Ji},\ and\ \citenamefont {Ceder}}]{Lun2021}%
  \BibitemOpen
  \bibfield  {author} {\bibinfo {author} {\bibfnamefont {Z.}~\bibnamefont
  {Lun}}, \bibinfo {author} {\bibfnamefont {B.}~\bibnamefont {Ouyang}},
  \bibinfo {author} {\bibfnamefont {D.-H.}\ \bibnamefont {Kwon}}, \bibinfo
  {author} {\bibfnamefont {Y.}~\bibnamefont {Ha}}, \bibinfo {author}
  {\bibfnamefont {E.~E.}\ \bibnamefont {Foley}}, \bibinfo {author}
  {\bibfnamefont {T.-Y.}\ \bibnamefont {Huang}}, \bibinfo {author}
  {\bibfnamefont {Z.}~\bibnamefont {Cai}}, \bibinfo {author} {\bibfnamefont
  {H.}~\bibnamefont {Kim}}, \bibinfo {author} {\bibfnamefont {M.}~\bibnamefont
  {Balasubramanian}}, \bibinfo {author} {\bibfnamefont {Y.}~\bibnamefont
  {Sun}}, \bibinfo {author} {\bibfnamefont {J.}~\bibnamefont {Huang}}, \bibinfo
  {author} {\bibfnamefont {Y.}~\bibnamefont {Tian}}, \bibinfo {author}
  {\bibfnamefont {H.}~\bibnamefont {Kim}}, \bibinfo {author} {\bibfnamefont
  {B.~D.}\ \bibnamefont {McCloskey}}, \bibinfo {author} {\bibfnamefont
  {W.}~\bibnamefont {Yang}}, \bibinfo {author} {\bibfnamefont {R.~J.}\
  \bibnamefont {Cl{\'{e}}ment}}, \bibinfo {author} {\bibfnamefont
  {H.}~\bibnamefont {Ji}},\ and\ \bibinfo {author} {\bibfnamefont
  {G.}~\bibnamefont {Ceder}},\ }\bibfield  {title} {\bibinfo {title}
  {{Cation-disordered rocksalt-type high-entropy cathodes for Li-ion
  batteries}},\ }\href {https://doi.org/10.1038/s41563-020-00816-0} {\bibfield
  {journal} {\bibinfo  {journal} {Nature Materials}\ }\textbf {\bibinfo
  {volume} {20}},\ \bibinfo {pages} {214} (\bibinfo {year} {2021})}\BibitemShut
  {NoStop}%
\bibitem [{\citenamefont {Sharma}\ \emph {et~al.}(2020)\citenamefont {Sharma},
  \citenamefont {Zheng}, \citenamefont {Mazza}, \citenamefont {Skoropata},
  \citenamefont {Heitmann}, \citenamefont {Gai}, \citenamefont {Musico},
  \citenamefont {Miceli}, \citenamefont {Sales}, \citenamefont {Keppens},
  \citenamefont {Brahlek},\ and\ \citenamefont {Ward}}]{Sharma2020}%
  \BibitemOpen
  \bibfield  {author} {\bibinfo {author} {\bibfnamefont {Y.}~\bibnamefont
  {Sharma}}, \bibinfo {author} {\bibfnamefont {Q.}~\bibnamefont {Zheng}},
  \bibinfo {author} {\bibfnamefont {A.~R.}\ \bibnamefont {Mazza}}, \bibinfo
  {author} {\bibfnamefont {E.}~\bibnamefont {Skoropata}}, \bibinfo {author}
  {\bibfnamefont {T.}~\bibnamefont {Heitmann}}, \bibinfo {author}
  {\bibfnamefont {Z.}~\bibnamefont {Gai}}, \bibinfo {author} {\bibfnamefont
  {B.}~\bibnamefont {Musico}}, \bibinfo {author} {\bibfnamefont {P.~F.}\
  \bibnamefont {Miceli}}, \bibinfo {author} {\bibfnamefont {B.~C.}\
  \bibnamefont {Sales}}, \bibinfo {author} {\bibfnamefont {V.}~\bibnamefont
  {Keppens}}, \bibinfo {author} {\bibfnamefont {M.}~\bibnamefont {Brahlek}},\
  and\ \bibinfo {author} {\bibfnamefont {T.~Z.}\ \bibnamefont {Ward}},\
  }\bibfield  {title} {\bibinfo {title} {Magnetic anisotropy in single-crystal
  high-entropy perovskite oxide
  $\mathrm{La}(\mathrm{C}{\mathrm{r}}_{0.2}\mathrm{M}{\mathrm{n}}_{0.2}\mathrm{F}{\mathrm{e}}_{0.2}\mathrm{C}{\mathrm{o}}_{0.2}\mathrm{N}{\mathrm{i}}_{0.2}){\mathrm{o}}_{3}$
  films},\ }\href {https://doi.org/10.1103/PhysRevMaterials.4.014404}
  {\bibfield  {journal} {\bibinfo  {journal} {Phys. Rev. Materials}\ }\textbf
  {\bibinfo {volume} {4}},\ \bibinfo {pages} {014404} (\bibinfo {year}
  {2020})}\BibitemShut {NoStop}%
\bibitem [{\citenamefont {Ding}\ \emph {et~al.}(2019)\citenamefont {Ding},
  \citenamefont {Zhang}, \citenamefont {Chen}, \citenamefont {Fu},
  \citenamefont {Chen}, \citenamefont {Chen}, \citenamefont {Gu}, \citenamefont
  {Wei}, \citenamefont {Bei}, \citenamefont {Gao}, \citenamefont {Wen},
  \citenamefont {Li}, \citenamefont {Zhang}, \citenamefont {Zhu}, \citenamefont
  {Ritchie},\ and\ \citenamefont {Yu}}]{Ding2019HEA}%
  \BibitemOpen
  \bibfield  {author} {\bibinfo {author} {\bibfnamefont {Q.}~\bibnamefont
  {Ding}}, \bibinfo {author} {\bibfnamefont {Y.}~\bibnamefont {Zhang}},
  \bibinfo {author} {\bibfnamefont {X.}~\bibnamefont {Chen}}, \bibinfo {author}
  {\bibfnamefont {X.}~\bibnamefont {Fu}}, \bibinfo {author} {\bibfnamefont
  {D.}~\bibnamefont {Chen}}, \bibinfo {author} {\bibfnamefont {S.}~\bibnamefont
  {Chen}}, \bibinfo {author} {\bibfnamefont {L.}~\bibnamefont {Gu}}, \bibinfo
  {author} {\bibfnamefont {F.}~\bibnamefont {Wei}}, \bibinfo {author}
  {\bibfnamefont {H.}~\bibnamefont {Bei}}, \bibinfo {author} {\bibfnamefont
  {Y.}~\bibnamefont {Gao}}, \bibinfo {author} {\bibfnamefont {M.}~\bibnamefont
  {Wen}}, \bibinfo {author} {\bibfnamefont {J.}~\bibnamefont {Li}}, \bibinfo
  {author} {\bibfnamefont {Z.}~\bibnamefont {Zhang}}, \bibinfo {author}
  {\bibfnamefont {T.}~\bibnamefont {Zhu}}, \bibinfo {author} {\bibfnamefont
  {R.~O.}\ \bibnamefont {Ritchie}},\ and\ \bibinfo {author} {\bibfnamefont
  {Q.}~\bibnamefont {Yu}},\ }\bibfield  {title} {\bibinfo {title} {Tuning
  element distribution, structure and properties by composition in high-entropy
  alloys},\ }\href {https://doi.org/10.1038/s41586-019-1617-1} {\bibfield
  {journal} {\bibinfo  {journal} {Nature}\ }\textbf {\bibinfo {volume} {574}},\
  \bibinfo {pages} {223} (\bibinfo {year} {2019})}\BibitemShut {NoStop}%
\bibitem [{\citenamefont {Zhang}\ \emph {et~al.}(2020)\citenamefont {Zhang},
  \citenamefont {Zhao}, \citenamefont {Ding}, \citenamefont {Chong},
  \citenamefont {Jia}, \citenamefont {Ophus}, \citenamefont {Asta},
  \citenamefont {Ritchie},\ and\ \citenamefont {Minor}}]{Zhang2020Nat}%
  \BibitemOpen
  \bibfield  {author} {\bibinfo {author} {\bibfnamefont {R.}~\bibnamefont
  {Zhang}}, \bibinfo {author} {\bibfnamefont {S.}~\bibnamefont {Zhao}},
  \bibinfo {author} {\bibfnamefont {J.}~\bibnamefont {Ding}}, \bibinfo {author}
  {\bibfnamefont {Y.}~\bibnamefont {Chong}}, \bibinfo {author} {\bibfnamefont
  {T.}~\bibnamefont {Jia}}, \bibinfo {author} {\bibfnamefont {C.}~\bibnamefont
  {Ophus}}, \bibinfo {author} {\bibfnamefont {M.}~\bibnamefont {Asta}},
  \bibinfo {author} {\bibfnamefont {R.~O.}\ \bibnamefont {Ritchie}},\ and\
  \bibinfo {author} {\bibfnamefont {A.~M.}\ \bibnamefont {Minor}},\ }\bibfield
  {title} {\bibinfo {title} {Short-range order and its impact on the crconi
  medium-entropy alloy},\ }\href {https://doi.org/10.1038/s41586-020-2275-z}
  {\bibfield  {journal} {\bibinfo  {journal} {Nature}\ }\textbf {\bibinfo
  {volume} {581}},\ \bibinfo {pages} {283} (\bibinfo {year}
  {2020})}\BibitemShut {NoStop}%
\bibitem [{\citenamefont {Navrotsky}\ and\ \citenamefont
  {Kleppa}(1967)}]{NAVROTSKY1967}%
  \BibitemOpen
  \bibfield  {author} {\bibinfo {author} {\bibfnamefont {A.}~\bibnamefont
  {Navrotsky}}\ and\ \bibinfo {author} {\bibfnamefont {O.~J.}\ \bibnamefont
  {Kleppa}},\ }\bibfield  {title} {\bibinfo {title} {{The thermodynamics of
  cation distributions in simple spinels}},\ }\href
  {https://doi.org/https://doi.org/10.1016/0022-1902(67)80008-3} {\bibfield
  {journal} {\bibinfo  {journal} {Journal of Inorganic and Nuclear Chemistry}\
  }\textbf {\bibinfo {volume} {29}},\ \bibinfo {pages} {2701} (\bibinfo {year}
  {1967})}\BibitemShut {NoStop}%
\bibitem [{\citenamefont {D{\c{a}}browa}\ \emph {et~al.}(2018)\citenamefont
  {D{\c{a}}browa}, \citenamefont {Stygar}, \citenamefont {Miku{\l}a},
  \citenamefont {Knapik}, \citenamefont {Mroczka}, \citenamefont {Tejchman},
  \citenamefont {Danielewski},\ and\ \citenamefont {Martin}}]{Dabrowa2018}%
  \BibitemOpen
  \bibfield  {author} {\bibinfo {author} {\bibfnamefont {J.}~\bibnamefont
  {D{\c{a}}browa}}, \bibinfo {author} {\bibfnamefont {M.}~\bibnamefont
  {Stygar}}, \bibinfo {author} {\bibfnamefont {A.}~\bibnamefont {Miku{\l}a}},
  \bibinfo {author} {\bibfnamefont {A.}~\bibnamefont {Knapik}}, \bibinfo
  {author} {\bibfnamefont {K.}~\bibnamefont {Mroczka}}, \bibinfo {author}
  {\bibfnamefont {W.}~\bibnamefont {Tejchman}}, \bibinfo {author}
  {\bibfnamefont {M.}~\bibnamefont {Danielewski}},\ and\ \bibinfo {author}
  {\bibfnamefont {M.}~\bibnamefont {Martin}},\ }\bibfield  {title} {\bibinfo
  {title} {{Synthesis and microstructure of the (Co,Cr,Fe,Mn,Ni)3O4high entropy
  oxide characterized by spinel structure}},\ }\href
  {https://doi.org/10.1016/j.matlet.2017.12.148} {\bibfield  {journal}
  {\bibinfo  {journal} {Mater. Lett.}\ }\textbf {\bibinfo {volume} {216}},\
  \bibinfo {pages} {32} (\bibinfo {year} {2018})}\BibitemShut {NoStop}%
\bibitem [{\citenamefont {Albedwawi}\ \emph {et~al.}(2021)\citenamefont
  {Albedwawi}, \citenamefont {AlJaberi}, \citenamefont {Haidemenopoulos},\ and\
  \citenamefont {Polychronopoulou}}]{ALBEDWAWI2021109534}%
  \BibitemOpen
  \bibfield  {author} {\bibinfo {author} {\bibfnamefont {S.~H.}\ \bibnamefont
  {Albedwawi}}, \bibinfo {author} {\bibfnamefont {A.}~\bibnamefont {AlJaberi}},
  \bibinfo {author} {\bibfnamefont {G.~N.}\ \bibnamefont {Haidemenopoulos}},\
  and\ \bibinfo {author} {\bibfnamefont {K.}~\bibnamefont {Polychronopoulou}},\
  }\bibfield  {title} {\bibinfo {title} {High entropy oxides-exploring a
  paradigm of promising catalysts: A review},\ }\href
  {https://doi.org/https://doi.org/10.1016/j.matdes.2021.109534} {\bibfield
  {journal} {\bibinfo  {journal} {Materials \& Design}\ }\textbf {\bibinfo
  {volume} {202}},\ \bibinfo {pages} {109534} (\bibinfo {year}
  {2021})}\BibitemShut {NoStop}%
\bibitem [{\citenamefont {Salian}\ and\ \citenamefont
  {Mandal}(2021)}]{Salian21}%
  \BibitemOpen
  \bibfield  {author} {\bibinfo {author} {\bibfnamefont {A.}~\bibnamefont
  {Salian}}\ and\ \bibinfo {author} {\bibfnamefont {S.}~\bibnamefont
  {Mandal}},\ }\bibfield  {title} {\bibinfo {title} {Entropy stabilized
  multicomponent oxides with diverse functionality – a review},\ }\href
  {https://doi.org/10.1080/10408436.2021.1886047} {\bibfield  {journal}
  {\bibinfo  {journal} {Critical Reviews in Solid State and Materials
  Sciences}\ }\textbf {\bibinfo {volume} {0}},\ \bibinfo {pages} {1} (\bibinfo
  {year} {2021})},\ \Eprint
  {https://arxiv.org/abs/https://doi.org/10.1080/10408436.2021.1886047}
  {https://doi.org/10.1080/10408436.2021.1886047} \BibitemShut {NoStop}%
\bibitem [{\citenamefont {Fracchia}\ \emph {et~al.}(2020)\citenamefont
  {Fracchia}, \citenamefont {Manzoli}, \citenamefont {Anselmi-Tamburini},\ and\
  \citenamefont {Ghigna}}]{FRACCHIA2020}%
  \BibitemOpen
  \bibfield  {author} {\bibinfo {author} {\bibfnamefont {M.}~\bibnamefont
  {Fracchia}}, \bibinfo {author} {\bibfnamefont {M.}~\bibnamefont {Manzoli}},
  \bibinfo {author} {\bibfnamefont {U.}~\bibnamefont {Anselmi-Tamburini}},\
  and\ \bibinfo {author} {\bibfnamefont {P.}~\bibnamefont {Ghigna}},\
  }\bibfield  {title} {\bibinfo {title} {{A new eight-cation inverse high
  entropy spinel with large configurational entropy in both tetrahedral and
  octahedral sites: Synthesis and cation distribution by X-ray absorption
  spectroscopy}},\ }\href
  {https://doi.org/https://doi.org/10.1016/j.scriptamat.2020.07.002} {\bibfield
   {journal} {\bibinfo  {journal} {Scripta Materialia}\ }\textbf {\bibinfo
  {volume} {188}},\ \bibinfo {pages} {26} (\bibinfo {year} {2020})}\BibitemShut
  {NoStop}%
\bibitem [{\citenamefont {Mao}\ \emph {et~al.}(2020)\citenamefont {Mao},
  \citenamefont {Xiang}, \citenamefont {Zhang}, \citenamefont {Kuramoto},
  \citenamefont {Zhang},\ and\ \citenamefont {Jia}}]{Mao2020}%
  \BibitemOpen
  \bibfield  {author} {\bibinfo {author} {\bibfnamefont {A.}~\bibnamefont
  {Mao}}, \bibinfo {author} {\bibfnamefont {H.-Z.}\ \bibnamefont {Xiang}},
  \bibinfo {author} {\bibfnamefont {Z.-G.}\ \bibnamefont {Zhang}}, \bibinfo
  {author} {\bibfnamefont {K.}~\bibnamefont {Kuramoto}}, \bibinfo {author}
  {\bibfnamefont {H.}~\bibnamefont {Zhang}},\ and\ \bibinfo {author}
  {\bibfnamefont {Y.}~\bibnamefont {Jia}},\ }\bibfield  {title} {\bibinfo
  {title} {{A new class of spinel high-entropy oxides with controllable
  magnetic properties}},\ }\href {https://doi.org/10.1016/j.jmmm.2019.165884}
  {\bibfield  {journal} {\bibinfo  {journal} {Journal of Magnetism and Magnetic
  Materials}\ }\textbf {\bibinfo {volume} {497}},\ \bibinfo {pages} {165884}
  (\bibinfo {year} {2020})}\BibitemShut {NoStop}%
\bibitem [{\citenamefont {Music{\'{o}}}\ \emph {et~al.}(2019)\citenamefont
  {Music{\'{o}}}, \citenamefont {Wright}, \citenamefont {Ward}, \citenamefont
  {Grutter}, \citenamefont {Arenholz}, \citenamefont {Gilbert}, \citenamefont
  {Mandrus},\ and\ \citenamefont {Keppens}}]{Musico2019}%
  \BibitemOpen
  \bibfield  {author} {\bibinfo {author} {\bibfnamefont {B.}~\bibnamefont
  {Music{\'{o}}}}, \bibinfo {author} {\bibfnamefont {Q.}~\bibnamefont
  {Wright}}, \bibinfo {author} {\bibfnamefont {T.~Z.}\ \bibnamefont {Ward}},
  \bibinfo {author} {\bibfnamefont {A.}~\bibnamefont {Grutter}}, \bibinfo
  {author} {\bibfnamefont {E.}~\bibnamefont {Arenholz}}, \bibinfo {author}
  {\bibfnamefont {D.}~\bibnamefont {Gilbert}}, \bibinfo {author} {\bibfnamefont
  {D.}~\bibnamefont {Mandrus}},\ and\ \bibinfo {author} {\bibfnamefont
  {V.}~\bibnamefont {Keppens}},\ }\bibfield  {title} {\bibinfo {title}
  {{Tunable magnetic ordering through cation selection in entropic spinel
  oxides}},\ }\href {https://doi.org/10.1103/PhysRevMaterials.3.104416}
  {\bibfield  {journal} {\bibinfo  {journal} {Physical Review Materials}\
  }\textbf {\bibinfo {volume} {3}},\ \bibinfo {pages} {104416} (\bibinfo {year}
  {2019})}\BibitemShut {NoStop}%
\bibitem [{\citenamefont {Wang}\ \emph
  {et~al.}(2020{\natexlab{a}})\citenamefont {Wang}, \citenamefont {Jiang},
  \citenamefont {Duan}, \citenamefont {Mao}, \citenamefont {Dong},
  \citenamefont {Dong}, \citenamefont {Wang}, \citenamefont {Luo},
  \citenamefont {Liu},\ and\ \citenamefont {Qi}}]{Wang2020s}%
  \BibitemOpen
  \bibfield  {author} {\bibinfo {author} {\bibfnamefont {D.}~\bibnamefont
  {Wang}}, \bibinfo {author} {\bibfnamefont {S.}~\bibnamefont {Jiang}},
  \bibinfo {author} {\bibfnamefont {C.}~\bibnamefont {Duan}}, \bibinfo {author}
  {\bibfnamefont {J.}~\bibnamefont {Mao}}, \bibinfo {author} {\bibfnamefont
  {Y.}~\bibnamefont {Dong}}, \bibinfo {author} {\bibfnamefont {K.}~\bibnamefont
  {Dong}}, \bibinfo {author} {\bibfnamefont {Z.}~\bibnamefont {Wang}}, \bibinfo
  {author} {\bibfnamefont {S.}~\bibnamefont {Luo}}, \bibinfo {author}
  {\bibfnamefont {Y.}~\bibnamefont {Liu}},\ and\ \bibinfo {author}
  {\bibfnamefont {X.}~\bibnamefont {Qi}},\ }\bibfield  {title} {\bibinfo
  {title} {{Spinel-structured high entropy oxide (FeCoNiCrMn)3O4 as anode
  towards superior lithium storage performance}},\ }\href
  {https://doi.org/10.1016/j.jallcom.2020.156158} {\bibfield  {journal}
  {\bibinfo  {journal} {Journal of Alloys and Compounds}\ }\textbf {\bibinfo
  {volume} {844}},\ \bibinfo {pages} {156158} (\bibinfo {year}
  {2020}{\natexlab{a}})}\BibitemShut {NoStop}%
\bibitem [{\citenamefont {Nguyen}\ \emph
  {et~al.}(2020{\natexlab{a}})\citenamefont {Nguyen}, \citenamefont {Patra},
  \citenamefont {Chang},\ and\ \citenamefont {Ting}}]{Nguyen2020}%
  \BibitemOpen
  \bibfield  {author} {\bibinfo {author} {\bibfnamefont {T.~X.}\ \bibnamefont
  {Nguyen}}, \bibinfo {author} {\bibfnamefont {J.}~\bibnamefont {Patra}},
  \bibinfo {author} {\bibfnamefont {J.-K.}\ \bibnamefont {Chang}},\ and\
  \bibinfo {author} {\bibfnamefont {J.-M.}\ \bibnamefont {Ting}},\ }\bibfield
  {title} {\bibinfo {title} {{High entropy spinel oxide nanoparticles for
  superior lithiation–delithiation performance}},\ }\href
  {https://doi.org/10.1039/D0TA04844E} {\bibfield  {journal} {\bibinfo
  {journal} {J. Mater. Chem. A}\ ,\ } (\bibinfo {year}
  {2020}{\natexlab{a}})}\BibitemShut {NoStop}%
\bibitem [{\citenamefont {Lin}\ \emph {et~al.}(2021)\citenamefont {Lin},
  \citenamefont {Chang}, \citenamefont {Kaun},\ and\ \citenamefont
  {Su}}]{Lin_cryst21}%
  \BibitemOpen
  \bibfield  {author} {\bibinfo {author} {\bibfnamefont {C.-C.}\ \bibnamefont
  {Lin}}, \bibinfo {author} {\bibfnamefont {C.-W.}\ \bibnamefont {Chang}},
  \bibinfo {author} {\bibfnamefont {C.-C.}\ \bibnamefont {Kaun}},\ and\
  \bibinfo {author} {\bibfnamefont {Y.-H.}\ \bibnamefont {Su}},\ }\bibfield
  {title} {\bibinfo {title} {Stepwise evolution of photocatalytic
  spinel-structured (co,cr,fe,mn,ni)3o4 high entropy oxides from
  first-principles calculations to machine learning},\ }\bibfield  {journal}
  {\bibinfo  {journal} {Crystals}\ }\textbf {\bibinfo {volume} {11}},\ \href
  {https://doi.org/10.3390/cryst11091035} {10.3390/cryst11091035} (\bibinfo
  {year} {2021})\BibitemShut {NoStop}%
\bibitem [{\citenamefont {Nguyen}\ \emph
  {et~al.}(2020{\natexlab{b}})\citenamefont {Nguyen}, \citenamefont {Su},
  \citenamefont {Hattrick-Simpers}, \citenamefont {Joress}, \citenamefont
  {Nagata}, \citenamefont {Chang}, \citenamefont {Sarker}, \citenamefont
  {Mehta},\ and\ \citenamefont {Ting}}]{Nguyen20_Comb}%
  \BibitemOpen
  \bibfield  {author} {\bibinfo {author} {\bibfnamefont {T.~X.}\ \bibnamefont
  {Nguyen}}, \bibinfo {author} {\bibfnamefont {Y.-H.}\ \bibnamefont {Su}},
  \bibinfo {author} {\bibfnamefont {J.}~\bibnamefont {Hattrick-Simpers}},
  \bibinfo {author} {\bibfnamefont {H.}~\bibnamefont {Joress}}, \bibinfo
  {author} {\bibfnamefont {T.}~\bibnamefont {Nagata}}, \bibinfo {author}
  {\bibfnamefont {K.-S.}\ \bibnamefont {Chang}}, \bibinfo {author}
  {\bibfnamefont {S.}~\bibnamefont {Sarker}}, \bibinfo {author} {\bibfnamefont
  {A.}~\bibnamefont {Mehta}},\ and\ \bibinfo {author} {\bibfnamefont {J.-M.}\
  \bibnamefont {Ting}},\ }\bibfield  {title} {\bibinfo {title} {Exploring the
  first high-entropy thin film libraries: Composition spread-controlled
  crystalline structure},\ }\href {https://doi.org/10.1021/acscombsci.0c00159}
  {\bibfield  {journal} {\bibinfo  {journal} {ACS Combinatorial Science}\
  }\textbf {\bibinfo {volume} {22}},\ \bibinfo {pages} {858} (\bibinfo {year}
  {2020}{\natexlab{b}})},\ \bibinfo {note} {pMID: 33146510},\ \Eprint
  {https://arxiv.org/abs/https://doi.org/10.1021/acscombsci.0c00159}
  {https://doi.org/10.1021/acscombsci.0c00159} \BibitemShut {NoStop}%
\bibitem [{\citenamefont {Talluri}\ \emph {et~al.}(2021)\citenamefont
  {Talluri}, \citenamefont {Aparna}, \citenamefont {Sreenivasulu},
  \citenamefont {Bhattacharya},\ and\ \citenamefont
  {Thomas}}]{TALLURI2021103004}%
  \BibitemOpen
  \bibfield  {author} {\bibinfo {author} {\bibfnamefont {B.}~\bibnamefont
  {Talluri}}, \bibinfo {author} {\bibfnamefont {M.}~\bibnamefont {Aparna}},
  \bibinfo {author} {\bibfnamefont {N.}~\bibnamefont {Sreenivasulu}}, \bibinfo
  {author} {\bibfnamefont {S.}~\bibnamefont {Bhattacharya}},\ and\ \bibinfo
  {author} {\bibfnamefont {T.}~\bibnamefont {Thomas}},\ }\bibfield  {title}
  {\bibinfo {title} {High entropy spinel metal oxide (cocrfemnni)3o4
  nanoparticles as a high-performance supercapacitor electrode material},\
  }\href {https://doi.org/https://doi.org/10.1016/j.est.2021.103004} {\bibfield
   {journal} {\bibinfo  {journal} {Journal of Energy Storage}\ }\textbf
  {\bibinfo {volume} {42}},\ \bibinfo {pages} {103004} (\bibinfo {year}
  {2021})}\BibitemShut {NoStop}%
\bibitem [{\citenamefont {Huang}\ \emph {et~al.}(2021)\citenamefont {Huang},
  \citenamefont {Huang}, \citenamefont {Wu}, \citenamefont {Patra},
  \citenamefont {{Xuyen Nguyen}}, \citenamefont {Chang}, \citenamefont
  {Clemens}, \citenamefont {Ting}, \citenamefont {Li}, \citenamefont {Chang},\
  and\ \citenamefont {Wu}}]{HUANG2021129838}%
  \BibitemOpen
  \bibfield  {author} {\bibinfo {author} {\bibfnamefont {C.-Y.}\ \bibnamefont
  {Huang}}, \bibinfo {author} {\bibfnamefont {C.-W.}\ \bibnamefont {Huang}},
  \bibinfo {author} {\bibfnamefont {M.-C.}\ \bibnamefont {Wu}}, \bibinfo
  {author} {\bibfnamefont {J.}~\bibnamefont {Patra}}, \bibinfo {author}
  {\bibfnamefont {T.}~\bibnamefont {{Xuyen Nguyen}}}, \bibinfo {author}
  {\bibfnamefont {M.-T.}\ \bibnamefont {Chang}}, \bibinfo {author}
  {\bibfnamefont {O.}~\bibnamefont {Clemens}}, \bibinfo {author} {\bibfnamefont
  {J.-M.}\ \bibnamefont {Ting}}, \bibinfo {author} {\bibfnamefont
  {J.}~\bibnamefont {Li}}, \bibinfo {author} {\bibfnamefont {J.-K.}\
  \bibnamefont {Chang}},\ and\ \bibinfo {author} {\bibfnamefont {W.-W.}\
  \bibnamefont {Wu}},\ }\bibfield  {title} {\bibinfo {title} {Atomic-scale
  investigation of lithiation/delithiation mechanism in high-entropy spinel
  oxide with superior electrochemical performance},\ }\href
  {https://doi.org/https://doi.org/10.1016/j.cej.2021.129838} {\bibfield
  {journal} {\bibinfo  {journal} {Chemical Engineering Journal}\ }\textbf
  {\bibinfo {volume} {420}},\ \bibinfo {pages} {129838} (\bibinfo {year}
  {2021})}\BibitemShut {NoStop}%
\bibitem [{\citenamefont {Mao}\ \emph {et~al.}(2019)\citenamefont {Mao},
  \citenamefont {Quan}, \citenamefont {Xiang}, \citenamefont {Zhang},
  \citenamefont {Kuramoto},\ and\ \citenamefont {Xia}}]{MAO201911}%
  \BibitemOpen
  \bibfield  {author} {\bibinfo {author} {\bibfnamefont {A.}~\bibnamefont
  {Mao}}, \bibinfo {author} {\bibfnamefont {F.}~\bibnamefont {Quan}}, \bibinfo
  {author} {\bibfnamefont {H.-Z.}\ \bibnamefont {Xiang}}, \bibinfo {author}
  {\bibfnamefont {Z.-G.}\ \bibnamefont {Zhang}}, \bibinfo {author}
  {\bibfnamefont {K.}~\bibnamefont {Kuramoto}},\ and\ \bibinfo {author}
  {\bibfnamefont {A.-L.}\ \bibnamefont {Xia}},\ }\bibfield  {title} {\bibinfo
  {title} {Facile synthesis and ferrimagnetic property of spinel
  (cocrfemnni)3o4 high-entropy oxide nanocrystalline powder},\ }\href
  {https://doi.org/https://doi.org/10.1016/j.molstruc.2019.05.073} {\bibfield
  {journal} {\bibinfo  {journal} {Journal of Molecular Structure}\ }\textbf
  {\bibinfo {volume} {1194}},\ \bibinfo {pages} {11} (\bibinfo {year}
  {2019})}\BibitemShut {NoStop}%
\bibitem [{\citenamefont {Cieslak}\ \emph {et~al.}(2021)\citenamefont
  {Cieslak}, \citenamefont {Reissner}, \citenamefont {Berent}, \citenamefont
  {Dabrowa}, \citenamefont {Stygar}, \citenamefont {Mozdzierz},\ and\
  \citenamefont {Zajusz}}]{Cieslak2021}%
  \BibitemOpen
  \bibfield  {author} {\bibinfo {author} {\bibfnamefont {J.}~\bibnamefont
  {Cieslak}}, \bibinfo {author} {\bibfnamefont {M.}~\bibnamefont {Reissner}},
  \bibinfo {author} {\bibfnamefont {K.}~\bibnamefont {Berent}}, \bibinfo
  {author} {\bibfnamefont {J.}~\bibnamefont {Dabrowa}}, \bibinfo {author}
  {\bibfnamefont {M.}~\bibnamefont {Stygar}}, \bibinfo {author} {\bibfnamefont
  {M.}~\bibnamefont {Mozdzierz}},\ and\ \bibinfo {author} {\bibfnamefont
  {M.}~\bibnamefont {Zajusz}},\ }\bibfield  {title} {\bibinfo {title} {Magnetic
  properties and ionic distribution in high entropy spinels studied by
  mössbauer and ab initio methods},\ }\href
  {https://doi.org/https://doi.org/10.1016/j.actamat.2020.116600} {\bibfield
  {journal} {\bibinfo  {journal} {Acta Materialia}\ }\textbf {\bibinfo {volume}
  {206}},\ \bibinfo {pages} {116600} (\bibinfo {year} {2021})}\BibitemShut
  {NoStop}%
\bibitem [{\citenamefont {Grzesik}\ \emph {et~al.}(2020)\citenamefont
  {Grzesik}, \citenamefont {Smoła}, \citenamefont {Miszczak}, \citenamefont
  {Stygar}, \citenamefont {Dąbrowa}, \citenamefont {Zajusz}, \citenamefont
  {Świerczek},\ and\ \citenamefont {Danielewski}}]{GRZESIK2020835}%
  \BibitemOpen
  \bibfield  {author} {\bibinfo {author} {\bibfnamefont {Z.}~\bibnamefont
  {Grzesik}}, \bibinfo {author} {\bibfnamefont {G.}~\bibnamefont {Smoła}},
  \bibinfo {author} {\bibfnamefont {M.}~\bibnamefont {Miszczak}}, \bibinfo
  {author} {\bibfnamefont {M.}~\bibnamefont {Stygar}}, \bibinfo {author}
  {\bibfnamefont {J.}~\bibnamefont {Dabrowa}}, \bibinfo {author}
  {\bibfnamefont {M.}~\bibnamefont {Zajusz}}, \bibinfo {author} {\bibfnamefont
  {K.}~\bibnamefont {Świerczek}},\ and\ \bibinfo {author} {\bibfnamefont
  {M.}~\bibnamefont {Danielewski}},\ }\bibfield  {title} {\bibinfo {title}
  {Defect structure and transport properties of (co,cr,fe,mn,ni)3o4
  spinel-structured high entropy oxide},\ }\href
  {https://doi.org/https://doi.org/10.1016/j.jeurceramsoc.2019.10.026}
  {\bibfield  {journal} {\bibinfo  {journal} {Journal of the European Ceramic
  Society}\ }\textbf {\bibinfo {volume} {40}},\ \bibinfo {pages} {835}
  (\bibinfo {year} {2020})}\BibitemShut {NoStop}%
\bibitem [{\citenamefont {Nguyen}\ \emph {et~al.}(2021)\citenamefont {Nguyen},
  \citenamefont {Su}, \citenamefont {Lin}, \citenamefont {Ruan},\ and\
  \citenamefont {Ting}}]{Nguyen2021}%
  \BibitemOpen
  \bibfield  {author} {\bibinfo {author} {\bibfnamefont {T.~X.}\ \bibnamefont
  {Nguyen}}, \bibinfo {author} {\bibfnamefont {Y.-H.}\ \bibnamefont {Su}},
  \bibinfo {author} {\bibfnamefont {C.-C.}\ \bibnamefont {Lin}}, \bibinfo
  {author} {\bibfnamefont {J.}~\bibnamefont {Ruan}},\ and\ \bibinfo {author}
  {\bibfnamefont {J.-M.}\ \bibnamefont {Ting}},\ }\bibfield  {title} {\bibinfo
  {title} {A new high entropy glycerate for high performance oxygen evolution
  reaction},\ }\href {https://doi.org/https://doi.org/10.1002/advs.202002446}
  {\bibfield  {journal} {\bibinfo  {journal} {Advanced Science}\ }\textbf
  {\bibinfo {volume} {8}},\ \bibinfo {pages} {2002446} (\bibinfo {year}
  {2021})}\BibitemShut {NoStop}%
\bibitem [{\citenamefont {Stygar}\ \emph {et~al.}(2020)\citenamefont {Stygar},
  \citenamefont {Dąbrowa}, \citenamefont {Moździerz}, \citenamefont {Zajusz},
  \citenamefont {Skubida}, \citenamefont {Mroczka}, \citenamefont {Berent},
  \citenamefont {Świerczek},\ and\ \citenamefont
  {Danielewski}}]{STYGAR20201644}%
  \BibitemOpen
  \bibfield  {author} {\bibinfo {author} {\bibfnamefont {M.}~\bibnamefont
  {Stygar}}, \bibinfo {author} {\bibfnamefont {J.}~\bibnamefont {Dabrowa}},
  \bibinfo {author} {\bibfnamefont {M.}~\bibnamefont {Mozdzierz}}, \bibinfo
  {author} {\bibfnamefont {M.}~\bibnamefont {Zajusz}}, \bibinfo {author}
  {\bibfnamefont {W.}~\bibnamefont {Skubida}}, \bibinfo {author} {\bibfnamefont
  {K.}~\bibnamefont {Mroczka}}, \bibinfo {author} {\bibfnamefont
  {K.}~\bibnamefont {Berent}}, \bibinfo {author} {\bibfnamefont
  {K.}~\bibnamefont {Świerczek}},\ and\ \bibinfo {author} {\bibfnamefont
  {M.}~\bibnamefont {Danielewski}},\ }\bibfield  {title} {\bibinfo {title}
  {Formation and properties of high entropy oxides in co-cr-fe-mg-mn-ni-o
  system: Novel (cr,fe,mg,mn,ni)3o4 and (co,cr,fe,mg,mn)3o4 high entropy
  spinels},\ }\href
  {https://doi.org/https://doi.org/10.1016/j.jeurceramsoc.2019.11.030}
  {\bibfield  {journal} {\bibinfo  {journal} {Journal of the European Ceramic
  Society}\ }\textbf {\bibinfo {volume} {40}},\ \bibinfo {pages} {1644}
  (\bibinfo {year} {2020})}\BibitemShut {NoStop}%
\bibitem [{Top()}]{Topas2015}%
  \BibitemOpen
  \href@noop {} {\emph {\bibinfo {title} {{Topas V5, General profile and
  structure analysis software for powder diffraction data, User's Manual}}}}\
  (\bibinfo  {publisher} {Bruker AXS},\ \bibinfo {address}
  {Karlsruhe})\BibitemShut {NoStop}%
\bibitem [{\citenamefont {Welter}\ \emph {et~al.}(2019)\citenamefont {Welter},
  \citenamefont {Chernikov}, \citenamefont {Herrmann},\ and\ \citenamefont
  {Nemausat}}]{Welter2019}%
  \BibitemOpen
  \bibfield  {author} {\bibinfo {author} {\bibfnamefont {E.}~\bibnamefont
  {Welter}}, \bibinfo {author} {\bibfnamefont {R.}~\bibnamefont {Chernikov}},
  \bibinfo {author} {\bibfnamefont {M.}~\bibnamefont {Herrmann}},\ and\
  \bibinfo {author} {\bibfnamefont {R.}~\bibnamefont {Nemausat}},\ }\bibfield
  {title} {\bibinfo {title} {{A beamline for bulk sample x-ray absorption
  spectroscopy at the high brilliance storage ring PETRA III}},\ }\href
  {https://doi.org/10.1063/1.5084603} {\bibfield  {journal} {\bibinfo
  {journal} {AIP Conference Proceedings}\ }\textbf {\bibinfo {volume} {2054}},\
  \bibinfo {pages} {2} (\bibinfo {year} {2019})}\BibitemShut {NoStop}%
\bibitem [{\citenamefont {Weschke}\ and\ \citenamefont
  {Schierle}(2018)}]{Weschke2018}%
  \BibitemOpen
  \bibfield  {author} {\bibinfo {author} {\bibfnamefont {E.}~\bibnamefont
  {Weschke}}\ and\ \bibinfo {author} {\bibfnamefont {E.}~\bibnamefont
  {Schierle}},\ }\bibfield  {title} {\bibinfo {title} {{The UE46 PGM-1 beamline
  at BESSY II}},\ }\href {https://doi.org/10.17815/jlsrf-4-77} {\bibfield
  {journal} {\bibinfo  {journal} {Journal of large-scale research facilities
  JLSRF}\ }\textbf {\bibinfo {volume} {4}},\ \bibinfo {pages} {1} (\bibinfo
  {year} {2018})}\BibitemShut {NoStop}%
\bibitem [{\citenamefont {Retegan}(2019)}]{retegan_crispy}%
  \BibitemOpen
  \bibfield  {author} {\bibinfo {author} {\bibfnamefont {M.}~\bibnamefont
  {Retegan}},\ }\href {https://doi.org/10.5281/zenodo.1008184} {\bibinfo
  {title} {Crispy: v0.7.3}} (\bibinfo {year} {2019})\BibitemShut {NoStop}%
\bibitem [{\citenamefont {Stavitski}\ and\ \citenamefont
  {de~Groot}(2010)}]{CTM4XAS}%
  \BibitemOpen
  \bibfield  {author} {\bibinfo {author} {\bibfnamefont {E.}~\bibnamefont
  {Stavitski}}\ and\ \bibinfo {author} {\bibfnamefont {F.~M.}\ \bibnamefont
  {de~Groot}},\ }\bibfield  {title} {\bibinfo {title} {The {CTM}4xas program
  for {EELS} and {XAS} spectral shape analysis of transition metal l edges},\
  }\href {https://doi.org/10.1016/j.micron.2010.06.005} {\bibfield  {journal}
  {\bibinfo  {journal} {Micron}\ }\textbf {\bibinfo {volume} {41}},\ \bibinfo
  {pages} {687} (\bibinfo {year} {2010})}\BibitemShut {NoStop}%
\bibitem [{\citenamefont {Rák}\ and\ \citenamefont {Brenner}(2020)}]{Rak2020}%
  \BibitemOpen
  \bibfield  {author} {\bibinfo {author} {\bibfnamefont {Z.}~\bibnamefont
  {Rák}}\ and\ \bibinfo {author} {\bibfnamefont {D.~W.}\ \bibnamefont
  {Brenner}},\ }\bibfield  {title} {\bibinfo {title} {Exchange interactions and
  long-range magnetic order in the (mg,co,cu,ni,zn)o entropy-stabilized oxide:
  A theoretical investigation},\ }\href {https://doi.org/10.1063/5.0008258}
  {\bibfield  {journal} {\bibinfo  {journal} {Journal of Applied Physics}\
  }\textbf {\bibinfo {volume} {127}},\ \bibinfo {pages} {185108} (\bibinfo
  {year} {2020})}\BibitemShut {NoStop}%
\bibitem [{\citenamefont {Demirci}\ \emph {et~al.}(2017)\citenamefont
  {Demirci}, \citenamefont {Manna}, \citenamefont {Wroczynskyj}, \citenamefont
  {Akt{\"{u}}rk},\ and\ \citenamefont {van Lierop}}]{Demirci2017}%
  \BibitemOpen
  \bibfield  {author} {\bibinfo {author} {\bibfnamefont {{\c{C}}.~E.}\
  \bibnamefont {Demirci}}, \bibinfo {author} {\bibfnamefont {P.~K.}\
  \bibnamefont {Manna}}, \bibinfo {author} {\bibfnamefont {Y.}~\bibnamefont
  {Wroczynskyj}}, \bibinfo {author} {\bibfnamefont {S.}~\bibnamefont
  {Akt{\"{u}}rk}},\ and\ \bibinfo {author} {\bibfnamefont {J.}~\bibnamefont
  {van Lierop}},\ }\bibfield  {title} {\bibinfo {title} {{A comparison of the
  magnetism of cobalt-, manganese-, and nickel-ferrite nanoparticles}},\ }\href
  {https://doi.org/10.1088/1361-6463/aa9d79} {\bibfield  {journal} {\bibinfo
  {journal} {Journal of Physics D: Applied Physics}\ }\textbf {\bibinfo
  {volume} {51}},\ \bibinfo {pages} {25003} (\bibinfo {year}
  {2017})}\BibitemShut {NoStop}%
\bibitem [{\citenamefont {Clemens}\ \emph {et~al.}(2014)\citenamefont
  {Clemens}, \citenamefont {Gröting}, \citenamefont {Witte}, \citenamefont
  {Perez-Mato}, \citenamefont {Loho}, \citenamefont {Berry}, \citenamefont
  {Kruk}, \citenamefont {Knight}, \citenamefont {Wright}, \citenamefont
  {Hahn},\ and\ \citenamefont {Slater}}]{Clemens2014}%
  \BibitemOpen
  \bibfield  {author} {\bibinfo {author} {\bibfnamefont {O.}~\bibnamefont
  {Clemens}}, \bibinfo {author} {\bibfnamefont {M.}~\bibnamefont {Gröting}},
  \bibinfo {author} {\bibfnamefont {R.}~\bibnamefont {Witte}}, \bibinfo
  {author} {\bibfnamefont {J.~M.}\ \bibnamefont {Perez-Mato}}, \bibinfo
  {author} {\bibfnamefont {C.}~\bibnamefont {Loho}}, \bibinfo {author}
  {\bibfnamefont {F.~J.}\ \bibnamefont {Berry}}, \bibinfo {author}
  {\bibfnamefont {R.}~\bibnamefont {Kruk}}, \bibinfo {author} {\bibfnamefont
  {K.~S.}\ \bibnamefont {Knight}}, \bibinfo {author} {\bibfnamefont {A.~J.}\
  \bibnamefont {Wright}}, \bibinfo {author} {\bibfnamefont {H.}~\bibnamefont
  {Hahn}},\ and\ \bibinfo {author} {\bibfnamefont {P.~R.}\ \bibnamefont
  {Slater}},\ }\bibfield  {title} {\bibinfo {title} {Crystallographic and
  magnetic structure of the perovskite-type compound bafeo2.5: Unrivaled
  complexity in oxygen vacancy ordering},\ }\href
  {https://doi.org/10.1021/ic402988y} {\bibfield  {journal} {\bibinfo
  {journal} {Inorganic Chemistry}\ }\textbf {\bibinfo {volume} {53}},\ \bibinfo
  {pages} {5911} (\bibinfo {year} {2014})}\BibitemShut {NoStop}%
\bibitem [{\citenamefont {Chappert}\ and\ \citenamefont
  {Frankel}(1967)}]{ChappertPRL1967}%
  \BibitemOpen
  \bibfield  {author} {\bibinfo {author} {\bibfnamefont {J.}~\bibnamefont
  {Chappert}}\ and\ \bibinfo {author} {\bibfnamefont {R.~B.}\ \bibnamefont
  {Frankel}},\ }\bibfield  {title} {\bibinfo {title} {M\"ossbauer study of
  ferrimagnetic ordering in nickel ferrite and chromium-substituted nickel
  ferrite},\ }\href {https://doi.org/10.1103/PhysRevLett.19.570} {\bibfield
  {journal} {\bibinfo  {journal} {Phys. Rev. Lett.}\ }\textbf {\bibinfo
  {volume} {19}},\ \bibinfo {pages} {570} (\bibinfo {year} {1967})}\BibitemShut
  {NoStop}%
\bibitem [{\citenamefont {de~Groot}(2001)}]{Groot2001CR}%
  \BibitemOpen
  \bibfield  {author} {\bibinfo {author} {\bibfnamefont {F.}~\bibnamefont
  {de~Groot}},\ }\bibfield  {title} {\bibinfo {title} {High-resolution x-ray
  emission and x-ray absorption spectroscopy},\ }\href
  {https://doi.org/10.1021/cr9900681} {\bibfield  {journal} {\bibinfo
  {journal} {Chemical Reviews}\ }\textbf {\bibinfo {volume} {101}},\ \bibinfo
  {pages} {1779} (\bibinfo {year} {2001})}\BibitemShut {NoStop}%
\bibitem [{\citenamefont {de~Groot}\ \emph {et~al.}(2009)\citenamefont
  {de~Groot}, \citenamefont {Vank{\'{o}}},\ and\ \citenamefont
  {Glatzel}}]{de_Groot_2009}%
  \BibitemOpen
  \bibfield  {author} {\bibinfo {author} {\bibfnamefont {F.}~\bibnamefont
  {de~Groot}}, \bibinfo {author} {\bibfnamefont {G.}~\bibnamefont
  {Vank{\'{o}}}},\ and\ \bibinfo {author} {\bibfnamefont {P.}~\bibnamefont
  {Glatzel}},\ }\bibfield  {title} {\bibinfo {title} {The 1s x-ray absorption
  pre-edge structures in transition metal oxides},\ }\href
  {https://doi.org/10.1088/0953-8984/21/10/104207} {\bibfield  {journal}
  {\bibinfo  {journal} {Journal of Physics: Condensed Matter}\ }\textbf
  {\bibinfo {volume} {21}},\ \bibinfo {pages} {104207} (\bibinfo {year}
  {2009})}\BibitemShut {NoStop}%
\bibitem [{\citenamefont {Hunault}\ \emph {et~al.}(2014)\citenamefont
  {Hunault}, \citenamefont {Calas}, \citenamefont {Galoisy}, \citenamefont
  {Lelong},\ and\ \citenamefont {Newville}}]{Hunault2013}%
  \BibitemOpen
  \bibfield  {author} {\bibinfo {author} {\bibfnamefont {M.}~\bibnamefont
  {Hunault}}, \bibinfo {author} {\bibfnamefont {G.}~\bibnamefont {Calas}},
  \bibinfo {author} {\bibfnamefont {L.}~\bibnamefont {Galoisy}}, \bibinfo
  {author} {\bibfnamefont {G.}~\bibnamefont {Lelong}},\ and\ \bibinfo {author}
  {\bibfnamefont {M.}~\bibnamefont {Newville}},\ }\bibfield  {title} {\bibinfo
  {title} {{Local Ordering Around Tetrahedral Co2+ in Silicate Glasses}},\
  }\href {https://doi.org/10.1111/jace.12709} {\bibfield  {journal} {\bibinfo
  {journal} {Journal of the American Ceramic Society}\ }\textbf {\bibinfo
  {volume} {97}},\ \bibinfo {pages} {60} (\bibinfo {year} {2014})}\BibitemShut
  {NoStop}%
\bibitem [{\citenamefont {Moen}\ \emph {et~al.}(1997)\citenamefont {Moen},
  \citenamefont {Nicholson}, \citenamefont {Rnning}, \citenamefont {Lamble},
  \citenamefont {Lee},\ and\ \citenamefont {Emerich}}]{Moen1997}%
  \BibitemOpen
  \bibfield  {author} {\bibinfo {author} {\bibfnamefont {A.}~\bibnamefont
  {Moen}}, \bibinfo {author} {\bibfnamefont {D.~G.}\ \bibnamefont {Nicholson}},
  \bibinfo {author} {\bibfnamefont {M.}~\bibnamefont {Rnning}}, \bibinfo
  {author} {\bibfnamefont {G.~M.}\ \bibnamefont {Lamble}}, \bibinfo {author}
  {\bibfnamefont {J.-F.}\ \bibnamefont {Lee}},\ and\ \bibinfo {author}
  {\bibfnamefont {H.}~\bibnamefont {Emerich}},\ }\bibfield  {title} {\bibinfo
  {title} {{X-Ray absorption spectroscopic study at the cobalt K-edge on the
  calcination and reduction of the microporous cobalt silicoaluminophosphate
  catalyst CoSAPO-34}},\ }\href {https://doi.org/10.1039/a704488g} {\bibfield
  {journal} {\bibinfo  {journal} {Journal of the Chemical Society, Faraday
  Transactions}\ }\textbf {\bibinfo {volume} {93}},\ \bibinfo {pages} {4071}
  (\bibinfo {year} {1997})}\BibitemShut {NoStop}%
\bibitem [{\citenamefont {Chen}\ \emph {et~al.}(2013)\citenamefont {Chen},
  \citenamefont {Wu}, \citenamefont {Cui}, \citenamefont {Chu}, \citenamefont
  {Chen},\ and\ \citenamefont {Wu}}]{ChenZrCrO4}%
  \BibitemOpen
  \bibfield  {author} {\bibinfo {author} {\bibfnamefont {S.}~\bibnamefont
  {Chen}}, \bibinfo {author} {\bibfnamefont {Y.}~\bibnamefont {Wu}}, \bibinfo
  {author} {\bibfnamefont {P.}~\bibnamefont {Cui}}, \bibinfo {author}
  {\bibfnamefont {W.}~\bibnamefont {Chu}}, \bibinfo {author} {\bibfnamefont
  {X.}~\bibnamefont {Chen}},\ and\ \bibinfo {author} {\bibfnamefont
  {Z.}~\bibnamefont {Wu}},\ }\bibfield  {title} {\bibinfo {title} {Cation
  distribution in zncr2o4 nanocrystals investigated by x-ray absorption fine
  structure spectroscopy},\ }\href {https://doi.org/10.1021/jp404984y}
  {\bibfield  {journal} {\bibinfo  {journal} {The Journal of Physical Chemistry
  C}\ }\textbf {\bibinfo {volume} {117}},\ \bibinfo {pages} {25019} (\bibinfo
  {year} {2013})},\ \Eprint
  {https://arxiv.org/abs/https://doi.org/10.1021/jp404984y}
  {https://doi.org/10.1021/jp404984y} \BibitemShut {NoStop}%
\bibitem [{\citenamefont {Miyano}\ \emph {et~al.}(1997)\citenamefont {Miyano},
  \citenamefont {Woicik}, \citenamefont {{Sujatha Devi}},\ and\ \citenamefont
  {Gafney}}]{Miyano1997}%
  \BibitemOpen
  \bibfield  {author} {\bibinfo {author} {\bibfnamefont {K.~E.}\ \bibnamefont
  {Miyano}}, \bibinfo {author} {\bibfnamefont {J.~C.}\ \bibnamefont {Woicik}},
  \bibinfo {author} {\bibfnamefont {P.}~\bibnamefont {{Sujatha Devi}}},\ and\
  \bibinfo {author} {\bibfnamefont {H.~D.}\ \bibnamefont {Gafney}},\ }\bibfield
   {title} {\bibinfo {title} {{Cr K edge x-ray absorption study of Cr dopants
  in Mg2SiO4 and Ca2GeO4}},\ }\href {https://doi.org/10.1063/1.119615}
  {\bibfield  {journal} {\bibinfo  {journal} {Applied Physics Letters}\
  }\textbf {\bibinfo {volume} {71}},\ \bibinfo {pages} {1168} (\bibinfo {year}
  {1997})}\BibitemShut {NoStop}%
\bibitem [{\citenamefont {Dubrail}\ and\ \citenamefont
  {Farges}(2009)}]{Dubrail_2009}%
  \BibitemOpen
  \bibfield  {author} {\bibinfo {author} {\bibfnamefont {J.}~\bibnamefont
  {Dubrail}}\ and\ \bibinfo {author} {\bibfnamefont {F.}~\bibnamefont
  {Farges}},\ }\bibfield  {title} {\bibinfo {title} {Not all chromates show the
  same pre-edge feature. implications for the modelling of the speciation of cr
  in environmental systems},\ }\href
  {https://doi.org/10.1088/1742-6596/190/1/012176} {\bibfield  {journal}
  {\bibinfo  {journal} {Journal of Physics: Conference Series}\ }\textbf
  {\bibinfo {volume} {190}},\ \bibinfo {pages} {012176} (\bibinfo {year}
  {2009})}\BibitemShut {NoStop}%
\bibitem [{\citenamefont {Zimmermann}\ \emph {et~al.}(2018)\citenamefont
  {Zimmermann}, \citenamefont {Bouldi}, \citenamefont {Hunault}, \citenamefont
  {Sikora}, \citenamefont {Ablett}, \citenamefont {Rueff}, \citenamefont
  {Lebert}, \citenamefont {Sainctavit}, \citenamefont {{de Groot}},\ and\
  \citenamefont {Juhin}}]{ZIMMERMANN201874}%
  \BibitemOpen
  \bibfield  {author} {\bibinfo {author} {\bibfnamefont {P.}~\bibnamefont
  {Zimmermann}}, \bibinfo {author} {\bibfnamefont {N.}~\bibnamefont {Bouldi}},
  \bibinfo {author} {\bibfnamefont {M.~O.}\ \bibnamefont {Hunault}}, \bibinfo
  {author} {\bibfnamefont {M.}~\bibnamefont {Sikora}}, \bibinfo {author}
  {\bibfnamefont {J.~M.}\ \bibnamefont {Ablett}}, \bibinfo {author}
  {\bibfnamefont {J.-P.}\ \bibnamefont {Rueff}}, \bibinfo {author}
  {\bibfnamefont {B.}~\bibnamefont {Lebert}}, \bibinfo {author} {\bibfnamefont
  {P.}~\bibnamefont {Sainctavit}}, \bibinfo {author} {\bibfnamefont {F.~M.}\
  \bibnamefont {{de Groot}}},\ and\ \bibinfo {author} {\bibfnamefont
  {A.}~\bibnamefont {Juhin}},\ }\bibfield  {title} {\bibinfo {title} {1s2p
  resonant inelastic x-ray scattering magnetic circular dichroism as a probe
  for the local and non-local orbitals in cro2},\ }\href
  {https://doi.org/https://doi.org/10.1016/j.elspec.2017.08.004} {\bibfield
  {journal} {\bibinfo  {journal} {Journal of Electron Spectroscopy and Related
  Phenomena}\ }\textbf {\bibinfo {volume} {222}},\ \bibinfo {pages} {74}
  (\bibinfo {year} {2018})}\BibitemShut {NoStop}%
\bibitem [{\citenamefont {Carta}\ \emph {et~al.}(2007)\citenamefont {Carta},
  \citenamefont {Mountjoy}, \citenamefont {Navarra}, \citenamefont {Casula},
  \citenamefont {Loche}, \citenamefont {Marras},\ and\ \citenamefont
  {Corrias}}]{Carta2007JPCC}%
  \BibitemOpen
  \bibfield  {author} {\bibinfo {author} {\bibfnamefont {D.}~\bibnamefont
  {Carta}}, \bibinfo {author} {\bibfnamefont {G.}~\bibnamefont {Mountjoy}},
  \bibinfo {author} {\bibfnamefont {G.}~\bibnamefont {Navarra}}, \bibinfo
  {author} {\bibfnamefont {M.~F.}\ \bibnamefont {Casula}}, \bibinfo {author}
  {\bibfnamefont {D.}~\bibnamefont {Loche}}, \bibinfo {author} {\bibfnamefont
  {S.}~\bibnamefont {Marras}},\ and\ \bibinfo {author} {\bibfnamefont
  {A.}~\bibnamefont {Corrias}},\ }\bibfield  {title} {\bibinfo {title} {X-ray
  absorption investigation of the formation of cobalt ferrite nanoparticles in
  an aerogel silica matrix},\ }\href {https://doi.org/10.1021/jp0708805}
  {\bibfield  {journal} {\bibinfo  {journal} {The Journal of Physical Chemistry
  C}\ }\textbf {\bibinfo {volume} {111}},\ \bibinfo {pages} {6308} (\bibinfo
  {year} {2007})},\ \Eprint
  {https://arxiv.org/abs/https://doi.org/10.1021/jp0708805}
  {https://doi.org/10.1021/jp0708805} \BibitemShut {NoStop}%
\bibitem [{\citenamefont {Carta}\ \emph {et~al.}(2008)\citenamefont {Carta},
  \citenamefont {Loche}, \citenamefont {Mountjoy}, \citenamefont {Navarra},\
  and\ \citenamefont {Corrias}}]{Carta2008}%
  \BibitemOpen
  \bibfield  {author} {\bibinfo {author} {\bibfnamefont {D.}~\bibnamefont
  {Carta}}, \bibinfo {author} {\bibfnamefont {D.}~\bibnamefont {Loche}},
  \bibinfo {author} {\bibfnamefont {G.}~\bibnamefont {Mountjoy}}, \bibinfo
  {author} {\bibfnamefont {G.}~\bibnamefont {Navarra}},\ and\ \bibinfo {author}
  {\bibfnamefont {A.}~\bibnamefont {Corrias}},\ }\bibfield  {title} {\bibinfo
  {title} {Nife2o4 nanoparticles dispersed in an aerogel silica matrix: An
  x-ray absorption study},\ }\href {https://doi.org/10.1021/jp803982k}
  {\bibfield  {journal} {\bibinfo  {journal} {The Journal of Physical Chemistry
  C}\ }\textbf {\bibinfo {volume} {112}},\ \bibinfo {pages} {15623} (\bibinfo
  {year} {2008})},\ \Eprint
  {https://arxiv.org/abs/https://doi.org/10.1021/jp803982k}
  {https://doi.org/10.1021/jp803982k} \BibitemShut {NoStop}%
\bibitem [{\citenamefont {Chalmin}\ \emph {et~al.}(2009)\citenamefont
  {Chalmin}, \citenamefont {Farges},\ and\ \citenamefont
  {Brown}}]{chalmin2009XANES}%
  \BibitemOpen
  \bibfield  {author} {\bibinfo {author} {\bibfnamefont {E.}~\bibnamefont
  {Chalmin}}, \bibinfo {author} {\bibfnamefont {F.}~\bibnamefont {Farges}},\
  and\ \bibinfo {author} {\bibfnamefont {G.~E.}\ \bibnamefont {Brown}},\
  }\bibfield  {title} {\bibinfo {title} {A pre-edge analysis of mn k-edge xanes
  spectra to help determine the speciation of manganese in minerals and
  glasses},\ }\href {https://doi.org/10.1007/s00410-008-0323-z} {\bibfield
  {journal} {\bibinfo  {journal} {Contributions to Mineralogy and Petrology}\
  }\textbf {\bibinfo {volume} {157}},\ \bibinfo {pages} {111} (\bibinfo {year}
  {2009})}\BibitemShut {NoStop}%
\bibitem [{\citenamefont {Liu}\ \emph {et~al.}(2013)\citenamefont {Liu},
  \citenamefont {Shan}, \citenamefont {Lian}, \citenamefont {Xie},
  \citenamefont {Yang},\ and\ \citenamefont {He}}]{Liu2013Cat}%
  \BibitemOpen
  \bibfield  {author} {\bibinfo {author} {\bibfnamefont {F.}~\bibnamefont
  {Liu}}, \bibinfo {author} {\bibfnamefont {W.}~\bibnamefont {Shan}}, \bibinfo
  {author} {\bibfnamefont {Z.}~\bibnamefont {Lian}}, \bibinfo {author}
  {\bibfnamefont {L.}~\bibnamefont {Xie}}, \bibinfo {author} {\bibfnamefont
  {W.}~\bibnamefont {Yang}},\ and\ \bibinfo {author} {\bibfnamefont
  {H.}~\bibnamefont {He}},\ }\bibfield  {title} {\bibinfo {title} {{Novel MnWOx
  catalyst with remarkable performance for low temperature NH3-SCR of NOx}},\
  }\href {https://doi.org/10.1039/C3CY00326D} {\bibfield  {journal} {\bibinfo
  {journal} {Catal. Sci. Technol.}\ }\textbf {\bibinfo {volume} {3}},\ \bibinfo
  {pages} {2699} (\bibinfo {year} {2013})}\BibitemShut {NoStop}%
\bibitem [{\citenamefont {Anspoks}\ and\ \citenamefont
  {Kuzmin}(2011)}]{ANSPOKS2011}%
  \BibitemOpen
  \bibfield  {author} {\bibinfo {author} {\bibfnamefont {A.}~\bibnamefont
  {Anspoks}}\ and\ \bibinfo {author} {\bibfnamefont {A.}~\bibnamefont
  {Kuzmin}},\ }\bibfield  {title} {\bibinfo {title} {{Interpretation of the Ni
  K-edge EXAFS in nanocrystalline nickel oxide using molecular dynamics
  simulations}},\ }\href {https://doi.org/10.1016/j.jnoncrysol.2011.02.030}
  {\bibfield  {journal} {\bibinfo  {journal} {Journal of Non-Crystalline
  Solids}\ }\textbf {\bibinfo {volume} {357}},\ \bibinfo {pages} {2604}
  (\bibinfo {year} {2011})}\BibitemShut {NoStop}%
\bibitem [{\citenamefont {Chen}\ \emph {et~al.}(1995)\citenamefont {Chen},
  \citenamefont {Idzerda}, \citenamefont {Lin}, \citenamefont {Smith},
  \citenamefont {Meigs}, \citenamefont {Chaban}, \citenamefont {Ho},
  \citenamefont {Pellegrin},\ and\ \citenamefont {Sette}}]{Chen1995}%
  \BibitemOpen
  \bibfield  {author} {\bibinfo {author} {\bibfnamefont {C.~T.}\ \bibnamefont
  {Chen}}, \bibinfo {author} {\bibfnamefont {Y.~U.}\ \bibnamefont {Idzerda}},
  \bibinfo {author} {\bibfnamefont {H.-J.}\ \bibnamefont {Lin}}, \bibinfo
  {author} {\bibfnamefont {N.~V.}\ \bibnamefont {Smith}}, \bibinfo {author}
  {\bibfnamefont {G.}~\bibnamefont {Meigs}}, \bibinfo {author} {\bibfnamefont
  {E.}~\bibnamefont {Chaban}}, \bibinfo {author} {\bibfnamefont {G.~H.}\
  \bibnamefont {Ho}}, \bibinfo {author} {\bibfnamefont {E.}~\bibnamefont
  {Pellegrin}},\ and\ \bibinfo {author} {\bibfnamefont {F.}~\bibnamefont
  {Sette}},\ }\bibfield  {title} {\bibinfo {title} {{Experimental Confirmation
  of the X-Ray Magnetic Circular Dichroism Sum Rules for Iron and Cobalt}},\
  }\href {https://doi.org/10.1103/PhysRevLett.75.152} {\bibfield  {journal}
  {\bibinfo  {journal} {Physical Review Letters}\ }\textbf {\bibinfo {volume}
  {75}},\ \bibinfo {pages} {152} (\bibinfo {year} {1995})}\BibitemShut
  {NoStop}%
\bibitem [{\citenamefont {Carra}\ \emph {et~al.}(1993)\citenamefont {Carra},
  \citenamefont {Thole}, \citenamefont {Altarelli},\ and\ \citenamefont
  {Wang}}]{Carra1993}%
  \BibitemOpen
  \bibfield  {author} {\bibinfo {author} {\bibfnamefont {P.}~\bibnamefont
  {Carra}}, \bibinfo {author} {\bibfnamefont {B.~T.}\ \bibnamefont {Thole}},
  \bibinfo {author} {\bibfnamefont {M.}~\bibnamefont {Altarelli}},\ and\
  \bibinfo {author} {\bibfnamefont {X.}~\bibnamefont {Wang}},\ }\bibfield
  {title} {\bibinfo {title} {{X-Ray Circular Dichroism and Local Magnetic
  Fields}},\ }\href
  {https://journals.aps.org/prl/pdf/10.1103/PhysRevLett.70.694} {\bibfield
  {journal} {\bibinfo  {journal} {Phys. Rev. Lett.}\ }\textbf {\bibinfo
  {volume} {70}},\ \bibinfo {pages} {694} (\bibinfo {year} {1993})}\BibitemShut
  {NoStop}%
\bibitem [{\citenamefont {Wende}\ and\ \citenamefont
  {Antoniak}(2010)}]{Wende2010}%
  \BibitemOpen
  \bibfield  {author} {\bibinfo {author} {\bibfnamefont {H.}~\bibnamefont
  {Wende}}\ and\ \bibinfo {author} {\bibfnamefont {C.}~\bibnamefont
  {Antoniak}},\ }\bibinfo {title} {Magnetism and synchrotron radiation}\
  (\bibinfo  {publisher} {Springer, Berlin, Heidelberg},\ \bibinfo {year}
  {2010})\ Chap.\ \bibinfo {chapter} {{X-Ray Magnetic Dichroism}}, pp.\
  \bibinfo {pages} {145--167}\BibitemShut {NoStop}%
\bibitem [{\citenamefont {Piamonteze}\ \emph {et~al.}(2009)\citenamefont
  {Piamonteze}, \citenamefont {Miedema},\ and\ \citenamefont
  {de~Groot}}]{Piamonteze2009PRB}%
  \BibitemOpen
  \bibfield  {author} {\bibinfo {author} {\bibfnamefont {C.}~\bibnamefont
  {Piamonteze}}, \bibinfo {author} {\bibfnamefont {P.}~\bibnamefont
  {Miedema}},\ and\ \bibinfo {author} {\bibfnamefont {F.~M.~F.}\ \bibnamefont
  {de~Groot}},\ }\bibfield  {title} {\bibinfo {title} {Accuracy of the spin sum
  rule in xmcd for the transition-metal $l$ edges from manganese to copper},\
  }\href {https://doi.org/10.1103/PhysRevB.80.184410} {\bibfield  {journal}
  {\bibinfo  {journal} {Phys. Rev. B}\ }\textbf {\bibinfo {volume} {80}},\
  \bibinfo {pages} {184410} (\bibinfo {year} {2009})}\BibitemShut {NoStop}%
\bibitem [{\citenamefont {Yang}\ \emph {et~al.}(2017)\citenamefont {Yang},
  \citenamefont {Feng}, \citenamefont {Spence}, \citenamefont {Al~Hindawi},
  \citenamefont {Latimer}, \citenamefont {Ellis}, \citenamefont {Binns},
  \citenamefont {Peddis}, \citenamefont {Dhesi}, \citenamefont {Zhang},
  \citenamefont {Zhang}, \citenamefont {Trohidou}, \citenamefont {Vasilakaki},
  \citenamefont {Ntallis}, \citenamefont {MacLaren},\ and\ \citenamefont
  {de~Groot}}]{Shengfu2017}%
  \BibitemOpen
  \bibfield  {author} {\bibinfo {author} {\bibfnamefont {S.}~\bibnamefont
  {Yang}}, \bibinfo {author} {\bibfnamefont {C.}~\bibnamefont {Feng}}, \bibinfo
  {author} {\bibfnamefont {D.}~\bibnamefont {Spence}}, \bibinfo {author}
  {\bibfnamefont {A.~M. A.~A.}\ \bibnamefont {Al~Hindawi}}, \bibinfo {author}
  {\bibfnamefont {E.}~\bibnamefont {Latimer}}, \bibinfo {author} {\bibfnamefont
  {A.~M.}\ \bibnamefont {Ellis}}, \bibinfo {author} {\bibfnamefont
  {C.}~\bibnamefont {Binns}}, \bibinfo {author} {\bibfnamefont
  {D.}~\bibnamefont {Peddis}}, \bibinfo {author} {\bibfnamefont {S.~S.}\
  \bibnamefont {Dhesi}}, \bibinfo {author} {\bibfnamefont {L.}~\bibnamefont
  {Zhang}}, \bibinfo {author} {\bibfnamefont {Y.}~\bibnamefont {Zhang}},
  \bibinfo {author} {\bibfnamefont {K.~N.}\ \bibnamefont {Trohidou}}, \bibinfo
  {author} {\bibfnamefont {M.}~\bibnamefont {Vasilakaki}}, \bibinfo {author}
  {\bibfnamefont {N.}~\bibnamefont {Ntallis}}, \bibinfo {author} {\bibfnamefont
  {I.}~\bibnamefont {MacLaren}},\ and\ \bibinfo {author} {\bibfnamefont
  {F.~M.~F.}\ \bibnamefont {de~Groot}},\ }\bibfield  {title} {\bibinfo {title}
  {Robust ferromagnetism of chromium nanoparticles formed in superfluid
  helium},\ }\href {https://doi.org/https://doi.org/10.1002/adma.201604277}
  {\bibfield  {journal} {\bibinfo  {journal} {Advanced Materials}\ }\textbf
  {\bibinfo {volume} {29}},\ \bibinfo {pages} {1604277} (\bibinfo {year}
  {2017})}\BibitemShut {NoStop}%
\bibitem [{\citenamefont {Yang}\ \emph {et~al.}(2020)\citenamefont {Yang},
  \citenamefont {Seong}, \citenamefont {Lee}, \citenamefont {Ghanathe},
  \citenamefont {Kumar}, \citenamefont {Yusuf}, \citenamefont {Kim},\ and\
  \citenamefont {Kang}}]{Yang2020APL}%
  \BibitemOpen
  \bibfield  {author} {\bibinfo {author} {\bibfnamefont {M.~Y.}\ \bibnamefont
  {Yang}}, \bibinfo {author} {\bibfnamefont {S.}~\bibnamefont {Seong}},
  \bibinfo {author} {\bibfnamefont {E.}~\bibnamefont {Lee}}, \bibinfo {author}
  {\bibfnamefont {M.}~\bibnamefont {Ghanathe}}, \bibinfo {author}
  {\bibfnamefont {A.}~\bibnamefont {Kumar}}, \bibinfo {author} {\bibfnamefont
  {S.~M.}\ \bibnamefont {Yusuf}}, \bibinfo {author} {\bibfnamefont
  {Y.}~\bibnamefont {Kim}},\ and\ \bibinfo {author} {\bibfnamefont {J.-S.}\
  \bibnamefont {Kang}},\ }\bibfield  {title} {\bibinfo {title} {{Electronic
  structures and magnetization reversal in Li 0.5 FeCr 1.5 O 4}},\ }\href
  {https://doi.org/10.1063/5.0007411} {\bibfield  {journal} {\bibinfo
  {journal} {Applied Physics Letters}\ }\textbf {\bibinfo {volume} {116}},\
  \bibinfo {pages} {252401} (\bibinfo {year} {2020})}\BibitemShut {NoStop}%
\bibitem [{\citenamefont {Garcia-Barriocanal}\ \emph
  {et~al.}(2010)\citenamefont {Garcia-Barriocanal}, \citenamefont {Cezar},
  \citenamefont {Bruno}, \citenamefont {Thakur}, \citenamefont {Brookes},
  \citenamefont {Utfeld}, \citenamefont {Rivera-Calzada}, \citenamefont
  {Giblin}, \citenamefont {Taylor}, \citenamefont {Duffy} \emph
  {et~al.}}]{Garcia2010}%
  \BibitemOpen
  \bibfield  {author} {\bibinfo {author} {\bibfnamefont {J.}~\bibnamefont
  {Garcia-Barriocanal}}, \bibinfo {author} {\bibfnamefont {J.}~\bibnamefont
  {Cezar}}, \bibinfo {author} {\bibfnamefont {F.~Y.}\ \bibnamefont {Bruno}},
  \bibinfo {author} {\bibfnamefont {P.}~\bibnamefont {Thakur}}, \bibinfo
  {author} {\bibfnamefont {N.}~\bibnamefont {Brookes}}, \bibinfo {author}
  {\bibfnamefont {C.}~\bibnamefont {Utfeld}}, \bibinfo {author} {\bibfnamefont
  {A.}~\bibnamefont {Rivera-Calzada}}, \bibinfo {author} {\bibfnamefont
  {S.}~\bibnamefont {Giblin}}, \bibinfo {author} {\bibfnamefont
  {J.}~\bibnamefont {Taylor}}, \bibinfo {author} {\bibfnamefont
  {J.}~\bibnamefont {Duffy}}, \emph {et~al.},\ }\bibfield  {title} {\bibinfo
  {title} {Spin and orbital ti magnetism at lamno 3/srtio 3 interfaces},\
  }\href {https://doi.org/10.1038/ncomms1080} {\bibfield  {journal} {\bibinfo
  {journal} {Nature communications}\ }\textbf {\bibinfo {volume} {1}},\
  \bibinfo {pages} {1} (\bibinfo {year} {2010})}\BibitemShut {NoStop}%
\bibitem [{\citenamefont {Daff{\'{e}}}\ \emph {et~al.}(2018)\citenamefont
  {Daff{\'{e}}}, \citenamefont {Choueikani}, \citenamefont {Neveu},
  \citenamefont {Arrio}, \citenamefont {Juhin}, \citenamefont {Ohresser},
  \citenamefont {Dupuis},\ and\ \citenamefont {Sainctavit}}]{Daffe2018}%
  \BibitemOpen
  \bibfield  {author} {\bibinfo {author} {\bibfnamefont {N.}~\bibnamefont
  {Daff{\'{e}}}}, \bibinfo {author} {\bibfnamefont {F.}~\bibnamefont
  {Choueikani}}, \bibinfo {author} {\bibfnamefont {S.}~\bibnamefont {Neveu}},
  \bibinfo {author} {\bibfnamefont {M.-A.}\ \bibnamefont {Arrio}}, \bibinfo
  {author} {\bibfnamefont {A.}~\bibnamefont {Juhin}}, \bibinfo {author}
  {\bibfnamefont {P.}~\bibnamefont {Ohresser}}, \bibinfo {author}
  {\bibfnamefont {V.}~\bibnamefont {Dupuis}},\ and\ \bibinfo {author}
  {\bibfnamefont {P.}~\bibnamefont {Sainctavit}},\ }\bibfield  {title}
  {\bibinfo {title} {{Magnetic anisotropies and cationic distribution in
  CoFe2O4 nanoparticles prepared by co-precipitation route: Influence of
  particle size and stoichiometry}},\ }\href
  {https://doi.org/10.1016/j.jmmm.2018.03.041} {\bibfield  {journal} {\bibinfo
  {journal} {Journal of Magnetism and Magnetic Materials}\ }\textbf {\bibinfo
  {volume} {460}},\ \bibinfo {pages} {243} (\bibinfo {year}
  {2018})}\BibitemShut {NoStop}%
\bibitem [{\citenamefont {Vinai}\ \emph {et~al.}(2015)\citenamefont {Vinai},
  \citenamefont {Khare}, \citenamefont {Rana}, \citenamefont {{Di Gennaro}},
  \citenamefont {Gobaut}, \citenamefont {Moroni}, \citenamefont {Petrov},
  \citenamefont {{Scotti di Uccio}}, \citenamefont {Rossi}, \citenamefont
  {{Miletto Granozio}}, \citenamefont {Panaccione},\ and\ \citenamefont
  {Torelli}}]{Vinai2015}%
  \BibitemOpen
  \bibfield  {author} {\bibinfo {author} {\bibfnamefont {G.}~\bibnamefont
  {Vinai}}, \bibinfo {author} {\bibfnamefont {A.}~\bibnamefont {Khare}},
  \bibinfo {author} {\bibfnamefont {D.~S.}\ \bibnamefont {Rana}}, \bibinfo
  {author} {\bibfnamefont {E.}~\bibnamefont {{Di Gennaro}}}, \bibinfo {author}
  {\bibfnamefont {B.}~\bibnamefont {Gobaut}}, \bibinfo {author} {\bibfnamefont
  {R.}~\bibnamefont {Moroni}}, \bibinfo {author} {\bibfnamefont {A.~Y.}\
  \bibnamefont {Petrov}}, \bibinfo {author} {\bibfnamefont {U.}~\bibnamefont
  {{Scotti di Uccio}}}, \bibinfo {author} {\bibfnamefont {G.}~\bibnamefont
  {Rossi}}, \bibinfo {author} {\bibfnamefont {F.}~\bibnamefont {{Miletto
  Granozio}}}, \bibinfo {author} {\bibfnamefont {G.}~\bibnamefont
  {Panaccione}},\ and\ \bibinfo {author} {\bibfnamefont {P.}~\bibnamefont
  {Torelli}},\ }\bibfield  {title} {\bibinfo {title} {{Unraveling the magnetic
  properties of BiFe 0.5 Cr 0.5 O 3 thin films}},\ }\href
  {https://doi.org/10.1063/1.4935618} {\bibfield  {journal} {\bibinfo
  {journal} {APL Materials}\ }\textbf {\bibinfo {volume} {3}},\ \bibinfo
  {pages} {116107} (\bibinfo {year} {2015})}\BibitemShut {NoStop}%
\bibitem [{\citenamefont {Pattrick}\ \emph {et~al.}(2002)\citenamefont
  {Pattrick}, \citenamefont {Van Der~Laan}, \citenamefont {Henderson},
  \citenamefont {Kuiper}, \citenamefont {Dudzik},\ and\ \citenamefont
  {Vaughan}}]{Pattrick2002}%
  \BibitemOpen
  \bibfield  {author} {\bibinfo {author} {\bibfnamefont {R.~A.}\ \bibnamefont
  {Pattrick}}, \bibinfo {author} {\bibfnamefont {G.}~\bibnamefont {Van
  Der~Laan}}, \bibinfo {author} {\bibfnamefont {C.~M.~B.}\ \bibnamefont
  {Henderson}}, \bibinfo {author} {\bibfnamefont {P.}~\bibnamefont {Kuiper}},
  \bibinfo {author} {\bibfnamefont {E.}~\bibnamefont {Dudzik}},\ and\ \bibinfo
  {author} {\bibfnamefont {D.~J.}\ \bibnamefont {Vaughan}},\ }\bibfield
  {title} {\bibinfo {title} {Cation site occupancy in spinel ferrites studied
  by x-ray magnetic circular dichroism: developing a method for
  mineralogists},\ }\href {https://doi.org/10.1127/0935-1221/2002/0014-1095}
  {\bibfield  {journal} {\bibinfo  {journal} {European Journal of Mineralogy}\
  }\textbf {\bibinfo {volume} {14}},\ \bibinfo {pages} {1095} (\bibinfo {year}
  {2002})}\BibitemShut {NoStop}%
\bibitem [{\citenamefont {Gilbert}\ \emph {et~al.}(2003)\citenamefont
  {Gilbert}, \citenamefont {Frazer}, \citenamefont {Belz}, \citenamefont
  {Conrad}, \citenamefont {Nealson}, \citenamefont {Haskel}, \citenamefont
  {Lang}, \citenamefont {Srajer},\ and\ \citenamefont {{De
  Stasio}}}]{Gilbert2003}%
  \BibitemOpen
  \bibfield  {author} {\bibinfo {author} {\bibfnamefont {B.}~\bibnamefont
  {Gilbert}}, \bibinfo {author} {\bibfnamefont {B.~H.}\ \bibnamefont {Frazer}},
  \bibinfo {author} {\bibfnamefont {A.}~\bibnamefont {Belz}}, \bibinfo {author}
  {\bibfnamefont {P.~G.}\ \bibnamefont {Conrad}}, \bibinfo {author}
  {\bibfnamefont {K.~H.}\ \bibnamefont {Nealson}}, \bibinfo {author}
  {\bibfnamefont {D.}~\bibnamefont {Haskel}}, \bibinfo {author} {\bibfnamefont
  {J.~C.}\ \bibnamefont {Lang}}, \bibinfo {author} {\bibfnamefont
  {G.}~\bibnamefont {Srajer}},\ and\ \bibinfo {author} {\bibfnamefont
  {G.}~\bibnamefont {{De Stasio}}},\ }\bibfield  {title} {\bibinfo {title}
  {{Multiple Scattering Calculations of Bonding and X-ray Absorption
  Spectroscopy of Manganese Oxides}},\ }\href
  {https://doi.org/10.1021/jp021493s} {\bibfield  {journal} {\bibinfo
  {journal} {The Journal of Physical Chemistry A}\ }\textbf {\bibinfo {volume}
  {107}},\ \bibinfo {pages} {2839} (\bibinfo {year} {2003})}\BibitemShut
  {NoStop}%
\bibitem [{\citenamefont {Kang}\ \emph {et~al.}(2008)\citenamefont {Kang},
  \citenamefont {Kim}, \citenamefont {Lee}, \citenamefont {Kim}, \citenamefont
  {Kim}, \citenamefont {Shim}, \citenamefont {Lee}, \citenamefont {Lee},
  \citenamefont {Kim}, \citenamefont {Kim},\ and\ \citenamefont
  {Min}}]{Kang2008}%
  \BibitemOpen
  \bibfield  {author} {\bibinfo {author} {\bibfnamefont {J.-S.}\ \bibnamefont
  {Kang}}, \bibinfo {author} {\bibfnamefont {G.}~\bibnamefont {Kim}}, \bibinfo
  {author} {\bibfnamefont {H.~J.}\ \bibnamefont {Lee}}, \bibinfo {author}
  {\bibfnamefont {D.~H.}\ \bibnamefont {Kim}}, \bibinfo {author} {\bibfnamefont
  {H.~S.}\ \bibnamefont {Kim}}, \bibinfo {author} {\bibfnamefont {J.~H.}\
  \bibnamefont {Shim}}, \bibinfo {author} {\bibfnamefont {S.}~\bibnamefont
  {Lee}}, \bibinfo {author} {\bibfnamefont {H.}~\bibnamefont {Lee}}, \bibinfo
  {author} {\bibfnamefont {J.-Y.}\ \bibnamefont {Kim}}, \bibinfo {author}
  {\bibfnamefont {B.~H.}\ \bibnamefont {Kim}},\ and\ \bibinfo {author}
  {\bibfnamefont {B.~I.}\ \bibnamefont {Min}},\ }\bibfield  {title} {\bibinfo
  {title} {Soft x-ray absorption spectroscopy and magnetic circular dichroism
  study of the valence and spin states in spinel
  $\mathrm{Mn}{\mathrm{fe}}_{2}{\mathrm{o}}_{4}$},\ }\href
  {https://doi.org/10.1103/PhysRevB.77.035121} {\bibfield  {journal} {\bibinfo
  {journal} {Phys. Rev. B}\ }\textbf {\bibinfo {volume} {77}},\ \bibinfo
  {pages} {035121} (\bibinfo {year} {2008})}\BibitemShut {NoStop}%
\bibitem [{\citenamefont {Pollert}(2000)}]{POLLERT2000661}%
  \BibitemOpen
  \bibfield  {author} {\bibinfo {author} {\bibfnamefont {E.}~\bibnamefont
  {Pollert}},\ }\bibfield  {title} {\bibinfo {title} {Influence of mn3+ ions on
  ordering in magnetic oxides},\ }\href
  {https://doi.org/https://doi.org/10.1016/S1466-6049(00)00066-0} {\bibfield
  {journal} {\bibinfo  {journal} {International Journal of Inorganic
  Materials}\ }\textbf {\bibinfo {volume} {2}},\ \bibinfo {pages} {661}
  (\bibinfo {year} {2000})},\ \bibinfo {note} {dedicated to Prof Raveau on the
  occasion of his 60th Birthday}\BibitemShut {NoStop}%
\bibitem [{\citenamefont {Ma}\ \emph {et~al.}(2020)\citenamefont {Ma},
  \citenamefont {Molokeev}, \citenamefont {Zhu}, \citenamefont {Zhao},
  \citenamefont {Han}, \citenamefont {Wu}, \citenamefont {Liu}, \citenamefont
  {Tyson}, \citenamefont {Croft},\ and\ \citenamefont {Li}}]{Ma2019Mn}%
  \BibitemOpen
  \bibfield  {author} {\bibinfo {author} {\bibfnamefont {Y.}~\bibnamefont
  {Ma}}, \bibinfo {author} {\bibfnamefont {M.~S.}\ \bibnamefont {Molokeev}},
  \bibinfo {author} {\bibfnamefont {C.}~\bibnamefont {Zhu}}, \bibinfo {author}
  {\bibfnamefont {S.}~\bibnamefont {Zhao}}, \bibinfo {author} {\bibfnamefont
  {Y.}~\bibnamefont {Han}}, \bibinfo {author} {\bibfnamefont {M.}~\bibnamefont
  {Wu}}, \bibinfo {author} {\bibfnamefont {S.}~\bibnamefont {Liu}}, \bibinfo
  {author} {\bibfnamefont {T.~A.}\ \bibnamefont {Tyson}}, \bibinfo {author}
  {\bibfnamefont {M.}~\bibnamefont {Croft}},\ and\ \bibinfo {author}
  {\bibfnamefont {M.-R.}\ \bibnamefont {Li}},\ }\bibfield  {title} {\bibinfo
  {title} {Magnetic transitions in exotic perovskites stabilized by chemical
  and physical pressure},\ }\href {https://doi.org/10.1039/C9TC06976C}
  {\bibfield  {journal} {\bibinfo  {journal} {J. Mater. Chem. C}\ }\textbf
  {\bibinfo {volume} {8}},\ \bibinfo {pages} {5082} (\bibinfo {year}
  {2020})}\BibitemShut {NoStop}%
\bibitem [{\citenamefont {Kang}\ \emph {et~al.}(2011)\citenamefont {Kang},
  \citenamefont {Kim}, \citenamefont {Hwang}, \citenamefont {Lee},
  \citenamefont {Nozaki}, \citenamefont {Hayashi}, \citenamefont {Kajitani},
  \citenamefont {Park}, \citenamefont {Kim},\ and\ \citenamefont
  {Min}}]{Kang2011APLA}%
  \BibitemOpen
  \bibfield  {author} {\bibinfo {author} {\bibfnamefont {J.-S.}\ \bibnamefont
  {Kang}}, \bibinfo {author} {\bibfnamefont {D.}~\bibnamefont {Kim}}, \bibinfo
  {author} {\bibfnamefont {J.}~\bibnamefont {Hwang}}, \bibinfo {author}
  {\bibfnamefont {E.}~\bibnamefont {Lee}}, \bibinfo {author} {\bibfnamefont
  {T.}~\bibnamefont {Nozaki}}, \bibinfo {author} {\bibfnamefont
  {K.}~\bibnamefont {Hayashi}}, \bibinfo {author} {\bibfnamefont
  {T.}~\bibnamefont {Kajitani}}, \bibinfo {author} {\bibfnamefont {B.-G.}\
  \bibnamefont {Park}}, \bibinfo {author} {\bibfnamefont {J.-Y.}\ \bibnamefont
  {Kim}},\ and\ \bibinfo {author} {\bibfnamefont {B.}~\bibnamefont {Min}},\
  }\bibfield  {title} {\bibinfo {title} {Phase separation in thermoelectric
  delafossite cufe1- x ni x o2 observed by soft x-ray magnetic circular
  dichroism},\ }\href@noop {} {\bibfield  {journal} {\bibinfo  {journal}
  {Applied Physics Letters}\ }\textbf {\bibinfo {volume} {99}},\ \bibinfo
  {pages} {012108} (\bibinfo {year} {2011})}\BibitemShut {NoStop}%
\bibitem [{\citenamefont {Parida}\ \emph {et~al.}(2020)\citenamefont {Parida},
  \citenamefont {Karati}, \citenamefont {Guruvidyathri}, \citenamefont
  {Murty},\ and\ \citenamefont {Markandeyulu}}]{PARIDA2020513}%
  \BibitemOpen
  \bibfield  {author} {\bibinfo {author} {\bibfnamefont {T.}~\bibnamefont
  {Parida}}, \bibinfo {author} {\bibfnamefont {A.}~\bibnamefont {Karati}},
  \bibinfo {author} {\bibfnamefont {K.}~\bibnamefont {Guruvidyathri}}, \bibinfo
  {author} {\bibfnamefont {B.}~\bibnamefont {Murty}},\ and\ \bibinfo {author}
  {\bibfnamefont {G.}~\bibnamefont {Markandeyulu}},\ }\bibfield  {title}
  {\bibinfo {title} {Novel rare-earth and transition metal-based entropy
  stabilized oxides with spinel structure},\ }\href
  {https://doi.org/https://doi.org/10.1016/j.scriptamat.2019.12.027} {\bibfield
   {journal} {\bibinfo  {journal} {Scripta Materialia}\ }\textbf {\bibinfo
  {volume} {178}},\ \bibinfo {pages} {513} (\bibinfo {year}
  {2020})}\BibitemShut {NoStop}%
\bibitem [{\citenamefont {Seko}\ \emph {et~al.}(2010)\citenamefont {Seko},
  \citenamefont {Oba},\ and\ \citenamefont {Tanaka}}]{Seko2010}%
  \BibitemOpen
  \bibfield  {author} {\bibinfo {author} {\bibfnamefont {A.}~\bibnamefont
  {Seko}}, \bibinfo {author} {\bibfnamefont {F.}~\bibnamefont {Oba}},\ and\
  \bibinfo {author} {\bibfnamefont {I.}~\bibnamefont {Tanaka}},\ }\bibfield
  {title} {\bibinfo {title} {{Classification of spinel structures based on
  first-principles cluster expansion analysis}},\ }\href
  {https://doi.org/10.1103/PhysRevB.81.054114} {\bibfield  {journal} {\bibinfo
  {journal} {Phys. Rev. B}\ }\textbf {\bibinfo {volume} {81}},\ \bibinfo
  {pages} {54114} (\bibinfo {year} {2010})}\BibitemShut {NoStop}%
\bibitem [{\citenamefont {Miracle}\ and\ \citenamefont
  {Senkov}(2017)}]{Miracle2017}%
  \BibitemOpen
  \bibfield  {author} {\bibinfo {author} {\bibfnamefont {D.}~\bibnamefont
  {Miracle}}\ and\ \bibinfo {author} {\bibfnamefont {O.}~\bibnamefont
  {Senkov}},\ }\bibfield  {title} {\bibinfo {title} {{A critical review of high
  entropy alloys and related concepts}},\ }\href
  {https://doi.org/10.1016/J.ACTAMAT.2016.08.081} {\bibfield  {journal}
  {\bibinfo  {journal} {Acta Mater.}\ }\textbf {\bibinfo {volume} {122}},\
  \bibinfo {pages} {448} (\bibinfo {year} {2017})},\ \Eprint
  {https://arxiv.org/abs/arXiv:1507.02142v2} {arXiv:arXiv:1507.02142v2}
  \BibitemShut {NoStop}%
\bibitem [{\citenamefont {Dippo}\ and\ \citenamefont
  {Vecchio}(2021)}]{DIPPO2021113974}%
  \BibitemOpen
  \bibfield  {author} {\bibinfo {author} {\bibfnamefont {O.~F.}\ \bibnamefont
  {Dippo}}\ and\ \bibinfo {author} {\bibfnamefont {K.~S.}\ \bibnamefont
  {Vecchio}},\ }\bibfield  {title} {\bibinfo {title} {A universal
  configurational entropy metric for high-entropy materials},\ }\href
  {https://doi.org/https://doi.org/10.1016/j.scriptamat.2021.113974} {\bibfield
   {journal} {\bibinfo  {journal} {Scripta Materialia}\ }\textbf {\bibinfo
  {volume} {201}},\ \bibinfo {pages} {113974} (\bibinfo {year}
  {2021})}\BibitemShut {NoStop}%
\bibitem [{\citenamefont {Rozenberg}\ \emph {et~al.}(2007)\citenamefont
  {Rozenberg}, \citenamefont {Amiel}, \citenamefont {Xu}, \citenamefont
  {Pasternak}, \citenamefont {Jeanloz}, \citenamefont {Hanfland},\ and\
  \citenamefont {Taylor}}]{Rozenberg2007}%
  \BibitemOpen
  \bibfield  {author} {\bibinfo {author} {\bibfnamefont {G.~K.}\ \bibnamefont
  {Rozenberg}}, \bibinfo {author} {\bibfnamefont {Y.}~\bibnamefont {Amiel}},
  \bibinfo {author} {\bibfnamefont {W.~M.}\ \bibnamefont {Xu}}, \bibinfo
  {author} {\bibfnamefont {M.~P.}\ \bibnamefont {Pasternak}}, \bibinfo {author}
  {\bibfnamefont {R.}~\bibnamefont {Jeanloz}}, \bibinfo {author} {\bibfnamefont
  {M.}~\bibnamefont {Hanfland}},\ and\ \bibinfo {author} {\bibfnamefont
  {R.~D.}\ \bibnamefont {Taylor}},\ }\bibfield  {title} {\bibinfo {title}
  {Structural characterization of temperature- and pressure-induced
  inverse$\ensuremath{\leftrightarrow}$normal spinel transformation in
  magnetite},\ }\href {https://doi.org/10.1103/PhysRevB.75.020102} {\bibfield
  {journal} {\bibinfo  {journal} {Phys. Rev. B}\ }\textbf {\bibinfo {volume}
  {75}},\ \bibinfo {pages} {020102} (\bibinfo {year} {2007})}\BibitemShut
  {NoStop}%
\bibitem [{\citenamefont {Zhao}\ \emph {et~al.}(2020)\citenamefont {Zhao},
  \citenamefont {Ding}, \citenamefont {Lu}, \citenamefont {Chen},\ and\
  \citenamefont {Hu}}]{Zhao2019HEO-Na_Lay}%
  \BibitemOpen
  \bibfield  {author} {\bibinfo {author} {\bibfnamefont {C.}~\bibnamefont
  {Zhao}}, \bibinfo {author} {\bibfnamefont {F.}~\bibnamefont {Ding}}, \bibinfo
  {author} {\bibfnamefont {Y.}~\bibnamefont {Lu}}, \bibinfo {author}
  {\bibfnamefont {L.}~\bibnamefont {Chen}},\ and\ \bibinfo {author}
  {\bibfnamefont {Y.-S.}\ \bibnamefont {Hu}},\ }\bibfield  {title} {\bibinfo
  {title} {High-entropy layered oxide cathodes for sodium-ion batteries},\
  }\href {https://doi.org/https://doi.org/10.1002/anie.201912171} {\bibfield
  {journal} {\bibinfo  {journal} {Angewandte Chemie International Edition}\
  }\textbf {\bibinfo {volume} {59}},\ \bibinfo {pages} {264} (\bibinfo {year}
  {2020})}\BibitemShut {NoStop}%
\bibitem [{\citenamefont {Wang}\ \emph
  {et~al.}(2020{\natexlab{b}})\citenamefont {Wang}, \citenamefont {Cui},
  \citenamefont {Wang}, \citenamefont {Wang}, \citenamefont {Huang},
  \citenamefont {Stenzel}, \citenamefont {Sarkar}, \citenamefont {Azmi},
  \citenamefont {Bergfeldt}, \citenamefont {Bhattacharya} \emph
  {et~al.}}]{J_Wang2020}%
  \BibitemOpen
  \bibfield  {author} {\bibinfo {author} {\bibfnamefont {J.}~\bibnamefont
  {Wang}}, \bibinfo {author} {\bibfnamefont {Y.}~\bibnamefont {Cui}}, \bibinfo
  {author} {\bibfnamefont {Q.}~\bibnamefont {Wang}}, \bibinfo {author}
  {\bibfnamefont {K.}~\bibnamefont {Wang}}, \bibinfo {author} {\bibfnamefont
  {X.}~\bibnamefont {Huang}}, \bibinfo {author} {\bibfnamefont
  {D.}~\bibnamefont {Stenzel}}, \bibinfo {author} {\bibfnamefont
  {A.}~\bibnamefont {Sarkar}}, \bibinfo {author} {\bibfnamefont
  {R.}~\bibnamefont {Azmi}}, \bibinfo {author} {\bibfnamefont {T.}~\bibnamefont
  {Bergfeldt}}, \bibinfo {author} {\bibfnamefont {S.~S.}\ \bibnamefont
  {Bhattacharya}}, \emph {et~al.},\ }\bibfield  {title} {\bibinfo {title}
  {Lithium containing layered high entropy oxide structures},\ }\href
  {https://doi.org/10.1038/s41598-020-75134-1} {\bibfield  {journal} {\bibinfo
  {journal} {Scientific reports}\ }\textbf {\bibinfo {volume} {10}},\ \bibinfo
  {pages} {1} (\bibinfo {year} {2020}{\natexlab{b}})}\BibitemShut {NoStop}%
\bibitem [{\citenamefont {Kim}\ \emph {et~al.}(2021)\citenamefont {Kim},
  \citenamefont {Park}, \citenamefont {Conlin}, \citenamefont {Ashburn},
  \citenamefont {Cho}, \citenamefont {Yu}, \citenamefont {Shapiro},
  \citenamefont {Maglia}, \citenamefont {Kim}, \citenamefont {Lamp},
  \citenamefont {Yoon},\ and\ \citenamefont {Sun}}]{KimEES2021}%
  \BibitemOpen
  \bibfield  {author} {\bibinfo {author} {\bibfnamefont {U.-H.}\ \bibnamefont
  {Kim}}, \bibinfo {author} {\bibfnamefont {G.-T.}\ \bibnamefont {Park}},
  \bibinfo {author} {\bibfnamefont {P.}~\bibnamefont {Conlin}}, \bibinfo
  {author} {\bibfnamefont {N.}~\bibnamefont {Ashburn}}, \bibinfo {author}
  {\bibfnamefont {K.}~\bibnamefont {Cho}}, \bibinfo {author} {\bibfnamefont
  {Y.-S.}\ \bibnamefont {Yu}}, \bibinfo {author} {\bibfnamefont {D.~A.}\
  \bibnamefont {Shapiro}}, \bibinfo {author} {\bibfnamefont {F.}~\bibnamefont
  {Maglia}}, \bibinfo {author} {\bibfnamefont {S.-J.}\ \bibnamefont {Kim}},
  \bibinfo {author} {\bibfnamefont {P.}~\bibnamefont {Lamp}}, \bibinfo {author}
  {\bibfnamefont {C.~S.}\ \bibnamefont {Yoon}},\ and\ \bibinfo {author}
  {\bibfnamefont {Y.-K.}\ \bibnamefont {Sun}},\ }\bibfield  {title} {\bibinfo
  {title} {Cation ordered ni-rich layered cathode for ultra-long battery
  life},\ }\href {https://doi.org/10.1039/D0EE03774E} {\bibfield  {journal}
  {\bibinfo  {journal} {Energy Environ. Sci.}\ }\textbf {\bibinfo {volume}
  {14}},\ \bibinfo {pages} {1573} (\bibinfo {year} {2021})}\BibitemShut
  {NoStop}%
\end{thebibliography}
%

\end{document}